\documentclass[aps,twocolumn,showpacs,pra,superscriptaddress,preprintnumbers,amsmath,amssymb]{revtex4-2}
\usepackage[utf8]{inputenc}
\usepackage[T1]{fontenc}
\usepackage{graphicx}
\usepackage{dcolumn}
\usepackage{appendix}
\usepackage{bm}
\usepackage[dvipsnames]{xcolor}
\usepackage{txfonts}

\usepackage{subfigure}
\usepackage{bbm}
\usepackage{enumerate}

\usepackage[
  bookmarks=true,
  colorlinks,
  linkcolor=black,
  urlcolor=black,
  citecolor=black,
  plainpages=false,
  pdfpagelabels,
  final,
  breaklinks=true
]{hyperref}

\usepackage{titlesec}

\usepackage{array}

\usepackage{physics}

\begin{document}
\allowdisplaybreaks

\title{Light-matter entanglement after above-threshold ionization processes in atoms}

\author{J.~Rivera-Dean}
\email{javier.rivera@icfo.eu}
\affiliation{ICFO -- Institut de Ciencies Fotoniques, The Barcelona Institute of Science and Technology, 08860 Castelldefels (Barcelona)}

\author{P. Stammer}
\affiliation{ICFO -- Institut de Ciencies Fotoniques, The Barcelona Institute of Science and Technology, 08860 Castelldefels (Barcelona)}

\author{A. S. Maxwell}
\affiliation{Department of Physics and Astronomy, Aarhus University, DK-8000 Aarhus C, Denmark}

\author{Th. Lamprou}
\affiliation{Foundation for Research and Technology-Hellas, Institute of Electronic Structure \& Laser, GR-70013 Heraklion (Crete), Greece}
\affiliation{Department of Physics, University of Crete, P.O. Box 2208, GR-70013 Heraklion (Crete), Greece}

\author{P. Tzallas}
\affiliation{Foundation for Research and Technology-Hellas, Institute of Electronic Structure \& Laser, GR-70013 Heraklion (Crete), Greece}
\affiliation{ELI-ALPS, ELI-Hu Non-Profit Ltd., Dugonics tér 13, H-6720 Szeged, Hungary}

\author{M. Lewenstein}
\affiliation{ICFO -- Institut de Ciencies Fotoniques, The Barcelona Institute of Science and Technology, 08860 Castelldefels (Barcelona)}
\affiliation{ICREA, Pg. Llu\'{\i}s Companys 23, 08010 Barcelona, Spain}

\author{M. F. Ciappina}
\email{marcelo.ciappina@gtiit.edu.cn}
\affiliation{Physics Program, Guangdong Technion--Israel Institute of Technology, Shantou, Guangdong 515063, China}
\affiliation{Technion -- Israel Institute of Technology, Haifa, 32000, Israel}
\affiliation{Guangdong Provincial Key Laboratory of Materials and Technologies for Energy Conversion, Guangdong Technion – Israel Institute of Technology, Shantou, Guangdong 515063, China}

\date{\today}

\begin{abstract}
    Light-matter entanglement plays a fundamental role in many applications of quantum information science. Thus, finding processes where it can be observed is an important task. Here, we address this matter by theoretically investigating the entanglement between light and electrons generated in above-threshold ionization (ATI) process. The study is based on the back-action of the ATI process on the quantum optical state of the system, and its dependence on the kinetic energy and direction of the emitted photoelectrons. Taking into account the dynamics of the process, we demonstrate the creation of hybrid entangled states. The amount of entanglement has been studied in terms of the entropy of entanglement. 
    Additionally, we use the Wigner function of the driving field mode to motivate the entanglement characterization when considering electrons propagating in opposite directions.
\end{abstract}
\maketitle

\section{INTRODUCTION}

Above-threshold ionization (ATI), firstly observed in 1979 \cite{agostini_free-free_1979}, has been one of the most studied processes in strong laser-field physics (see c.f. \cite{milosevic_above-threshold_2006,delone_multiphoton_2000,agostini_chapter_2012} and references therein). In ATI, a bound electron is released from the parent system due to its interaction with an intense electromagnetic field (typically of intensity $I \gtrsim 10^{13}$ W/cm$^2$). This interaction leads to the production of photoelectrons with kinetic momentum pointing towards the direction of the driving field polarization \cite{milosevic_above-threshold_2006}, and with kinetic energies corresponding to photon absorption above the ionization threshold of the system. The final kinetic energy, as well as the momentum distribution of these electrons, is determined by the ionization time within a cycle of the driving field, and on the sign of the laser electric field at that moment. Depending on whether the electron gets ionized at a maxima or a minima of the applied field, it is driven in opposite directions along the polarization direction, which are denoted as forward and backward photoelectrons respectively.

From a theoretical perspective, the numerical analysis of ATI processes can become ponderous, due to the different time, length and energy scales that are involved in the problem (c.f. \cite{AnneML2}). Instead, one can rely on the strong-field approximation (SFA), which introduces some approximations based on the highly intense nature of the driving field, and greatly simplifies the analysis of the time-dependent Schrödinger equation (TDSE). This approach, which was initially proposed in \cite{Keldysh}, has been widely used in the literature for the study of ATI \cite{lewenstein_rings_1995}, and for a wide plethora of strong-field processes such as high-harmonic generation (HHG) \cite{lewenstein_theory_1994}, where high harmonics of the driving frequency are generated upon the recombination of the freed-electron with its parent ion. Moreover, apart from being very successful in describing the experimental observations, the SFA provides an extension to the classical \emph{three-step} model or \emph{simple man's} model \cite{corkum_plasma_1993,krause_high-order_1992,Kulander1993}, in terms of quantum trajectories that are followed by the laser-ionized electron \cite{lewenstein_theory_1994,salieres_feynmans_2001,olga_simpleman}.

Most of the theoretical analysis that has been done so far considers a semiclassical framework, where the interacting system is treated quantum mechanically, while the field is described classically \cite{amini_symphony_2019}. However, in the recent years, particular attention has been paid to the interplay between quantum optics and strong-field physics. From an experimental perspective, the first measurement of quantum optical signatures in strong-field physics was obtained upon conditioning on HHG processes, in particular when looking at the photon number statistics of the driving field after its interaction with the atomic medium \cite{gonoskov_quantum_2016,tsatrafyllis_high-order_2017}. Later on, other experiments have studied the photon counting statistics of the harmonics generated in HHG \cite{fuchs_photon_2022}. From a theoretical perspective, different analysis have been proposed that are related to the study of the particular effects that arise when taking into account the quantum nature of the field in this strongly driven laser-matter interactions
\cite{gorlach_quantum-optical_2020,rivera_lightmatter_2020,varro_quantum_2021,foldi_describing_2021,gombkoto_quantum-optical_2021,rivera-dean_strong_2022,Even_ATTO}. In the intersection between theory and experiment, we highlight the set of works
\cite{lewenstein_generation_2021,rivera-dean_strong_2022,stammer_quantum_2022,stammer_high_2022,Paris_ATTO,Maciej_ATTO}, which show the generation of highly non-classical states of light, in the form of coherent state superpositions, when conditioning to HHG and ATI processes. Specifically, and regarding ATI, in ref.~\cite{rivera-dean_strong_2022} it was shown that ATI processes induce a displacement in the quantum optical state of the field, and in ref.~\cite{stammer_quantum_2022} it was found that the generated displacement depends on the final kinetic energy of the photoelectron, as well as on its propagation direction. All these studies have opened the door for the interface between strong-field physics and quantum optics towards applications in quantum information science.

In most of the applications of quantum information science, the presence of entanglement in a quantum state that is shared between two or more parties, plays a fundamental role \cite{nielsen_quantum_2010}. For instance, it is crucial in quantum teleportation protocols \cite{bennett_teleporting_1993,bouwmeester_experimental_1997,boschi_experimental_1998}, and is a necessary, although not sufficient, resource for sharing nonlocal correlations between two or more systems, and which therefore allows them to perform quantum communication in a secure way \cite{gisin_quantum_2007}. In strong laser-field physics, the existence of electron-electron \cite{liu_correlation_1999,christov_phase-dependent_1999,christov_phase-dependent_2000,omiste_effects_2019,maxwell_entanglement_2021}, electron-ion \cite{spanner_coherent_2007,spanner_entanglement_2007,czirjak_emergence_2013,majorosi_quantum_2017,ruberti_quantum_2021,vrakking_control_2021,koll_experimental_2022,Koll_ATTO,Shobeiry_ATTO} and atom-atom \cite{eckart_ultrafast_2021,Eckart_ATTO} entanglement in ultrashort time scales (femtosecond and attosecond regimes) has been studied over the years within a semiclassical framework. However, recently the quantum optical treatment of the electromagnetic field was included to show that, intense laser-atom interactions can lead to the generation of entangled states between the different optical field modes \cite{stammer_high_2022,stammer_theory_2022}.

In this work, we aim to study the light-matter entanglement between photoelectrons generated in ATI processes and the electromagnetic field modes, which gets displaced differently depending on the final kinetic momentum of the electron. We first study the regime of laser parameters for which quantum optical effects are visible at the single-atom level. We then proceed to study the quantum optical properties of the driving electromagnetic field after ATI processes by means of its displacement in phase-space, and the corresponding Wigner function of the respective field state. We further consider a phenomenological treatment of many-atoms to take into account more realistic experimental conditions. Finally, we perform an entanglement characterization for the single-atom case by means of the entropy of entanglement \cite{plenio_introduction_2007,nielsen_quantum_2010}.

The article is organized as follows. In Section~\ref{Sec:Theory:background}, we present the theoretical background, where we also study the effect of the electronic motion on the electromagnetic field modes. In Section~\ref{Sec:Results}, we present our results, where we discuss the regime of laser parameters for which we get non-negligible quantum optical effects over the electromagnetic field modes at the single-atom level. With this, we compute the Wigner function of the quantum optical states, and use it to motivate the entanglement characterization between the field modes and the generated photoelectrons. Finally, we end with the conclusions and a brief outlook in Section~\ref{Sec:Conclusions}.

\section{THEORETICAL BACKGROUND}\label{Sec:Theory:background}
In this section, we describe the theoretical model used in this manuscript for characterizing the final state of the total system after the interaction with the strong-laser field. We study the light-matter interaction in the so-called \emph{length} gauge form, and within the single active electron (SAE) and dipole approximations. More details about how to derive this form of the Hamiltonian starting from the minimal coupling Hamiltonian can be found in \cite{stammer_quantum_2022}.

\subsection{Hamiltonian of the light-matter interaction}

The Hamiltonian characterizing the light-matter interaction within the SAE and dipole approximations is given by
\begin{equation}
	\hat{H} = \hat{H}_\text{at} + \hat{H}_\text{int} + \hat{H}_\text{field},
\end{equation}
where $\hat{H}_\text{at}\equiv \hbar^2\hat{\vb{P}}^2/(2m) + V(\hat{\vb{R}})$ is the atomic Hamiltonian with $m$ the electron's mass and $V(\hat{\vb{R}})$ the atomic potential, $\hat{H}_\text{field} \equiv \sum_{\vb{k},\mu} \hbar \omega_k \hat{a}_{\vb{k},\mu}^\dagger \hat{a}_{\vb{k},\mu}$ is the electromagnetic free-field Hamiltonian with $\hat{a}_{\vb{k},\mu}$ ($\hat{a}^\dagger_{\vb{k},\mu}$) the annihilation (creation) operator acting over the mode with wavevector $\vb{k}$ and polarization $\mu$, and  $\hat{H}_\text{int} \equiv e \hat{\vb{R}}\cdot \hat{\vb{E}}$ is the interaction Hamiltonian within the so-called \emph{length gauge}, with $e$ the absolute value of the electronic charge and $\hat{\vb{E}}$ the electric field operator. In the following, we consider a discrete-mode version for the electric field operator
\begin{equation}
	\hat{\vb{E}} 
		= -i\sum_{\vb{k},\mu}
			\sqrt{\dfrac{\hbar c \abs{\vb{k}}}{2\epsilon_0 V}}
			\boldsymbol{\epsilon}_{\vb{k},\mu} \
				\big(
		\hat{a}^\dagger_{\vb{k},\mu} - \hat{a}_{\vb{k},\mu} 
				\big),
\end{equation}
where $V$ is the quantization volume, $c$ the speed of light and $\epsilon_0$ the vacuum permittivity. Although the above Hamiltonian can be generalized to tackle the interaction with other systems, such as molecules or solids, we restrict ourselves to the case of gases and, for this reason, we consider a linear dispersive relation $\omega_k = c \abs{\vb{k}}$, where $\omega_k$ is the frequency of the field mode.

\subsection{Solving the time-dependent Schrödinger equation}
In the most common strong-field experimental realization, an intense low-frequency laser field, usually in the infrared (IR) spectral region, interacts with an atomic medium which is initially in the ground state $\lvert \text{g}\rangle$. Thus, we describe the initial state of the system by
\begin{equation}\label{Eq:Init:state}
	\ket{\Psi(t=t_0)} 
		= \ket{\text{g}}
			 \bigotimes_{\vb{k},\mu \in  \text{IR}}\ket{\alpha_{\vb{k},\mu}} 
			 \bigotimes_{\vb{k},\mu \in \text{HH}} \ket{0_{\vb{k},\mu}},
\end{equation}
where we denote the IR modes belonging to the laser pulse, and that are initially in a coherent state of amplitude $\alpha_{\vb{k},\mu}$, with the label IR. Note that the amplitude $\alpha_{\vb{k},\mu}$ is a function of the mode $\vb{k}$ and polarization $\mu$, and hence describes the spectral behavior of the employed laser pulse~\cite{stammer_quantum_2022}. On the other hand, all the other modes that could be potentially excited by means of strong-field processes, but initially lie in a vacuum state $\lvert 0_{\vb{k},\mu}\rangle$, are denoted with the label HH.

The time-dependent Schrödinger equation describing the dynamics of this system is given by
\begin{equation}
	i \hbar \pdv{\ket{\Psi(t)}}{t}
		= \big(
			\hat{H}_\text{at} + \hat{H}_\text{int} + \hat{H}_{\text{field}}
		  \big) \ket{\Psi(t)},
\end{equation}
and in order to solve it, we: (i) move to the interaction picture with respect to the free-field term $\hat{H}_\text{field}$, such that the electric field becomes time-dependent and given by
\begin{equation}\label{eq:time:dep:field}
	\hat{\vb{E}}(t)
	= -i\sum_{\vb{k},\mu}
		\sqrt{\dfrac{\hbar \omega_k}{2\epsilon_0 V}}
		\boldsymbol{\epsilon}_{\vb{k},\mu} \
			\big( 
				\hat{a}^\dagger_{\vb{k},\mu}e^{i    \omega_k t}
				- \hat{a}_{\vb{k},\mu} e^{-i\omega_k t} 
			\big);
\end{equation}
and (ii) work in the displaced frame of reference with respect to the input IR field, so that the electric field operator splits into a \emph{classical} term $\vb{E}_{\text{cl}}(t)$ describing the mean value of the field
\begin{equation}
    \vb{E}_{\text{cl}}(t)
		= -i\sum_{\vb{k},\mu}
			\sqrt{\dfrac{\hbar \omega_k}{2\epsilon_0 V}}
			\boldsymbol{\epsilon}_{\vb{k},\mu} \
				\big( 
					\alpha^*_{\vb{k},\mu}e^{i\omega_k t}
					- \alpha_{\vb{k},\mu} e^{-i\omega_k t} 
				\big),
\end{equation}
and another term $\hat{\vb{E}}(t)$ describing the quantum fluctuations (see Eq.~\eqref{eq:time:dep:field}). Thus, the Schrödinger equation reads
\begin{equation}\label{schrodinger_1}
	i \hbar \pdv{\ket{\psi(t)}}{t}
		= \big(
			\hat{H}_\text{at} 
			+ e \hat{\vb{R}}\cdot \vb{E}_{\text{cl}}(t)
			+ e \hat{\vb{R}}\cdot \hat{\vb{E}}(t)
		  \big) \ket{\psi(t)},
\end{equation}
where, under the considered transformations, the initial state of the above equation reads $\ket{\psi(t=t_0)} = \lvert\text{g}\rangle\bigotimes_{\vb{k},\mu} \lvert 0_{\vb{k},\mu}\rangle$, which we shall also refer to as $\ket{\psi(t=t_0)} = \lvert\text{g}\rangle\lvert\bar{0}\rangle$, with $\lvert\bar{0}\rangle$ representing the vacuum state in all the modes.

In the spirit of the semiclassical description of strongly driven laser-matter interactions with low frequency laser sources~\cite{lewenstein_theory_1994}, we solve the Schrödinger equation in Eq.~\eqref{schrodinger_1} by considering the following ansatz,
\begin{equation}\label{Eq:ansatz}
	\ket{\psi(t)}
		= a(t) \ket{\text{g}} \ket{\Phi_{\text{g}}(t)}
			+ \int \dd^3 v \ b(\vb{v},t) \ket{\vb{v}} \ket{\Phi(\vb{v},t)},
\end{equation}
where $a(t)$ describes the probability amplitude of finding the electron in the ground state at the end of the process, and $b(\vb{v},t)$ describes the probability amplitude of finding the electron in a continuum state $\ket{\vb{v}}$. The ansatz we propose here is based on the \emph{standard} SFA formulation, which considers that the strong laser field does not couple with any bound state apart from the ground state $\lvert\text{g}\rangle$ such that, together with the continuum (scattering) states $\ket{\vb{v}}$, they are the only states contributing to the dynamics \cite{lewenstein_theory_1994,amini_symphony_2019}. Therefore, the ansatz we consider in Eq.~\eqref{Eq:ansatz} represents the most general state one can have in light-matter interactions within the SFA approach.

In order to solve the differential equation in Eq.~\eqref{schrodinger_1} by means of the ansatz shown in Eq.~\eqref{Eq:ansatz}, we introduce some approximations. First, we consider that the depletion of the ground state population is almost negligible, i.e. $|a(t)| \simeq 1$~\footnote{The depletion of the ground state can be easily incorporated using, for instance, the ADK or PPT ionization models.}. Second, we assume that the quantum orbits followed by the electron when ionized are not affected by the quantum optical fluctuations of the applied field. Furthermore, in the present study we neglect rescattering events, which imposes a limit on the range of photoelectron kinetic energy $\mathcal{E}(p)=\hbar^2 p^2/(2m)$, in particular $\mathcal{E}(p)\lesssim 2.5U_p$ where $U_p = e^2E_0^2/(4m\omega_L^2)$ is the ponderomotive energy, i.e. the average kinetic energy of an electron that oscillates embedded in a laser field, $E_0$ being the peak amplitude of the laser electric field and $\omega_L$ the central frequency of the applied field. In this regime, high-order ATI processes (HATI) \cite{milosevic_above-threshold_2006,milosevic__quantum-orbit_2006} do not provide a significant contribution, and direct ionization processes are the dominant. Under these considerations, we find that the quantum state of the system is given by (see Appendix \ref{App:TDSE} for a detailed derivation)
\begin{widetext}
\begin{equation}\label{Eq:Final:total:state}
    \ket{\psi(t)}
		= e^{-\tfrac{i}{\hbar}I_p(t-t_0)}
			\ket{\text{g}}\ket{\bar{0}}
        - \dfrac{i}{\hbar} \int \dd^3 \vb{p} \int^t_{t_0} \dd t'
				e^{-\tfrac{i}{\hbar}S(\vb{p},t,t')}
				\Tilde{D}
				\big(
					\boldsymbol{\delta}(\vb{p},t,t')
				\big)
				\big(\vb{E}_{\text{cl}}(t') + \hat{\vb{E}}(t')\big)
				\cdot \vb{d}
					\Big(
						{\vb{p}+\dfrac{e}{c}\vb{A}(t')}
					\Big)
				\ket{\vb{p}+\dfrac{e}{c}\vb{A}(t)}\ket{\bar{0}},
\end{equation}
\end{widetext}
where $\vb{v} = \vb{p} + (e/c)\vb{A}(t)$ with $\vb{p}$ the canonical momentum, $\vb{d}(\vb{p}+(e/c)\vb{A}(t)) \equiv \langle{\vb{p}+(e/c)\vb{A}(t)}\rvert \hat{\vb{R}}\lvert \text{g}\rangle$,  $S(\vb{p},t,t')$ is the semiclassical action
\begin{equation}\label{Eq:semiclassical:action}
    S(\vb{p},t,t')
        = \int^t_{t'}
        \dd \tau
            \Bigg(
                \dfrac{1}{2m}
                \bigg[
                    \vb{p}
                    +\dfrac{e}{c}\vb{A}(\tau)
                \bigg]^2 + I_p
            \Bigg),
\end{equation}
with $I_p$ the ionization potential of the considered atom, and where
\begin{equation}
    \Tilde{D}
			\big(
				\boldsymbol{\delta}(\vb{p},t,t')
			\big)
		= \prod_{\vb{k},\mu} e^{i\varphi_{\vb{k},\mu}(\vb{p},t)}
			D\big(
				\delta_{\vb{k},\mu}(\vb{p},t,t')
			   \big).
\end{equation}

\begin{figure*}
	\centering
	\includegraphics[width=0.65\textwidth]{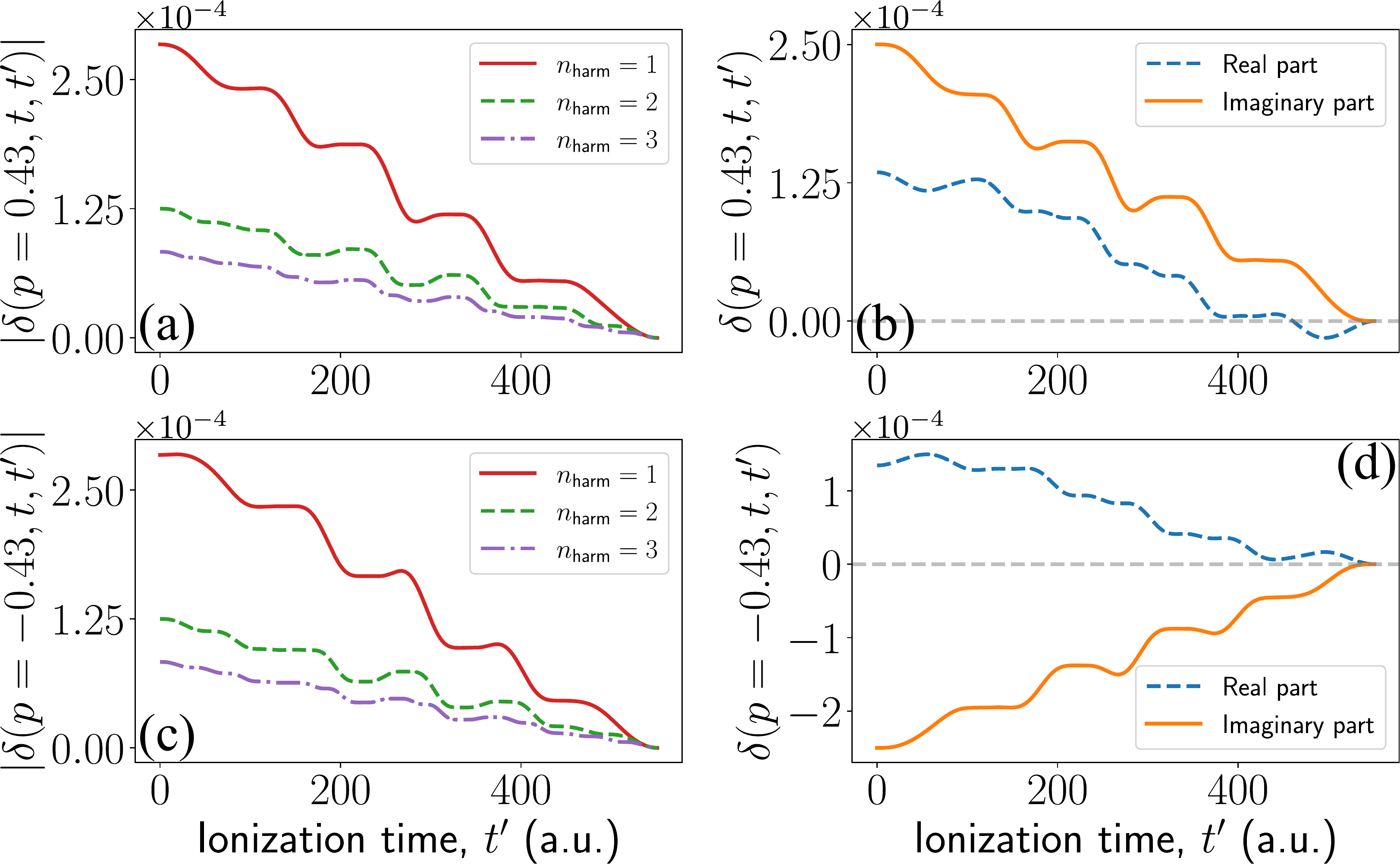}
	\caption{Behavior of $\delta_{\vb{k},\mu}(p,t,t')$. In plots (a) and (c) we show the norm of this quantity for $p=0.43$ a.u. and $p=-0.43$ a.u., respectively. The different curves correspond to distinct frequency modes, namely the red solid curve corresponds to the fundamental mode ($n_\text{harm}=1$), the green dashed curve to the second harmonic mode ($n_\text{harm} = 2$) and the purple dash-dotted curve to the third harmonic mode ($n_\text{harm}=3$). In plots (b) and (d) we show the real (blue dashed curve) and imaginary (orange solid curve) parts of $\delta_{\vb{k},\mu}(\vb{p},t,t')$ when considering the fundamental mode, for $p=0.43$ a.u. and $p=-0.43$ a.u. respectively.}
	\label{Fig:Behaviour:delta}
\end{figure*}

In this last expression, $\varphi_{\vb{k,\mu}}(\vb{p},t)$ is a phase prefactor that arises when solving the quantum optical part of the Schrödinger equation (see for instance \cite{stammer_quantum_2022}), and $\delta_{\vb{k},\mu}(\vb{p},t,t')$ is the Fourier transform of the electronic displacement in the continuum ($\Delta \vb{r}(\vb{p},\tau,t') = \nabla_{\vb{p}}S(\vb{p},t,t')$) from the ionization time $t'$ up to the final time $t$. Both quantities are explicitly given by
\begin{equation}\label{Eq:displacement}
	\delta_{\vb{k},\mu}(\vb{p},t,t')
		= -\dfrac{e}{\hbar} \sqrt{\dfrac{\hbar \omega_k}{2\epsilon_0 V}}
			\int^t_{t'} \dd \tau 
				\Delta\vb{r}(\vb{p},\tau,t')e^{i\omega_k \tau},
\end{equation}
\begin{equation}
    \begin{aligned}
    \varphi_{\vb{k},\mu}(\vb{p},t)
        = \dfrac{e^2}{\hbar^2}
            \dfrac{\hbar \omega_k}{2\epsilon_0 V}
            \int^t_{t'}\dd t_1
            \int^{t_1}_{t'}\dd t_2
            &\Big(
                \boldsymbol{\epsilon}_{\vb{k},\mu}\cdot \Delta \vb{r}(\vb{p},t_1,t')
            \Big)
            \\&\times
              \Big( \boldsymbol{\epsilon}_{\vb{k},\mu}\cdot \Delta \vb{r}(\vb{p},t_2,t')
            \Big)
            \\&\times
                \sin(\omega_k(t_1-t_2)).
    \end{aligned}
\end{equation}

In the context of this work, we are working within the strong-field regime, where the amplitude of the input electric field is on the order of $10^{7}$ V/cm or larger. For this reason, we expect the mean value of the field $\vb{E}_{\text{cl}}(t)$ to dominate over the quantum optical fluctuations $\hat{\vb{E}}(t)$. Consequently, we approximate Eq.~\eqref{Eq:Final:total:state} by
\begin{widetext}
\begin{equation}\label{Eq:ATI:state}
   \begin{aligned}
    \ket{\psi(t)}
		\simeq e^{-\tfrac{i}{\hbar}I_p(t-t_0)}
		&\ket{\text{g}}
			\bigotimes_{\vb{k},\mu\in\text{IR}} \ket{\alpha_{\vb{k},\mu}}
			\bigotimes_{\vb{k},\mu\in\text{HH}} \ket{0_{\vb{k},\mu}}
       \\
			 &\quad
				- \dfrac{i}{\hbar}
				\bigg[
				    \prod_{\vb{k},\mu}D(\alpha_{\vb{k},\mu})
				    \bigg]
				\int \dd^3 \vb{p} \int^t_{t_0} \dd t'
				e^{-\tfrac{i}{\hbar}S(\vb{p},t,t')}
				\vb{E}_{\text{cl}}(t')
				\cdot \vb{d}
					\Big(
						{\vb{p}+\dfrac{e}{c}\vb{A}(t')}
					\Big)
				\ket{\vb{p}+\dfrac{e}{c}\vb{A}(t)}
				\Tilde{D}
				\big(
					\boldsymbol{\delta}(\vb{p},t,t')
				\big)
				\ket{\bar{0}},
	\end{aligned}
\end{equation}
\end{widetext}
where we have further undone the displacement with respect to the input IR field.

Equation~\eqref{Eq:ATI:state} has two contributions: the first one describes the situation in which the electron is completely unaffected by the field, and in consequence the quantum optical state of the field does not experience any change \cite{Comment1}; the second one describes ionization processes, where the electron reaches the continuum at time $t'$ and accelerates. During the latter, the electron motion in the continuum leads to a displacement $\boldsymbol{\delta}(\vb{p},t,t')$ on the state of the electromagnetic field modes. In order to have an idea of how big this quantity is, let us consider the case of a linearly polarized laser field with a sinusoidal squared envelope that has $\omega_L = 0.057$~a.u. (corresponding to a wavelength $\lambda_L\approx800$ nm), 5 cycles of duration (corresponding to $ \Delta t \approx 15 $ fs for this frequency) and an electric field peak amplitude $E_0 = 0.053$~a.u. (corresponding to a laser intensity $I=1\times10^{14}$ W/cm$^2$). Moreover, we consider a 1D model for the atom (we used Hydrogen in the numerics). The typical value of $\lvert \alpha \rvert$ that we have for these fields is in the order of~$10^6$, which allows us to obtain an estimate of $V \sim 10^{14}$~a.u. for the quantization volume. Using these quantities, we show in Fig.~\ref{Fig:Behaviour:delta} (a) and (c) the absolute value of $\delta_{\vb{k},\mu}(p,t,t')$ for the fundamental mode, the second and the third harmonic modes when the final kinetic momentum of the electron (at the end of the pulse) is (a) $p = 0.43$ a.u. and (c) $p=-0.43$ a.u., these two values satisfying $\mathcal{E}(p) < 2.5 U_p$. On the other hand, in (b) and (d) we show the real and imaginary parts of $\delta_{\vb{k},\mu}(p,t,t')$ for the fundamental mode with the two values of momentum shown before. From these figures we see that, at the single atom level, the radiation generated during the electronic oscillation has very small amplitudes ($\sim\!10^{-4}$) and, hence, barely affects the initial coherent state of the field. On the other hand, an interesting feature is that, depending on the direction along which the electron ionizes, the imaginary part of $\delta_{\vb{k},\mu}(p,t,t')$ differs in a minus sign. We see that, the earlier the ionization time is, the bigger the contribution to the input field would be. This is an expected behavior as the electron spends more time in the continuum. Finally, we observe that the effect on the harmonic modes becomes smaller as the harmonic order increases.

\begin{figure}
    \centering
    \includegraphics[width=1\columnwidth]{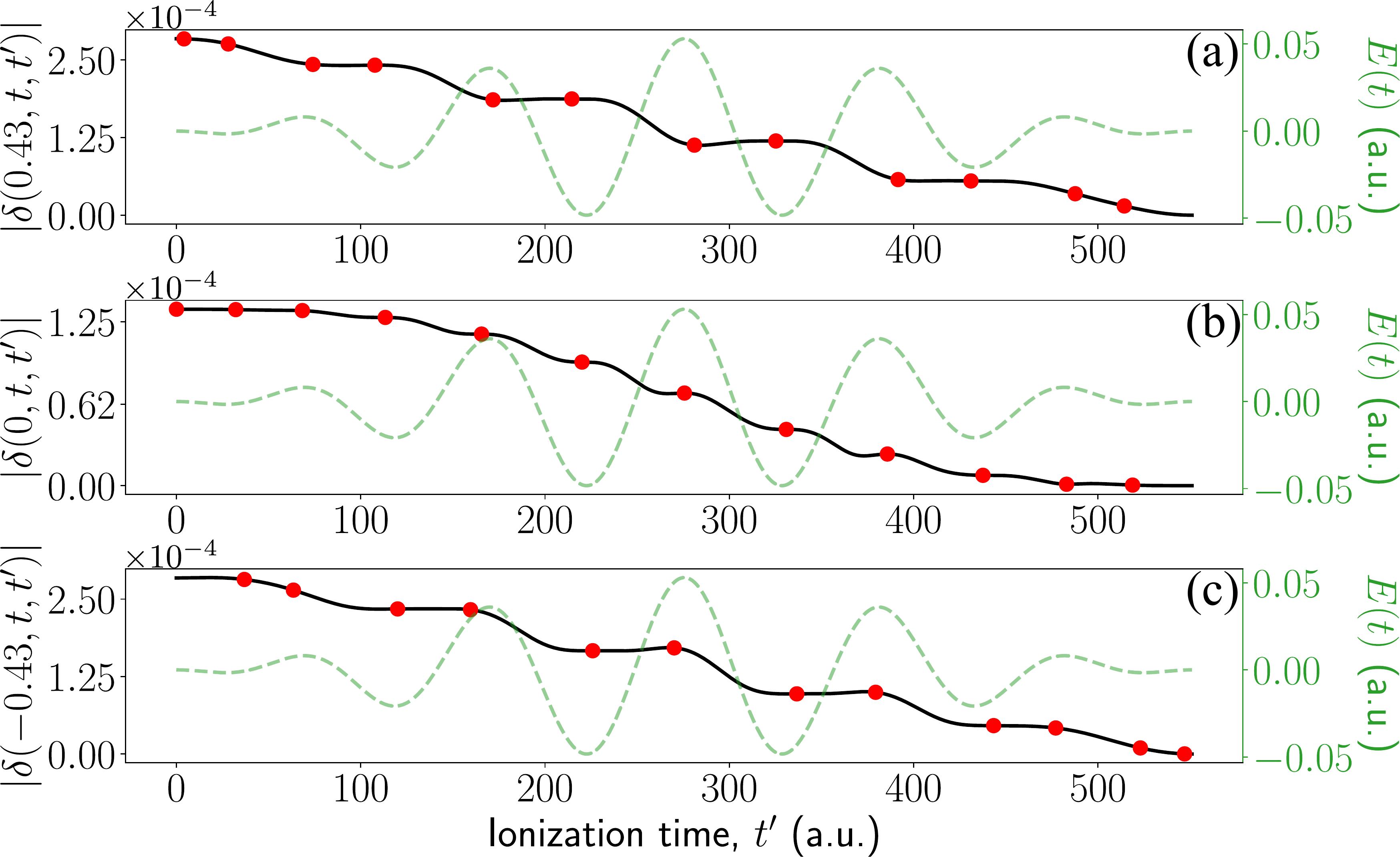}
    \caption{Norm of $\delta_{\vb{k},\mu}(p,t,t')$ (black solid line) for three different values of momentum: (a) $p=0.43$ a.u., (b) $p=0.00$ a.u. and (c) $p=-0.43$ a.u. shown in the different subplots. The red bullet points show the value of the corresponding function evaluated at the real part of the ionization time, which has been computed numerically using the semiclassical equations (see Appendix \ref{App:Ionization:Times}). For these calculations, we have considered a 1D hydrogen system ($I_p = 0.5$ a.u.) driven by a linearly polarized laser field (shown with the green dashed line) with a sinusoidal squared envelope, that has 5 cycles of duration, $\omega_L = 0.057$ a.u. for the central frequency and $E_0=0.053$ a.u. for the field's amplitude.}
    \label{Fig:Delta:and:Ionization}
\end{figure}

In Fig.~\ref{Fig:Delta:and:Ionization} we attempt to relate the behavior of $\lvert\delta_{\vb{k}_L,\mu}(p,t,t')\rvert$ of the fundamental with the characteristics of the electronic motion. More concretely, we consider three different values of $p$ ($p=0.43$ a.u., $p=0.00$ a.u. and $p=-0.43$ a.u. in (a), (b) and (c) respectively), where the real part of the ionization time, computed by solving the saddle-point equation, is shown with the red dots (see Appendix \ref{App:Ionization:Times} for details). As we can see, depending on the outgoing electron's momentum, the value of $\lvert\delta_{\vb{k}_L,\mu}(p,t,t')\rvert$ differs. For instance, for $p=0.00$ a.u. we see that, between each possible ionization time, there is a step in the value of $\lvert\delta_{\vb{k}_L,\mu}(p,t,t')\rvert$, which is not observed at all ionization times for the other two non-zero values of momentum shown in Figs.~\ref{Fig:Delta:and:Ionization}~(a) and (c). Furthermore, for the non-zero values of momentum, depending on the direction along which the outgoing electron propagates we might find that, two consecutive ionization times lead approximately to the same value of $\lvert\delta_{\vb{k}_L,\mu}(p,t,t')\rvert$. Using the two ionization times located at the right and left of $t' = 300$ a.u. in Fig.~\ref{Fig:Delta:and:Ionization}~(a) (the same discussion can be done for Fig.~\ref{Fig:Delta:and:Ionization}~(c)) we see that, if the electron ionizes close to a maximum of the field, its contribution to $\lvert\delta_{\vb{k}_L,\mu}(p,t,t')\rvert$ will be almost the same as if ionizes at the minimum of the field. We note that, for the ionization time placed at the left of $t' = 300$, the field is pointing in the same direction as that of the photoelectron momentum, since we are in a maximum and does not provide any significant contribution to $\lvert\delta_{\vb{k}_L,\mu}(p,t,t')\rvert$ compared to the ionization time place at the right hand side, for which field and momentum are pointing towards different directions. This contrasts with what happens for $p=0.00$ a.u. (Fig.~\ref{Fig:Delta:and:Ionization}~(b)), where for ionization times placed at maximum or minimum values of the field, the value of $|\delta_{\vb{k}_L,\mu}(p,t,t')|$ increases the earlier ionization takes place.


\section{RESULTS}\label{Sec:Results}
\subsection{Dependence of the light field displacement on the electronic motion}\label{Sec:Results:laser:params}
So far, in the plots we have presented in Figs.~\ref{Fig:Behaviour:delta} and \ref{Fig:Delta:and:Ionization}, we have worked with laser parameters for which the displacement $\delta_{\vb{k}_L,\mu}(p,t,t')$ generated over the quantum optical state is negligible ($\lvert\delta_{\vb{k}_L,\mu}(p,t,t')\rvert \sim 10^{-4}$). These effects could be naturally enhanced, for instance, by considering a many-body picture where more than one atom participates in the ATI process. In this subsection, we instead restrict ourselves to single-atom dynamics, and seek for a regime of parameters for which the generated displacement $\delta_{\vb{k}_L,\mu}(p,t,t')$ becomes non-negligible.

\begin{figure*}
    \centering
    \includegraphics[width=0.65\textwidth]{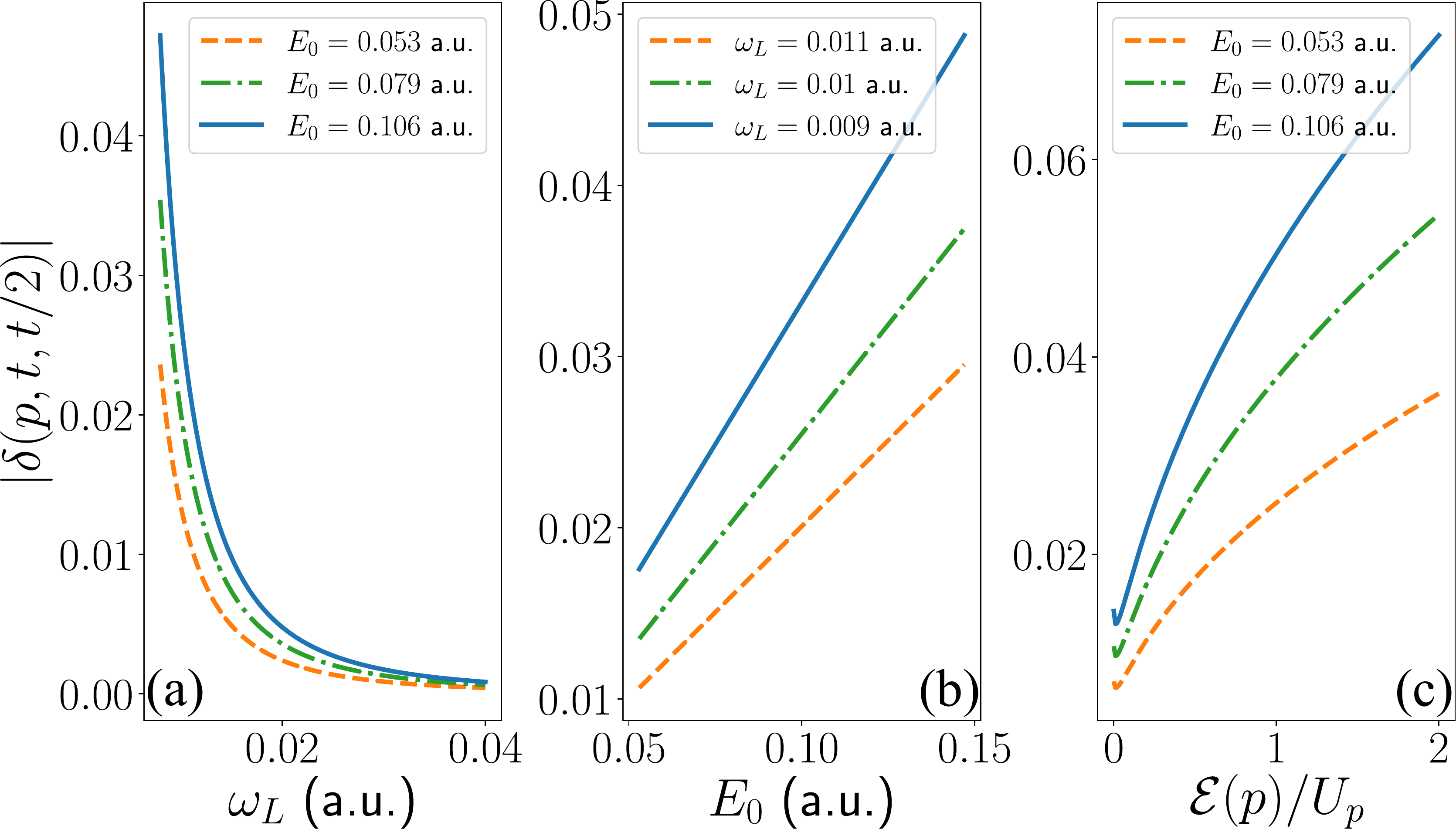}
    \caption{In (a) we show the dependence of $\lvert\delta_{\vb{k}_L,\mu}(p,t,t/2)\rvert$ with the central frequency of the employed laser field, for three different values of the electric field amplitude, in particular $E_0 = 0.053$ a.u. (orange dashed curve), $E_0 = 0.079$ a.u. (green dash-dotted curve) and $E_0 = 0.106$ a.u. (blue solid curve). In (b) we show instead the dependence of $\lvert\delta_{\vb{k}_L,\mu}(p,t,t/2)\rvert$ with the field's amplitude for three different frequencies: $\omega_L =0.011$ a.u. (orange dash-dotted curve), $\omega_L = 0.010$ a.u. (green dashed curve) and $\omega_L = 0.009$ a.u. (blue solid curve). In (a) and (b) we have set $p = \sqrt{2U_p}$ a.u. In (c) we show the dependence of $\lvert\delta_{\vb{k}_L,\mu}(p,t,t/2)\rvert$ with the final kinetic energy of the measured electron. Here, we have restricted to $p>0$. For these calculations, we have considered a 1D hydrogen system ($I_p = 0.5$ a.u.) driven by a linearly polarized laser field with a sinusoidal squared envelope, that has 5 cycles of duration. We have restricted to frequencies $\omega_L$ low enough such that tunneling events are the dominant ones. Furthermore, we have kept $E_0 < 0.147$ a.u. such that over-the-barrier ionization events are less likely than tunneling ones.}
    \label{Fig:Delta:Up:Momentum}
\end{figure*}

As it was mentioned before, the displacement defined in Eq.~\eqref{Eq:displacement} appears as a consequence of the coupling between the electronic motion and the electromagnetic field modes. Thus, the longer the trajectories of the electron in the continuum, the higher is the kinetic energy that the electron acquires, and hence the bigger the value of $\lvert\delta_{\vb{k}_L,\mu}(p,t,t')\rvert$. In consequence, we expect this quantity to depend on the ponderomotive energy $U_p\propto E_0^2/\omega_L^2$. In Fig.~\ref{Fig:Delta:Up:Momentum} (a) we show the behavior of $\lvert\delta_{\vb{k}_L,\mu}(p,t,t')\rvert$ on the frequency for $p=\sqrt{2U_p}$ a.u., $t' = t/2$ a.u., i.e. at the maximum value of the field, and three different field peak strengths $E_0$. We have restricted to the regime $E_0 < 0.147$ a.u. such that over-the-barrier ionization events are less likely than tunneling ones when considering hydrogenic atoms \cite{bauer2006theory}. Furthermore, we have used $\omega_L$ low enough ($\omega_L \in [0.008,0.04]$ a.u. corresponding to a wavelength in the range of $1-5$ $\mu$m) such that tunneling events are more likely than multiphoton ionization processes. Within this regime we see that, for a fixed value of the field peak strength, the displacement $\lvert\delta_{\vb{k}_L,\mu}(p,t,t')\rvert$ is, approximately, inversely proportional to $\omega_L^2$ unlike the electron displacement which is inversely proportional to $\omega_L$. This is a consequence of the fact that the displacement we get in the field is the Fourier transform of the trajectory that is followed by the electron, which follows an oscillatory movement with the frequency of the field. On the other hand, in Fig.~\ref{Fig:Delta:Up:Momentum} (b) we present the behavior of $\lvert\delta_{\vb{k}_L,\mu}(p,t,t')\rvert$ with respect to the field's peak amplitude for three different frequencies which belong to the MIR regime ($\lambda_L\sim 4-5$ $\mu$m). As we can see, the generated displacement is directly proportional to $E_0$ which, together with the previous plot, allows us to see that the displacement behaves as $\lvert\delta_{\vb{k}_L,\mu}(p,t,t')\rvert \propto \sqrt{U_p}/\omega_L$. Thus, the bigger the kinetic energy acquired by the electron in the continuum, the greater $\lvert\delta_{\vb{k}_L,\mu}(p,t,t')\rvert$ would be. Therefore, by increasing the intensity and reducing the frequency of the driving field, we enter in a regime where, for the same number of cycles in a laser field, the electron follows longer trajectories and acquires more kinetic energy. This translates into a greater impact onto the final quantum optical state of the system, leading to $\lvert\delta_{\vb{k}_L,\mu}(p,t,t')\rvert \sim 10^{-2}$ when working with $\omega \sim 0.01$ a.u., and whose effects could be, in principle, measured.

Apart from the laser parameters, the final value of $\lvert\delta_{\vb{k}_L,\mu}(p,t,t')\rvert$ is also determined by the kinetic energy with which the electron is found in the continuum. In this discussion, we assume the electron to be in the continuum at time $t'=t/2$ a.u., and that is found with momentum $p\geq0$ by the end of the pulse. In Fig.~\ref{Fig:Delta:Up:Momentum}~(c), we show how the displacement changes with the final kinetic energy of the electron for three different field peak strengths and for a fixed frequency, $\omega_L = 0.009$ a.u. ($\lambda_L \sim 5$ $\mu$m). As we can see, for increasing values of $E_0$ and $p$, the bigger is the effect on the final value of $\lvert\delta_{\vb{k}_L,\mu}(p,t,t')\rvert$. In particular, we observe a linear dependence with the electron momentum, i.e. with the square root of the photoelectron energy $\mathcal{E}(p)$. We note that at $p=0.00$ a.u., we still get non-zero contributions to the displacement, which are originated by the oscillation of the electron with the field. Finally, although this analysis has been done considering $p>0$, the same features are found for $p < 0$ as well.

In order to work with $\lvert\delta_{\vb{k}_L,\mu}(p,t,t')\rvert\sim 10^{-2}$, in the following we restrict ourselves to the use of MIR laser sources unless otherwise is stated. We note that, in the strong-field literature, this regime of parameters has been already implemented experimentally in the analysis of above-threshold ionization \cite{quan_classical_2009,blaga_strong-field_2009}. These studies have shown that, in the low energy region of the spectrum (around zero values of the electron kinetic energy), a spikelike low-energy structure (LES) appears and that is not captured by the SFA model. These effects have been attributed to a disturbance of the electron momentum due to the Coulomb potential soon after ionization takes place \cite{liu_origin_2010}. Although this effect may alter the final displacement obtained in the quantum optical state, we expect this perturbation to be relatively small given that, for small values of momentum, $\lvert\delta_{\vb{k}_L,\mu}(p,t,t')\rvert< 10^{-2}$ (see for instance Fig.~\ref{Fig:Delta:Up:Momentum}~(c)). Hence, hereupon we keep working under the SFA.

\subsection{Conditioning onto electrons with a fixed kinetic energy}
In Eq.~\eqref{Eq:Final:total:state}, we presented the state of the system after the interaction with the applied laser field. Here, we restrict our study to the characterization of the quantum state of the system after ATI processes. In particular, we want to study the light-matter entanglement for an electron that propagates either along the forward or the backward direction and the displacement it generates on the quantum optical state of the field. For simplicity, we restrict our analysis to electrons that have a fixed value of final kinetic energy. Additionally, we consider a 1D analysis and interaction with linearly polarized laser fields.

In order to restrict ourselves to ATI processes, we need the electron to be found in the continuum. We impose this constraint by means of the following projective operation
\begin{equation}
    \hat{P}_{\text{ATI}}
        = \int \dd p \dyad{p+\dfrac{e}{c}A(t)},
\end{equation}
such that the conditioned to ATI state reads
\begin{equation}\label{Eq:ATI:state:cond}
    \begin{aligned}
    \ket{\psi_{\text{ATI}}(t)}
        &= \hat{P}_{\text{ATI}}\ket{\psi(t)}
        \\
        &= \bigg[
                \prod_{\vb{k},\mu\in\text{IR}}
                D(\alpha_{\vb{k},\mu})
          \bigg]
          \int \dd p \int \dd t'
            M(p,t')
        \\
        &\quad\times
            \ket{p+\dfrac{e}{c}A(t)}
            \bigotimes_{\vb{k},\mu}
                e^{i\varphi_{\vb{k},\mu}(p,t')}
                \ket{\delta_{\vb{k},\mu}(p,t,t')},
    \end{aligned}
\end{equation}
where $M(\vb{p},t)$ corresponds to the integrand of the semiclassical probability amplitude of finding an electron in the continuum, which is given by
\begin{equation}
    \begin{aligned}
    M(\vb{p},t')
        &= e^{-\tfrac{i}{\hbar}S(p,t,t')}
            E(t')d\bigg(
                    p+\dfrac{e}{c}A(t')
                  \bigg).
    \end{aligned}
\end{equation}

Hereupon, we impose the measurement time $t$ to identify with the end of the laser pulse, such that $A(t) = 0$. Having this in mind, we now introduce the projector that restrict us to the situation where the electron is found with a given value of the kinetic energy
\begin{equation}
    \hat{P}(p) = \dyad{-p} + \dyad{p},
\end{equation}
where each term distinguishes between electrons propagating in the backward or forward direction, respectively, with kinetic energy $\hbar^2 p^2/(2m)$. When applying this operator on the state shown in Eq.~\eqref{Eq:ATI:state:cond} we get
\begin{equation}\label{Eq:ATI:fixed:p}
    \begin{aligned}
    \ket{\psi_\text{ATI}(p,t)}
        &= \hat{P}(p) \ket{\psi_{\text{ATI}}(t)}
        \\
        &=  \dfrac{1}{\sqrt{\mathcal{N}}}
            \bigg[
                \ket{p}\ket{\Phi(p,t)}
                + \ket{-p}\ket{\Phi(-p,t)}
            \bigg],
    \end{aligned}
\end{equation}
where $\mathcal{N}$ is a normalization factor, and we have defined
\begin{equation}\label{Eq:QO:parts}
    \begin{aligned}
    \ket{\Phi(\pm p,t)}
        &= \bigg[
                \prod_{\vb{k},\mu\in\text{IR}}
                D(\alpha_{\vb{k},\mu})
          \bigg]
          \int \dd t' M(\pm p,t')
            \\
            &\hspace{1.5cm}
            \bigotimes_{\vb{k},\mu}
                e^{i\varphi_{\vb{k},\mu}(\pm p,t')}
                \ket{\delta_{\vb{k},\mu}(\pm p,t,t')}.
    \end{aligned}
\end{equation}

We observe that the above quantum optical state is given as superposition where, for a fixed momentum, the amplitude of each of the coherent states appearing in the expression depends on the electron's ionization time $t'$. As we have seen before, the phase of $\delta_{\vb{k},\mu}(p,t,t')$ varies with the the direction along which the electron is moving. Thus, up to a more thorough entanglement characterization, the state in Eq.~\eqref{Eq:ATI:fixed:p} has the general form of an entangled state between the electronic momentum and the quantum optical state of the system.

\subsection{Wigner function characterization}
In order to gain intuition on how different the states $\{\lvert\Phi(p,t)\rangle,\lvert\Phi(-p,t)\rangle\}$ are, in this section we study their Wigner function representation. The Wigner function is a quasiprobability distribution of a wavefunction in phase space \cite{wigner_quantum_1932,Schleich_Book_2001}. We refer to it as a \emph{quasiprobability} distribution as there are some properties central to the definition of \emph{proper} probability distributions that the Wigner function does not satisfy. For instance, certain quantum states have associated Wigner functions which show negative values in some regions of the phase space \cite{hudson_when_1974}. Quantum states that show this kind of behavior are usually referred to as \emph{non-classical} states. In the field of quantum optics, Wigner functions have played a fundamental role for characterizing different kind of radiation sources \cite{smithey_measurement_1993}.

Following the approach shown in \cite{royer_wigner_1977}, the Wigner function of a quantum state $\hat{\rho}$ can be written as
\begin{equation}
    W(\beta)
        = \dfrac{2}{\pi} 
            \tr(D(\beta)\Pi D(-\beta)\hat{\rho}),
\end{equation}
where $\Pi$ is the parity operator. In our case, we are interested in looking at the Wigner function representation of the driving field mode when we look at electrons propagating either in the forward or backward direction. Therefore, we denote by $\rho_{+}$ ($\rho_{-}$) the quantum optical state of the system we get when the electron propagates in the forward (backward) direction, such that 
\begin{equation}\label{Eq:Wigner:pn}
    \hat{\rho}_{\pm} 
        = \dyad{\psi^{\pm}_\text{ATI}(p,t)},
\end{equation} 
where $\lvert\psi^{\pm}_\text{ATI}(p,t)\rangle = \langle\pm p\vert\psi_\text{ATI}(p,t)\rangle$. In the following, and in order to tackle the numerical calculations, we perform a single-mode approximation such that the input coherent state in Eq.~\eqref{Eq:Init:state} is written as $\ket{\alpha}$, and populates the mode of frequency $\omega_L$, i.e. the central frequency of the employed laser pulse. Thus, we can express the Wigner function of the considered states as
\begin{equation}\label{Eq:Final:Wigner:Expression}
    \begin{aligned}
	W(\Tilde{\beta},t)
		&= \int^t_{t_0} \dd t_1\int^t_{t_0} \dd t_2
			M^*(p,t_1)M(p,t_2)
			C_\text{HH}(p,t,t_1,t_2)
		    \\&\hspace{1cm}\times	e^{i(\varphi_{\vb{k}_L,\mu}(p,t_2)-\varphi_{\vb{k}_L,\mu}(p,t_1))}
		    e^{-\tfrac12 \lvert 2 \Tilde{\beta} - \delta_1-\delta_2\rvert^2}
			\\&\hspace{1cm}\times
			e^{\Tilde{\beta}^*(\delta_2 - \delta_1)-\Tilde{\beta}(\delta_2 - \delta_1)^*}
			e^{\tfrac12(\delta_1\delta_2^* -\delta_1^*\delta_2)},
	\end{aligned}
\end{equation}
where $\Tilde{\beta} = \beta-\alpha$ and $C_\text{HH}(p,t,t_1,t_2)$ is a function defined as the overlap between the coherent states in which the harmonics can be found, evaluated at different ionization times $t_1$ and $t_2$ (see Appendix \ref{App:Sec:Wigner} for details). Furthermore, we have used $\delta_i$ as a shortened notation for $\delta_{\vb{k}_L,\mu}(p,t,t_i)$. In order to compute these integrals, we have used the saddle-point approximation (see Appendix \ref{App:Sec:Wigner:Saddles} for more details).

\begin{figure}
    \centering
    \includegraphics[width = 1.\columnwidth]{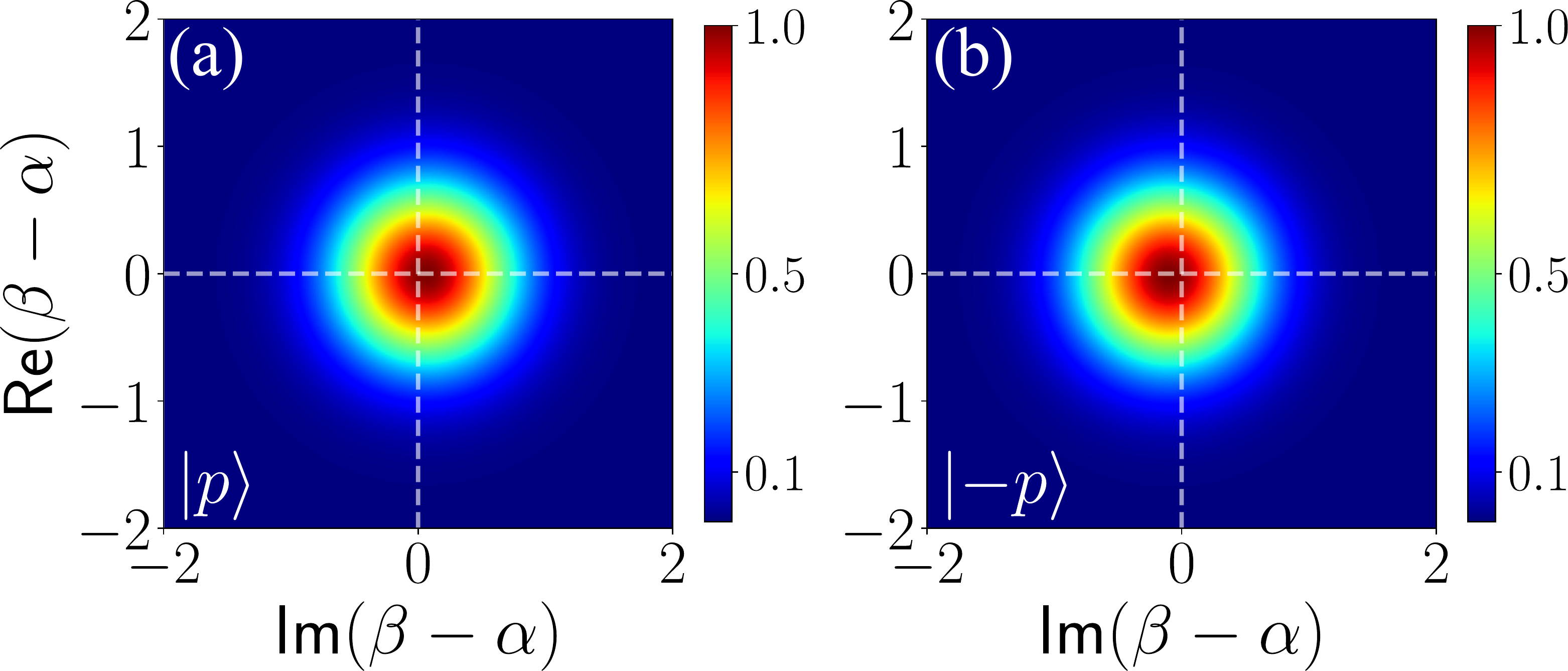}
    \caption{Wigner function representation of the driving field mode when conditioning on electrons propagating in different directions. Specifically, in (a) we have considered the projection onto $\ket{p}$, while in (b) we have considered the projection onto $\ket{-p}$. The intersection between the dashed lines points the origin of our phase-space frame of reference. We have used a laser source with a sinusoidal squared envelope, 5 cycles of duration, $\omega_L = 0.009$ a.u. for the central frequency and $E_0 =0.106$ a.u. for the field's amplitude. In these plots, the Wigner function has been normalized to its maximum value.}
    \label{Fig:Wigner:pn:oe}
\end{figure}

In Figs.~\ref{Fig:Wigner:pn:oe}~(a) and (b) we show the Wigner function computed from the state shown in Eq.~\eqref{Eq:Wigner:pn} when using $\mathcal{E}(p) = 2.2U_p$. Here, we have used a linearly polarized laser field of 5 cycles of duration, $\omega_L = 0.009$ a.u. for the central frequency and $E_0 = 0.106$ a.u. for the field's amplitude. The generated Wigner function presents a Gaussian-like behavior which lacks from non-classical signatures in terms of negative regions. However, there exists a difference between the generated distributions when considering electrons propagating in the forward or in the backward direction. In particular, we observe that for positive values of momentum (Fig.~\ref{Fig:Wigner:pn:oe}~(a)) the quasiprobability distribution is slightly shifted towards positive regions of the $x$ axis, while for negative values of momentum (Fig.~\ref{Fig:Wigner:pn:oe}~(b)) it is slightly shifted towards negative regions of the $x$ axis. We note that this lack of non-classical features in the Wigner function representation contrasts with what it was observed in \cite{rivera-dean_strong_2022}. This is due to the difference between the amplitudes of the coherent states appearing in the superposition state given in Eq.~\eqref{Eq:QO:parts}, which in the present case are much smaller compared to ref.~\cite{rivera-dean_strong_2022}.

\begin{figure*}
    \centering
    \includegraphics[width = 0.8\textwidth]{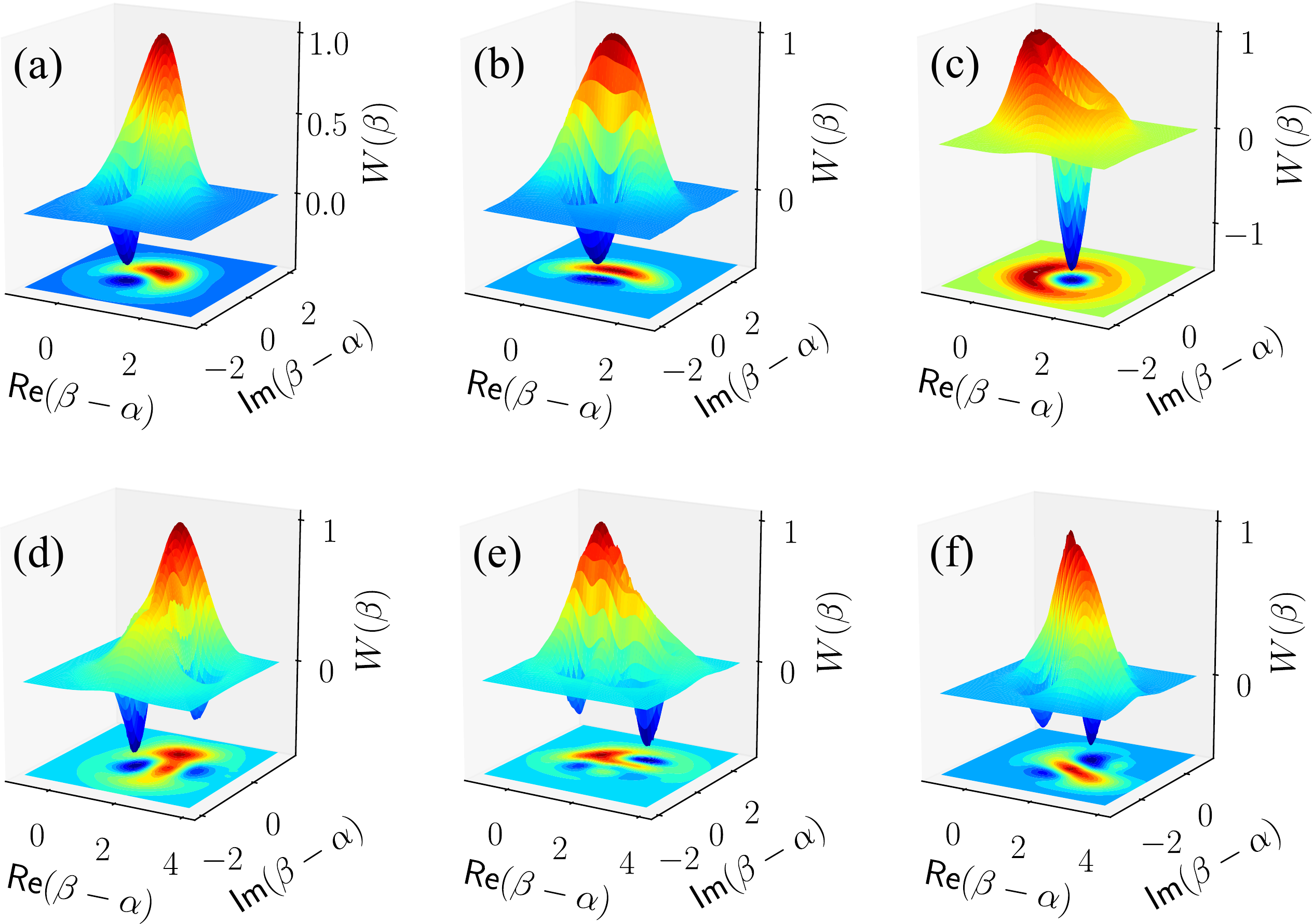}
    \caption{Wigner function of the state shown in Eq.~\eqref{Eq:ATI:ManyBody}. In the first row (plots (a)--(c)), we have set $N\sim10^4$ atoms while in the second row (plots (d)--(f)) we have used $N\sim2\times 10^4$. Each of the columns correspond to different values of the employed canonical momentum. In particular, we have used $p =0.00$ a.u. (plots (a) and (d)), $p=0.43$ a.u. (plots (b) and (e)) and $p=-0.43$ a.u. (plots (c) and (f)). For the numerical calculations we have used a linearly polarized electromagnetic field with a sinusoidal squared envelope of 5 cycles of duration, $E_0 = 0.053$ a.u. for the field's amplitude, $\omega_L = 0.057$ a.u. for the central frequency. For the atomic system, we used a 1D model of Hydrogen with $I_p = 0.5$ a.u. for the ionization potential. We have set in our calculations the final time to coincide with the end of the pulse. Furthermore, we have normalized the results to the maximum value found for the Wigner function. More details about the generation of the Wigner plots can be found in Appendix \ref{App:Sec:Wigner:Numerical}.}
    \label{Fig:Wigner:diff:mom}
\end{figure*}

In the many-body situation, where more than one atom participate in the ATI process, we expect to observe non-classical features in the Wigner function representation. In general, the many-body characterization becomes computationally demanding as the average number of atoms $N$ that one could get in the experimental gas jets can easily surpass the order of $10^{6}$. Thus, in order to perform a \emph{proof-of-principle} analysis, we instead consider a situation where all the atoms participate collectively in the process, such that we can take into account the many-body effects in a phenomenological way by multiplying the generated displacement $\delta_{\vb{k},\mu}(p,t,t')$ by a factor $N$. Thus, instead of working with Eq.~\eqref{Eq:QO:parts}, we consider now
\begin{equation}\label{Eq:ATI:ManyBody}
    \begin{aligned}
    \ket{\Phi^{(N)}(\pm p,t)}
        &= D(\alpha)
        \int \dd t' M(\pm p,t')
            \\
            &\hspace{0.5cm}
            \bigotimes_{\vb{k},\mu}
                e^{iN^2\varphi_{\vb{k},\mu}(\pm p,t')}
                \ket{N\delta_{\vb{k},\mu}(\pm p,t,t')}.
    \end{aligned}
\end{equation}

In Fig.~\ref{Fig:Wigner:diff:mom} we show the Wigner functions obtained from Eq.~\eqref{Eq:Final:Wigner:Expression} when considering that the total number of atoms that undergo ATI is $N \sim 10^4$ (upper row) and $N\sim2\times 10^4$ (lower row). On the other hand, in each of the columns we consider different values of the momentum $p$, in particular (from left to right): $p=0.00$ a.u. (first column), $p=0.43$ a.u. (second column) and $p=-0.43$ a.u. (third column). There are two main features to highlight in these plots. Firstly, already for $N\sim10^4$, highly non-classical behaviors can be found in the state, which are witnessed in terms of the Wigner function negativities. These negativities are a consequence of the quantum superposition between the different coherent states that contribute with a distinct amplitude depending on when the electrons ionize. For $N\sim10^4$, a small number of coherent states for which $\langle N\delta_{\vb{k}_L,\mu}(p,t,t_i)\vert N \delta_{\vb{k}_L,\mu}(p,t,t_j\rangle \to 0$, with $t_i\neq t_j$, appearing in the superposition and, in consequence, the final Wigner function depicts a behavior which can be reproduced by the unbalanced superposition of two coherent states with close, but yet different, amplitudes. As $N$ increases, the real and imaginary parts of $\delta_{\vb{k}_L,\mu}(p,t,t_i)$ cover a bigger range of values, and therefore we get more terms in the superposition for which $\langle\delta_{\vb{k}_L,\mu}(p,t,t_i)\vert\delta_{\vb{k}_L,\mu}(p,t,t_j)\rangle \to 0$. This translates in a more complicated structure for the obtained Wigner functions, with more minima and maxima distributed along the phase space.

On the other hand, and as the second main feature, we get differences between the distinct values of the canonical momentum. For $N\sim10^4$ (first row), we see that the differences are mainly due to the form of the Wigner function itself: its orientation and the depth of the obtained minimum. These differences are a consequence of the contribution of the different ionization times of the electron, which gives rise to distinct relative phase amplitudes among the different coherent states in the superposition. We note that, if a change of $\pi$ is implemented in the carrier-envelope phase (CEP), i.e. the phase difference between the envelope and the carrier wave, of the employed pulse the behavior for positive and negative momentum interchanges \cite{paulus_measurement_2003}. Furthermore, if we use instead a constant laser field with no envelope, then the differences regarding the form of the Wigner function for positive and negative momentum would vanish, as in this case the ionization times are, for the same value of the kinetic energy, symmetric with respect to the maximum values of the field's intensity. However, there is another main difference between the plots similar to the one observed in Fig.~\ref{Fig:Wigner:pn:oe}, which can be seen clearly for $N\sim2\times 10^4$ (second row), and involves the location of the Wigner function in phase space. In particular, we see that for positive values of momentum (Fig.~\ref{Fig:Wigner:diff:mom} (e)), the maximum peak is located in positive regions of the $\text{Im}(\beta-\alpha)$ axis, while for negative values it is located along the negative direction. Furthermore, for $p=0.00$ a.u. (Fig.~\ref{Fig:Wigner:diff:mom} (d)), it is centered around zero. All the behaviors we have discussed so far are a consequence of the radiation generated by the electron during its oscillation in the continuum, which depends on its final kinetic momentum. However, we stress that different contributions can be interchanged by implementing a modification of $\pi$ in the CEP \cite{paulus_measurement_2003}.

\subsection{Entanglement characterization}
The Wigner function characterization we have done in the previous section has allowed us to see that the quantum optical part of the state appearing in Eq.~\eqref{Eq:ATI:fixed:p} differs depending on the propagation direction of the electron. This implies that the aforementioned state cannot be written in general as a product state, and therefore is entangled. In particular, the structure this state presents is that of an \emph{hybrid entangled} state \cite{van_loock_optical_2011}, as we have the tensor product of an effective finite dimensional Hilbert space (spanned by $\{\ket{p},\ket{-p}\}$), and an infinite dimensional Hilbert space (spanned, for instance, by the Fock basis). 

The entanglement characterization of hybrid entangled states is in general an open problem, and has to be studied carefully depending on the particular form of the state, as may it involve the definition of specific entanglement witnesses \cite{van_loock_optical_2011,kreis_classifying_2012,masse_implementable_2020}. However, for the case of pure states, this entanglement characterization can be performed by means of the entropy of entanglement \cite{van_loock_optical_2011,nielsen_quantum_2010}. Instead of working with the states $\{\lvert\Tilde{\Phi}(p,t)\rangle,\lvert\Tilde{\Phi}(-p,t)\rangle\}$, where $\lvert\Tilde{\Phi}(\pm p,t)\rangle = \lvert\Phi(\pm p,t)\rangle/\sqrt{\mathcal{N}_\pm}$ with $\mathcal{N}_\pm$ the normalization of the state, which in general have a non-vanishing overlap, we instead work with the orthonormal set $\{\ket{u},\ket{v}\}$. This allow us to treat effectively our Hilbert space as being of dimension $2 \otimes 2$.

The relation between the orthonormal set $\{\ket{u},\ket{v}\}$ and $\{\lvert\Tilde{\Phi}(p,t)\rangle,\lvert\Tilde{\Phi}(-p,t)\rangle\}$ is given by
\begin{equation}
    \left\{
    \begin{aligned}
    & \ket{u} 
        = \dfrac{1}{2\mu}
            \Big(
                \ket{\Tilde{\Phi}(p,t)}
                + e^{-i \theta} 
                    \ket{\Tilde{\Phi}(-p,t)}
            \Big),\\
    & \ket{v} 
        = \dfrac{1}{2\nu}
            \Big(
                \ket{\Tilde{\Phi}(p,t)}
                - e^{-i \theta} 
                    \ket{\Tilde{\Phi}(-p,t)}
            \Big),
    \end{aligned}
    \right.
\end{equation}
where we have defined
\begin{equation}
    \left\{
    \begin{aligned}
    &\mu = \sqrt{(1+\lvert\langle\Tilde{\Phi}(p,t)\vert\Tilde{\Phi}(-p,t)\rangle\rvert)/2}, 
    \\
    &\nu = \sqrt{1-\mu^2},
    \\
    &\langle\Tilde{\Phi}(p,t)\vert\Tilde{\Phi}(-p,t)\rangle = e^{i\theta}\lvert\langle\Tilde{\Phi}(p,t)\vert\Tilde{\Phi}(-p,t)\rangle\rvert,
    \end{aligned}
    \right.
\end{equation}
such that in this new basis the state in Eq.~\eqref{Eq:ATI:fixed:p} can be rewritten as follows
\begin{equation}\label{Eq:State:Orthonormal}
    \begin{aligned}
    \ket{\psi_{\text{ATI}}(p,t)}
        &= \dfrac{1}{\sqrt{\mathcal{N}}}
            \bigg[
                \mu \ket{u}
                \Big(
                    \sqrt{\mathcal{N}_+}
                        \ket{p}
                    + e^{i\theta}\sqrt{\mathcal{N}_-}
                        \ket{-p}
                \Big)
                \\&\hspace{1.2cm}
                + \nu\ket{v}
                \Big(
                    \sqrt{\mathcal{N}_+}
                        \ket{p}
                    - e^{i\theta}\sqrt{\mathcal{N}_-}
                        \ket{-p}
                \Big)
            \bigg].
    \end{aligned}
\end{equation}

The entropy of entanglement is defined as $S \coloneqq -\Tr[\hat{\rho} \log\hat{\rho}]$ \cite{plenio_introduction_2007,nielsen_quantum_2010}, where $\hat{\rho}$ is the reduced density matrix obtained by doing the partial trace with respect to either the electron or the quantum optical degrees of freedom of $\lvert\psi_{\text{ATI}}(p,t)\rangle\!\langle\psi_{\text{ATI}}(p,t)\rvert$. From here, we get that the amount of entropy of entanglement our state has is given by (see Appendix \ref{App:Entanglement:Expressions})
\begin{equation}\label{Eq:Entanglement:Entropy}
    S(\hat{\rho})
        = - \lambda_+^2 \log_2 \lambda_+^2
            - \lambda_-^2 \log_2 \lambda_-^2,
\end{equation}
where we define
\begin{equation}\label{Eq:Eigvals:Single}
    \lambda_{\pm}
        = \dfrac12
            \Bigg[
                1 \pm \sqrt{1-4\bigg(1-\abs{\braket{\Tilde{\Phi}(p,t)}{\Tilde{\Phi}(-p,t)}}^2\bigg)\dfrac{\mathcal{N}_+\mathcal{N}_-}{\mathcal{N}}}
            \Bigg].
\end{equation}

Because of normalization conditions, we have that $\lambda_+ + \lambda_- = 1$ and, therefore, we say that a state is maximally entangled if $S(\hat{\rho}) = 1$, i.e. $\lambda_\pm = 1/2$, and it is a separable state if $S(\hat{\rho}) = 0$, i.e. $\lambda_\pm = 1$ and $\lambda_\mp = 0$. From Eq.~\eqref{Eq:Eigvals:Single}, we see that the degree of entanglement is determined by the amount of population $\mathcal{N}_+$ and $\mathcal{N}_-$ that we have respectively in the $\ket{p}$ and $\ket{-p}$ states, and on the overlap between their associated quantum optical contributions. First, we note that the populations $\mathcal{N}_+$ and $\mathcal{N}_-$ depend on the probability of an electron being ionized with momentum $p$, and on the overlaps $\lvert\langle\Tilde{\Phi}(\pm p,t)\vert\Tilde{\Phi}(\pm p,t)\rangle\rvert$. Since for multicycle laser fields the photoionization spectrum is symmetric against a change in sign of the momentum \cite{milosevic_above-threshold_2006,protopapas_atomic_1997}, we expect both populations to be almost identical and close to $1/2$ after adding the proper normalization factor. Thus, this leaves the overlap between the two quantum optical states in Eq.~\eqref{Eq:QO:parts} as the most important quantity in determining the final degree of entanglement. As we saw in Fig.~\ref{Fig:Wigner:pn:oe}, the bigger the value of $p$, the further away the two states are in phase space, and therefore the smaller we expect their overlap to be.

\begin{figure}
    \centering
    \includegraphics[width = 1.\columnwidth]{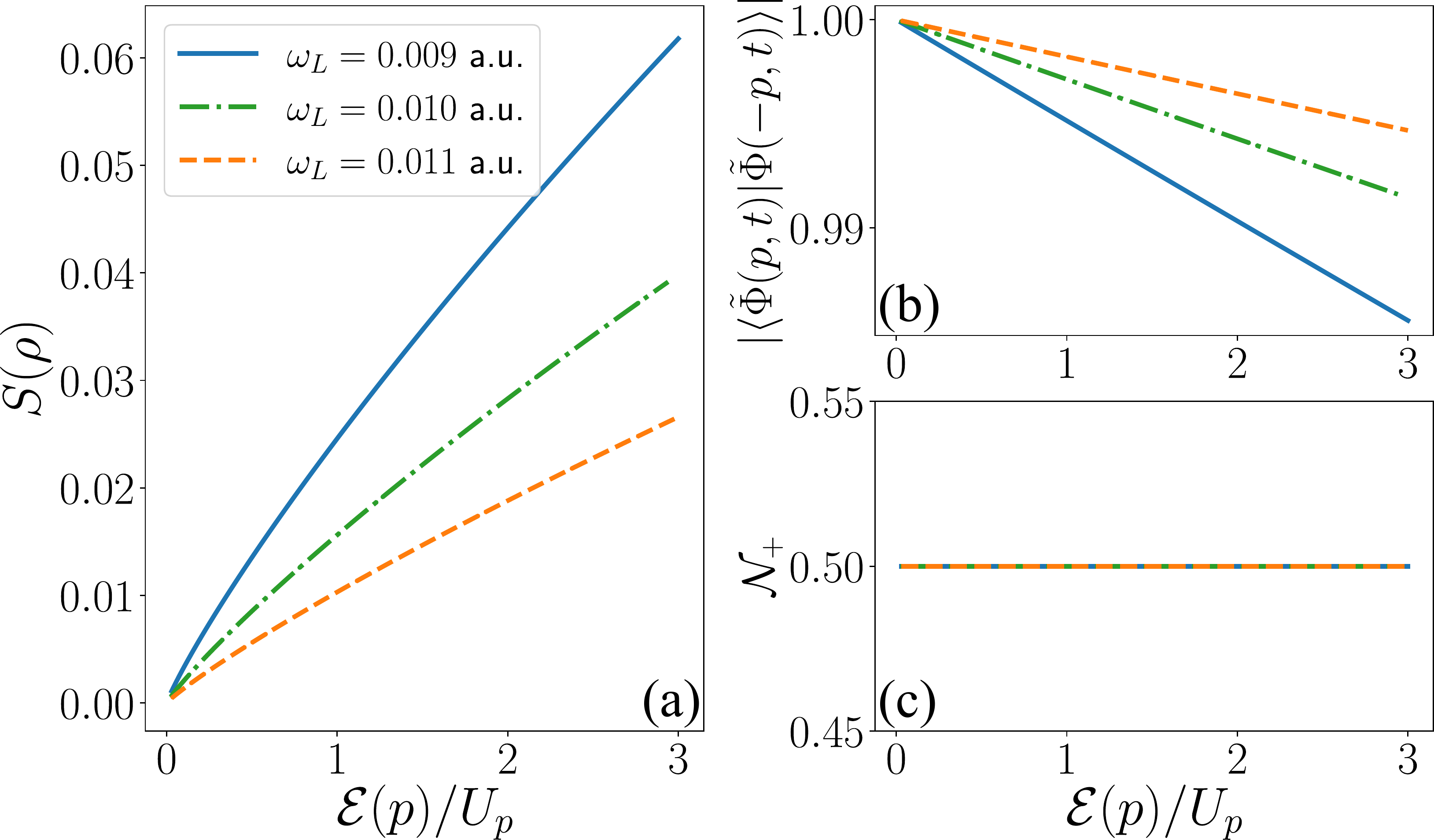}
    \caption{Behavior of (a) $S(\hat{\rho})$, (b) $\lvert\langle\Tilde{\Phi}(p,t)\vert\Tilde{\Phi}(-p,t)\rangle\rvert$ and (c) $\mathcal{N}_+$ with respect to the photoelectron energy for different frequencies. In particular, we considered $\omega_L = 0.009$ a.u., $\omega_L = 0.010$ a.u. and $\omega_L = 0.011$ a.u. which are respectively shown with the blue solid, green dash-dotted and orange dashed curves in the three plots. In order to do these calculations, we considered a linearly polarized electromagnetic field for the input, with a sinusoidal squared envelope of 5 cycles of duration and $E_0 = 0.106$ a.u. for the field's amplitude. For the atomic system, we used a 1D model of Hydrogen with $I_p = 0.5$ a.u. for the ionization potential.}
    \label{Fig:Entanglement:Entropy}
\end{figure}

In Fig.~\eqref{Fig:Entanglement:Entropy}~(a) we show the entropy of entanglement following Eq.~\eqref{Eq:Entanglement:Entropy} for a laser field of amplitude $E_0 = 0.106$ a.u., and three different frequencies, namely $\omega_L = 0.009$ a.u. (blue solid curve), $\omega_L = 0.010$ a.u. (green dash-dotted curve) and $\omega_L = 0.011$ a.u. (orange dashed curve). We observe that the entropy of entanglement is zero when $\mathcal{E}(p) = 0$, and becomes bigger for increasing values of the photoelectron energy. We also observe that, for smaller frequencies, the amount of entanglement increases as well. This is a consequence of how important the quantum optical displacement becomes when modifying the ponderomotive energy, as discussed in Sec.~\ref{Sec:Results:laser:params}.  In Figs.~\ref{Fig:Entanglement:Entropy}~(b) and (c), we present the overlap $\lvert\langle\Tilde{\Phi}( p,t)\vert\Tilde{\Phi}(- p,t)\rangle\rvert$ and the normalization constant $\mathcal{N}_+$, respectively, for the three frequencies we have considered. Note that the behavior of $\mathcal{N}_-$ can be obtained by considering $1-\mathcal{N}_+$. We see that, in all the cases, the normalization constant remains around 0.5 for all values of the photoelectron energy. As we mentioned, this is an expected feature since we are working with multicycle pulses, and the normalization constant is related to the probability of measuring an electron propagating along one of the possible directions. On the other hand, the overlap between the two possible states starts being unity, when the generated displacement is very small (see Fig.~\ref{Fig:Delta:Up:Momentum}~(c)), and decays for increasing values of the photoelectron energy. We therefore confirm the importance of this parameter in determining the amount of entanglement we find in the state. We remark that all the integrals involved in the calculation of the normalization constant and the overlap between the states, have been done under the saddle-point approximation (see Appendix \ref{App:Sec:Wigner:Saddles}).

From this analysis we have observed that the state in Eq.~\eqref{Eq:ATI:fixed:p} is entangled, although the amount of entanglement we find is small. We have seen that this quantity is mainly determined by the overlap between the quantum optical states appearing in the superposition, which in this case is close, but yet different, to unity. However, we note that the entropy of entanglement can be further increased by considering bigger values of the kinetic energy. Going beyond the range of values for the kinetic energy we consider here requires the introduction of rescattering effects in ATI, which are typically non-negligible for $p > 3U_p$, and that are out of the scope of the present work.


\section{CONCLUSIONS AND OUTLOOK}\label{Sec:Conclusions}
In this work, we have studied light-matter entanglement after ATI processes. We have studied the effects of the freed-electron's motion on the quantum optical state of the field, and found that in the MIR regime these effects can be observed at the single-atom level, with typical values for the field amplitude \cite{blaga_strong-field_2009}. In order to motivate the entanglement characterization, we have studied the Wigner function of the quantum optical state of the driving field mode when the generated electrons are either propagating in the forward and backward direction. We have also implemented a phenomenological many-body analysis to understand the regime at which non-classical features could be observed. Finally, we have used the entropy of entanglement as an entanglement witness, and show how it varies for different values of the photoelectron energy and frequency of the employed laser field.

While the generated coherent state superpositions in ATI processes are already of interest \emph{per se} for a wide variety of quantum technology applications \cite{gilchrist_schrodinger_2004,lvovsky_production_2020,ralph_quantum_2003,sanders_entangled_1992,jeong_quantum_2003,stobinska_violation_2007,munro_weak-force_2002}, we have checked that, when including as well aspects related to the electronic features of the state, ATI processes could potentially lead to hybrid entangled states which show a small but non-zero amount of entanglement even at the single-atom level. This feature, together with the attosecond time scales which are associated to strong-field processes, could extend the current applicability of hybrid entangled states \cite{van_loock_optical_2011} to unprecedented time scales, while using the same experimental architectures that have been used thus far in the strong-field community \cite{SchulzVrakking_Book}. Therefore, this work can be understood as a first step towards this direction.

\section*{ACKNOWLEDGMENTS}

ICFO group acknowledges support from: ERC AdG NOQIA; Agencia Estatal de Investigación (R$\&$D project CEX2019-000910-S, funded by MCIN/ AEI/10.13039/501100011033, Plan National FIDEUA PID2019-106901GB-I00, FPI, QUANTERA MAQS PCI2019-111828-2, QUANTERA DYNAMITE PCI2022-132919, Proyectos de I+D+I “Retos Colaboración” QUSPIN RTC2019-007196-7); Fundació Cellex; Fundació Mir-Puig; Generalitat de Catalunya through the European Social Fund FEDER and CERCA program (AGAUR Grant No. 2017 SGR 134, QuantumCAT \ U16-011424, co-funded by ERDF Operational Program of Catalonia 2014-2020); the computer resources and technical support at Barcelona Supercomputing Center MareNostrum (FI-2022-1-0042); EU Horizon 2020 FET-OPEN OPTOlogic (Grant No 899794); National Science Centre, Poland (Symfonia Grant No. 2016/20/W/ST4/00314); European Union’s Horizon 2020 research and innovation programme under the Marie-Skłodowska-Curie grant agreement No 101029393 (STREDCH) and No 847648 (“La Caixa” Junior Leaders fellowships ID100010434: LCF/BQ/PI19/11690013, LCF/BQ/PI20/11760031, LCF/BQ/PR20/11770012, LCF/BQ/PR21/11840013); the Government of Spain (FIS2020-TRANQI and Severo Ochoa CEX2019-000910-S).

P. Tzallas group at FORTH acknowledges LASERLABEUROPE V (H2020-EU.1.4.1.2 grant no.871124), FORTH Synergy Grant AgiIDA (grand no. 00133), the H2020 framework program for research and innovation under the NEP-Europe-Pilot project (no. 101007417). ELI-ALPS is supported by the European Union and co-financed by the European Regional Development Fund (GINOP Grant No. 2.3.6-15-2015-00001).

J.R-D. acknowledges support from the Secretaria d'Universitats i Recerca del Departament d'Empresa i Coneixement de la Generalitat de Catalunya, as well as the European Social Fund (L'FSE inverteix en el teu futur)--FEDER. 
P.S. acknowledges funding from the European Union’s Horizon 2020
research and innovation programme under the Marie Skłodowska-Curie grant agreement No 847517. 
A.S.M. acknowledges funding support from the European Union’s Horizon 2020 research and innovation programme under the Marie Sk\l odowska-Curie grant agreement, SSFI No.\ 887153.
M.F.C. acknowledges financial support from the Guangdong Province Science and Technology Major Project (Future functional materials under extreme conditions - 2021B0301030005).

\bibliography{references.bib}{}

\begin{thebibliography}{82}%
\makeatletter
\providecommand \@ifxundefined [1]{%
 \@ifx{#1\undefined}
}%
\providecommand \@ifnum [1]{%
 \ifnum #1\expandafter \@firstoftwo
 \else \expandafter \@secondoftwo
 \fi
}%
\providecommand \@ifx [1]{%
 \ifx #1\expandafter \@firstoftwo
 \else \expandafter \@secondoftwo
 \fi
}%
\providecommand \natexlab [1]{#1}%
\providecommand \enquote  [1]{``#1''}%
\providecommand \bibnamefont  [1]{#1}%
\providecommand \bibfnamefont [1]{#1}%
\providecommand \citenamefont [1]{#1}%
\providecommand \href@noop [0]{\@secondoftwo}%
\providecommand \href [0]{\begingroup \@sanitize@url \@href}%
\providecommand \@href[1]{\@@startlink{#1}\@@href}%
\providecommand \@@href[1]{\endgroup#1\@@endlink}%
\providecommand \@sanitize@url [0]{\catcode `\\12\catcode `\$12\catcode
  `\&12\catcode `\#12\catcode `\^12\catcode `\_12\catcode `\%12\relax}%
\providecommand \@@startlink[1]{}%
\providecommand \@@endlink[0]{}%
\providecommand \url  [0]{\begingroup\@sanitize@url \@url }%
\providecommand \@url [1]{\endgroup\@href {#1}{\urlprefix }}%
\providecommand \urlprefix  [0]{URL }%
\providecommand \Eprint [0]{\href }%
\providecommand \doibase [0]{https://doi.org/}%
\providecommand \selectlanguage [0]{\@gobble}%
\providecommand \bibinfo  [0]{\@secondoftwo}%
\providecommand \bibfield  [0]{\@secondoftwo}%
\providecommand \translation [1]{[#1]}%
\providecommand \BibitemOpen [0]{}%
\providecommand \bibitemStop [0]{}%
\providecommand \bibitemNoStop [0]{.\EOS\space}%
\providecommand \EOS [0]{\spacefactor3000\relax}%
\providecommand \BibitemShut  [1]{\csname bibitem#1\endcsname}%
\let\auto@bib@innerbib\@empty
\bibitem [{\citenamefont {Agostini}\ \emph {et~al.}(1979)\citenamefont
  {Agostini}, \citenamefont {Fabre}, \citenamefont {Mainfray}, \citenamefont
  {Petite},\ and\ \citenamefont {Rahman}}]{agostini_free-free_1979}%
  \BibitemOpen
  \bibfield  {author} {\bibinfo {author} {\bibfnamefont {P.}~\bibnamefont
  {Agostini}}, \bibinfo {author} {\bibfnamefont {F.}~\bibnamefont {Fabre}},
  \bibinfo {author} {\bibfnamefont {G.}~\bibnamefont {Mainfray}}, \bibinfo
  {author} {\bibfnamefont {G.}~\bibnamefont {Petite}},\ and\ \bibinfo {author}
  {\bibfnamefont {N.~K.}\ \bibnamefont {Rahman}},\ }\bibfield  {title}
  {\bibinfo {title} {Free-{Free} {Transitions} {Following} {Six}-{Photon}
  {Ionization} of {Xenon} {Atoms}},\ }\href
  {https://doi.org/10.1103/PhysRevLett.42.1127} {\bibfield  {journal} {\bibinfo
   {journal} {Physical Review Letters}\ }\textbf {\bibinfo {volume} {42}},\
  \bibinfo {pages} {1127} (\bibinfo {year} {1979})}\BibitemShut {NoStop}%
\bibitem [{\citenamefont {Milošević}\ \emph
  {et~al.}(2006{\natexlab{a}})\citenamefont {Milošević}, \citenamefont
  {Paulus}, \citenamefont {Bauer},\ and\ \citenamefont
  {Becker}}]{milosevic_above-threshold_2006}%
  \BibitemOpen
  \bibfield  {author} {\bibinfo {author} {\bibfnamefont {D.~B.}\ \bibnamefont
  {Milošević}}, \bibinfo {author} {\bibfnamefont {G.~G.}\ \bibnamefont
  {Paulus}}, \bibinfo {author} {\bibfnamefont {D.}~\bibnamefont {Bauer}},\ and\
  \bibinfo {author} {\bibfnamefont {W.}~\bibnamefont {Becker}},\ }\bibfield
  {title} {\bibinfo {title} {Above-threshold ionization by few-cycle pulses},\
  }\href {https://doi.org/10.1088/0953-4075/39/14/R01} {\bibfield  {journal}
  {\bibinfo  {journal} {Journal of Physics B: Atomic, Molecular and Optical
  Physics}\ }\textbf {\bibinfo {volume} {39}},\ \bibinfo {pages} {R203}
  (\bibinfo {year} {2006}{\natexlab{a}})}\BibitemShut {NoStop}%
\bibitem [{\citenamefont {Delone}\ and\ \citenamefont
  {Kraĭnov}(2000)}]{delone_multiphoton_2000}%
  \BibitemOpen
  \bibfield  {author} {\bibinfo {author} {\bibfnamefont {N.~B.}\ \bibnamefont
  {Delone}}\ and\ \bibinfo {author} {\bibfnamefont {V.~P.}\ \bibnamefont
  {Kraĭnov}},\ }\href@noop {} {\emph {\bibinfo {title} {Multiphoton
  {Processes} in {Atoms}: {Second} {Edition}}}}\ (\bibinfo  {publisher}
  {Springer Science \& Business Media},\ \bibinfo {year} {2000})\BibitemShut
  {NoStop}%
\bibitem [{\citenamefont {Agostini}\ and\ \citenamefont
  {DiMauro}(2012)}]{agostini_chapter_2012}%
  \BibitemOpen
  \bibfield  {author} {\bibinfo {author} {\bibfnamefont {P.}~\bibnamefont
  {Agostini}}\ and\ \bibinfo {author} {\bibfnamefont {L.~F.}\ \bibnamefont
  {DiMauro}},\ }\bibfield  {title} {\bibinfo {title} {Chapter 3 - {Atomic} and
  {Molecular} {Ionization} {Dynamics} in {Strong} {Laser} {Fields}: {From}
  {Optical} to {X}-rays},\ }in\ \href
  {https://doi.org/10.1016/B978-0-12-396482-3.00003-X} {\emph {\bibinfo
  {booktitle} {Advances {In} {Atomic}, {Molecular}, and {Optical}
  {Physics}}}},\ \bibinfo {series} {Advances in {Atomic}, {Molecular}, and
  {Optical} {Physics}}, Vol.~\bibinfo {volume} {61},\ \bibinfo {editor} {edited
  by\ \bibinfo {editor} {\bibfnamefont {P.}~\bibnamefont {Berman}}, \bibinfo
  {editor} {\bibfnamefont {E.}~\bibnamefont {Arimondo}},\ and\ \bibinfo
  {editor} {\bibfnamefont {C.}~\bibnamefont {Lin}}}\ (\bibinfo  {publisher}
  {Academic Press},\ \bibinfo {year} {2012})\ pp.\ \bibinfo {pages}
  {117--158}\BibitemShut {NoStop}%
\bibitem [{\citenamefont {Lewenstein}\ and\ \citenamefont
  {L'Huillier}(2009)}]{AnneML2}%
  \BibitemOpen
  \bibfield  {author} {\bibinfo {author} {\bibfnamefont {M.}~\bibnamefont
  {Lewenstein}}\ and\ \bibinfo {author} {\bibfnamefont {A.}~\bibnamefont
  {L'Huillier}},\ }\href@noop {} {\emph {\bibinfo {title} {Strong Field Laser
  Physics}}},\ edited by\ \bibinfo {editor} {\bibfnamefont {T.}~\bibnamefont
  {Brabec}}\ (\bibinfo  {publisher} {Springer, New York},\ \bibinfo {year}
  {2009})\ Chap.\ \bibinfo {chapter} {Principles of Single Atom Physics:
  High-Order Harmonic Generation, Above-Threshold Ionization and Non-Sequential
  Ionization}, pp.\ \bibinfo {pages} {147--183}\BibitemShut {NoStop}%
\bibitem [{\citenamefont {Keldysh}(1964)}]{Keldysh}%
  \BibitemOpen
  \bibfield  {author} {\bibinfo {author} {\bibfnamefont {L.~V.}\ \bibnamefont
  {Keldysh}},\ }\bibfield  {title} {\bibinfo {title} {Ionization in the field
  of a strong electromagnetic wave},\ }\bibfield  {journal} {\bibinfo
  {journal} {Sov. Phys. JETP}\ }\textbf {\bibinfo {volume} {20}},\ \href@noop
  {} {} (\bibinfo {year} {1964})\BibitemShut {NoStop}%
\bibitem [{\citenamefont {Lewenstein}\ \emph {et~al.}(1995)\citenamefont
  {Lewenstein}, \citenamefont {Kulander}, \citenamefont {Schafer},\ and\
  \citenamefont {Bucksbaum}}]{lewenstein_rings_1995}%
  \BibitemOpen
  \bibfield  {author} {\bibinfo {author} {\bibfnamefont {M.}~\bibnamefont
  {Lewenstein}}, \bibinfo {author} {\bibfnamefont {K.~C.}\ \bibnamefont
  {Kulander}}, \bibinfo {author} {\bibfnamefont {K.~J.}\ \bibnamefont
  {Schafer}},\ and\ \bibinfo {author} {\bibfnamefont {P.~H.}\ \bibnamefont
  {Bucksbaum}},\ }\bibfield  {title} {\bibinfo {title} {Rings in
  above-threshold ionization: {A} quasiclassical analysis},\ }\href
  {https://doi.org/10.1103/PhysRevA.51.1495} {\bibfield  {journal} {\bibinfo
  {journal} {Physical Review A}\ }\textbf {\bibinfo {volume} {51}},\ \bibinfo
  {pages} {1495} (\bibinfo {year} {1995})}\BibitemShut {NoStop}%
\bibitem [{\citenamefont {Lewenstein}\ \emph {et~al.}(1994)\citenamefont
  {Lewenstein}, \citenamefont {Balcou}, \citenamefont {Ivanov}, \citenamefont
  {L’Huillier},\ and\ \citenamefont {Corkum}}]{lewenstein_theory_1994}%
  \BibitemOpen
  \bibfield  {author} {\bibinfo {author} {\bibfnamefont {M.}~\bibnamefont
  {Lewenstein}}, \bibinfo {author} {\bibfnamefont {P.}~\bibnamefont {Balcou}},
  \bibinfo {author} {\bibfnamefont {M.~Y.}\ \bibnamefont {Ivanov}}, \bibinfo
  {author} {\bibfnamefont {A.}~\bibnamefont {L’Huillier}},\ and\ \bibinfo
  {author} {\bibfnamefont {P.~B.}\ \bibnamefont {Corkum}},\ }\bibfield  {title}
  {\bibinfo {title} {Theory of high-harmonic generation by low-frequency laser
  fields},\ }\href {https://doi.org/10.1103/PhysRevA.49.2117} {\bibfield
  {journal} {\bibinfo  {journal} {Physical Review A}\ }\textbf {\bibinfo
  {volume} {49}},\ \bibinfo {pages} {2117} (\bibinfo {year}
  {1994})}\BibitemShut {NoStop}%
\bibitem [{\citenamefont {Corkum}(1993)}]{corkum_plasma_1993}%
  \BibitemOpen
  \bibfield  {author} {\bibinfo {author} {\bibfnamefont {P.~B.}\ \bibnamefont
  {Corkum}},\ }\bibfield  {title} {\bibinfo {title} {Plasma perspective on
  strong field multiphoton ionization},\ }\href
  {https://doi.org/10.1103/PhysRevLett.71.1994} {\bibfield  {journal} {\bibinfo
   {journal} {Physical Review Letters}\ }\textbf {\bibinfo {volume} {71}},\
  \bibinfo {pages} {1994} (\bibinfo {year} {1993})}\BibitemShut {NoStop}%
\bibitem [{\citenamefont {Krause}\ \emph {et~al.}(1992)\citenamefont {Krause},
  \citenamefont {Schafer},\ and\ \citenamefont
  {Kulander}}]{krause_high-order_1992}%
  \BibitemOpen
  \bibfield  {author} {\bibinfo {author} {\bibfnamefont {J.~L.}\ \bibnamefont
  {Krause}}, \bibinfo {author} {\bibfnamefont {K.~J.}\ \bibnamefont
  {Schafer}},\ and\ \bibinfo {author} {\bibfnamefont {K.~C.}\ \bibnamefont
  {Kulander}},\ }\bibfield  {title} {\bibinfo {title} {High-order harmonic
  generation from atoms and ions in the high intensity regime},\ }\href
  {https://doi.org/10.1103/PhysRevLett.68.3535} {\bibfield  {journal} {\bibinfo
   {journal} {Physical Review Letters}\ }\textbf {\bibinfo {volume} {68}},\
  \bibinfo {pages} {3535} (\bibinfo {year} {1992})}\BibitemShut {NoStop}%
\bibitem [{\citenamefont {Kulander}\ \emph {et~al.}(1993)\citenamefont
  {Kulander}, \citenamefont {Schafer},\ and\ \citenamefont
  {Krause}}]{Kulander1993}%
  \BibitemOpen
  \bibfield  {author} {\bibinfo {author} {\bibfnamefont {K.~C.}\ \bibnamefont
  {Kulander}}, \bibinfo {author} {\bibfnamefont {K.~J.}\ \bibnamefont
  {Schafer}},\ and\ \bibinfo {author} {\bibfnamefont {J.~L.}\ \bibnamefont
  {Krause}},\ }\bibfield  {title} {\bibinfo {title} {Dynamics of short-pulse
  excitation, ionization and harmonic conversion},\ }in\ \href@noop {} {\emph
  {\bibinfo {booktitle}
  {\href{https://www.springer.com/gp/book/9780306445873}{Super-Intense Laser
  Atom Physics}}}},\ \bibinfo {series} {NATO Advanced Studies Institute Series
  B: Physics}, Vol.\ \bibinfo {volume} {316},\ \bibinfo {editor} {edited by\
  \bibinfo {editor} {\bibfnamefont {B.}~\bibnamefont {Piraux}}, \bibinfo
  {editor} {\bibfnamefont {A.}~\bibnamefont {L'Huillier}},\ and\ \bibinfo
  {editor} {\bibfnamefont {K.}~\bibnamefont {Rz\k{a}\.zewski}}}\ (\bibinfo
  {publisher} {Plenum},\ \bibinfo {address} {New York},\ \bibinfo {year}
  {1993})\ pp.\ \bibinfo {pages} {95--110}\BibitemShut {NoStop}%
\bibitem [{\citenamefont {Salières}\ \emph {et~al.}(2001)\citenamefont
  {Salières}, \citenamefont {Carré}, \citenamefont {Le~Déroff},
  \citenamefont {Grasbon}, \citenamefont {Paulus}, \citenamefont {Walther},
  \citenamefont {Kopold}, \citenamefont {Becker}, \citenamefont {Milošević},
  \citenamefont {Sanpera},\ and\ \citenamefont
  {Lewenstein}}]{salieres_feynmans_2001}%
  \BibitemOpen
  \bibfield  {author} {\bibinfo {author} {\bibfnamefont {P.}~\bibnamefont
  {Salières}}, \bibinfo {author} {\bibfnamefont {B.}~\bibnamefont {Carré}},
  \bibinfo {author} {\bibfnamefont {L.}~\bibnamefont {Le~Déroff}}, \bibinfo
  {author} {\bibfnamefont {F.}~\bibnamefont {Grasbon}}, \bibinfo {author}
  {\bibfnamefont {G.~G.}\ \bibnamefont {Paulus}}, \bibinfo {author}
  {\bibfnamefont {H.}~\bibnamefont {Walther}}, \bibinfo {author} {\bibfnamefont
  {R.}~\bibnamefont {Kopold}}, \bibinfo {author} {\bibfnamefont
  {W.}~\bibnamefont {Becker}}, \bibinfo {author} {\bibfnamefont {D.~B.}\
  \bibnamefont {Milošević}}, \bibinfo {author} {\bibfnamefont
  {A.}~\bibnamefont {Sanpera}},\ and\ \bibinfo {author} {\bibfnamefont
  {M.}~\bibnamefont {Lewenstein}},\ }\bibfield  {title} {\bibinfo {title}
  {Feynman's {Path}-{Integral} {Approach} for {Intense}-{Laser}-{Atom}
  {Interactions}},\ }\bibfield  {journal} {\bibinfo  {journal} {Science}\
  }\textbf {\bibinfo {volume} {292}},\ \href
  {https://doi.org/10.1126/science.108836} {10.1126/science.108836} (\bibinfo
  {year} {2001})\BibitemShut {NoStop}%
\bibitem [{\citenamefont {Smirnova}\ and\ \citenamefont
  {Ivanov}(2014)}]{olga_simpleman}%
  \BibitemOpen
  \bibfield  {author} {\bibinfo {author} {\bibfnamefont {O.}~\bibnamefont
  {Smirnova}}\ and\ \bibinfo {author} {\bibfnamefont {M.}~\bibnamefont
  {Ivanov}},\ }\bibinfo {title} {Multielectron high harmonic generation: Simple
  man on a complex plane},\ in\ \href
  {https://doi.org/https://doi.org/10.1002/9783527677689.ch7} {\emph {\bibinfo
  {booktitle} {Attosecond and XUV Physics}}}\ (\bibinfo  {publisher} {John
  Wiley \& Sons, Ltd},\ \bibinfo {year} {2014})\ Chap.~\bibinfo {chapter} {7},
  pp.\ \bibinfo {pages} {201--256}\BibitemShut {NoStop}%
\bibitem [{\citenamefont {Amini}\ \emph {et~al.}(2019)\citenamefont {Amini},
  \citenamefont {Biegert}, \citenamefont {Calegari}, \citenamefont {Chacón},
  \citenamefont {Ciappina}, \citenamefont {Dauphin}, \citenamefont {Efimov},
  \citenamefont {Faria}, \citenamefont {Giergiel}, \citenamefont {Gniewek},
  \citenamefont {Landsman}, \citenamefont {Lesiuk}, \citenamefont {Mandrysz},
  \citenamefont {Maxwell}, \citenamefont {Moszyński}, \citenamefont {Ortmann},
  \citenamefont {Pérez-Hernández}, \citenamefont {Picón}, \citenamefont
  {Pisanty}, \citenamefont {Prauzner-Bechcicki}, \citenamefont {Sacha},
  \citenamefont {Suárez}, \citenamefont {Zaïr}, \citenamefont {Zakrzewski},\
  and\ \citenamefont {Lewenstein}}]{amini_symphony_2019}%
  \BibitemOpen
  \bibfield  {author} {\bibinfo {author} {\bibfnamefont {K.}~\bibnamefont
  {Amini}}, \bibinfo {author} {\bibfnamefont {J.}~\bibnamefont {Biegert}},
  \bibinfo {author} {\bibfnamefont {F.}~\bibnamefont {Calegari}}, \bibinfo
  {author} {\bibfnamefont {A.}~\bibnamefont {Chacón}}, \bibinfo {author}
  {\bibfnamefont {M.~F.}\ \bibnamefont {Ciappina}}, \bibinfo {author}
  {\bibfnamefont {A.}~\bibnamefont {Dauphin}}, \bibinfo {author} {\bibfnamefont
  {D.~K.}\ \bibnamefont {Efimov}}, \bibinfo {author} {\bibfnamefont {C.~F.
  d.~M.}\ \bibnamefont {Faria}}, \bibinfo {author} {\bibfnamefont
  {K.}~\bibnamefont {Giergiel}}, \bibinfo {author} {\bibfnamefont
  {P.}~\bibnamefont {Gniewek}}, \bibinfo {author} {\bibfnamefont {A.~S.}\
  \bibnamefont {Landsman}}, \bibinfo {author} {\bibfnamefont {M.}~\bibnamefont
  {Lesiuk}}, \bibinfo {author} {\bibfnamefont {M.}~\bibnamefont {Mandrysz}},
  \bibinfo {author} {\bibfnamefont {A.~S.}\ \bibnamefont {Maxwell}}, \bibinfo
  {author} {\bibfnamefont {R.}~\bibnamefont {Moszyński}}, \bibinfo {author}
  {\bibfnamefont {L.}~\bibnamefont {Ortmann}}, \bibinfo {author} {\bibfnamefont
  {J.~A.}\ \bibnamefont {Pérez-Hernández}}, \bibinfo {author} {\bibfnamefont
  {A.}~\bibnamefont {Picón}}, \bibinfo {author} {\bibfnamefont
  {E.}~\bibnamefont {Pisanty}}, \bibinfo {author} {\bibfnamefont
  {J.}~\bibnamefont {Prauzner-Bechcicki}}, \bibinfo {author} {\bibfnamefont
  {K.}~\bibnamefont {Sacha}}, \bibinfo {author} {\bibfnamefont
  {N.}~\bibnamefont {Suárez}}, \bibinfo {author} {\bibfnamefont
  {A.}~\bibnamefont {Zaïr}}, \bibinfo {author} {\bibfnamefont
  {J.}~\bibnamefont {Zakrzewski}},\ and\ \bibinfo {author} {\bibfnamefont
  {M.}~\bibnamefont {Lewenstein}},\ }\bibfield  {title} {\bibinfo {title}
  {Symphony on strong field approximation},\ }\href
  {https://doi.org/10.1088/1361-6633/ab2bb1} {\bibfield  {journal} {\bibinfo
  {journal} {Reports on Progress in Physics}\ }\textbf {\bibinfo {volume}
  {82}},\ \bibinfo {pages} {116001} (\bibinfo {year} {2019})}\BibitemShut
  {NoStop}%
\bibitem [{\citenamefont {Gonoskov}\ \emph {et~al.}(2016)\citenamefont
  {Gonoskov}, \citenamefont {Tsatrafyllis}, \citenamefont {Kominis},\ and\
  \citenamefont {Tzallas}}]{gonoskov_quantum_2016}%
  \BibitemOpen
  \bibfield  {author} {\bibinfo {author} {\bibfnamefont {I.~A.}\ \bibnamefont
  {Gonoskov}}, \bibinfo {author} {\bibfnamefont {N.}~\bibnamefont
  {Tsatrafyllis}}, \bibinfo {author} {\bibfnamefont {I.~K.}\ \bibnamefont
  {Kominis}},\ and\ \bibinfo {author} {\bibfnamefont {P.}~\bibnamefont
  {Tzallas}},\ }\bibfield  {title} {\bibinfo {title} {Quantum optical
  signatures in strong-field laser physics: {Infrared} photon counting in
  high-order-harmonic generation},\ }\href {https://doi.org/10.1038/srep32821}
  {\bibfield  {journal} {\bibinfo  {journal} {Scientific Reports}\ }\textbf
  {\bibinfo {volume} {6}},\ \bibinfo {pages} {32821} (\bibinfo {year}
  {2016})}\BibitemShut {NoStop}%
\bibitem [{\citenamefont {Tsatrafyllis}\ \emph {et~al.}(2017)\citenamefont
  {Tsatrafyllis}, \citenamefont {Kominis}, \citenamefont {Gonoskov},\ and\
  \citenamefont {Tzallas}}]{tsatrafyllis_high-order_2017}%
  \BibitemOpen
  \bibfield  {author} {\bibinfo {author} {\bibfnamefont {N.}~\bibnamefont
  {Tsatrafyllis}}, \bibinfo {author} {\bibfnamefont {I.~K.}\ \bibnamefont
  {Kominis}}, \bibinfo {author} {\bibfnamefont {I.~A.}\ \bibnamefont
  {Gonoskov}},\ and\ \bibinfo {author} {\bibfnamefont {P.}~\bibnamefont
  {Tzallas}},\ }\bibfield  {title} {\bibinfo {title} {High-order harmonics
  measured by the photon statistics of the infrared driving-field exiting the
  atomic medium},\ }\href {https://doi.org/10.1038/ncomms15170} {\bibfield
  {journal} {\bibinfo  {journal} {Nature Communications}\ }\textbf {\bibinfo
  {volume} {8}},\ \bibinfo {pages} {15170} (\bibinfo {year}
  {2017})}\BibitemShut {NoStop}%
\bibitem [{\citenamefont {Fuchs}\ \emph {et~al.}(2022)\citenamefont {Fuchs},
  \citenamefont {Abel}, \citenamefont {Nathanael}, \citenamefont {Reinhard},
  \citenamefont {Wiesner}, \citenamefont {Wünsche}, \citenamefont
  {Skruszewicz}, \citenamefont {Rödel}, \citenamefont {Born}, \citenamefont
  {Schmidt},\ and\ \citenamefont {Paulus}}]{fuchs_photon_2022}%
  \BibitemOpen
  \bibfield  {author} {\bibinfo {author} {\bibfnamefont {S.}~\bibnamefont
  {Fuchs}}, \bibinfo {author} {\bibfnamefont {J.~J.}\ \bibnamefont {Abel}},
  \bibinfo {author} {\bibfnamefont {J.}~\bibnamefont {Nathanael}}, \bibinfo
  {author} {\bibfnamefont {J.}~\bibnamefont {Reinhard}}, \bibinfo {author}
  {\bibfnamefont {F.}~\bibnamefont {Wiesner}}, \bibinfo {author} {\bibfnamefont
  {M.}~\bibnamefont {Wünsche}}, \bibinfo {author} {\bibfnamefont
  {S.}~\bibnamefont {Skruszewicz}}, \bibinfo {author} {\bibfnamefont
  {C.}~\bibnamefont {Rödel}}, \bibinfo {author} {\bibfnamefont
  {D.}~\bibnamefont {Born}}, \bibinfo {author} {\bibfnamefont {H.}~\bibnamefont
  {Schmidt}},\ and\ \bibinfo {author} {\bibfnamefont {G.~G.}\ \bibnamefont
  {Paulus}},\ }\bibfield  {title} {\bibinfo {title} {Photon counting of extreme
  ultraviolet high harmonics using a superconducting nanowire single-photon
  detector},\ }\href {https://doi.org/10.1007/s00340-022-07754-6} {\bibfield
  {journal} {\bibinfo  {journal} {Applied Physics B}\ }\textbf {\bibinfo
  {volume} {128}},\ \bibinfo {pages} {26} (\bibinfo {year} {2022})}\BibitemShut
  {NoStop}%
\bibitem [{\citenamefont {Gorlach}\ \emph {et~al.}(2020)\citenamefont
  {Gorlach}, \citenamefont {Neufeld}, \citenamefont {Rivera}, \citenamefont
  {Cohen},\ and\ \citenamefont {Kaminer}}]{gorlach_quantum-optical_2020}%
  \BibitemOpen
  \bibfield  {author} {\bibinfo {author} {\bibfnamefont {A.}~\bibnamefont
  {Gorlach}}, \bibinfo {author} {\bibfnamefont {O.}~\bibnamefont {Neufeld}},
  \bibinfo {author} {\bibfnamefont {N.}~\bibnamefont {Rivera}}, \bibinfo
  {author} {\bibfnamefont {O.}~\bibnamefont {Cohen}},\ and\ \bibinfo {author}
  {\bibfnamefont {I.}~\bibnamefont {Kaminer}},\ }\bibfield  {title} {\bibinfo
  {title} {The quantum-optical nature of high harmonic generation},\ }\href
  {https://doi.org/10.1038/s41467-020-18218-w} {\bibfield  {journal} {\bibinfo
  {journal} {Nature Communications}\ }\textbf {\bibinfo {volume} {11}},\
  \bibinfo {pages} {4598} (\bibinfo {year} {2020})}\BibitemShut {NoStop}%
\bibitem [{\citenamefont {Rivera}\ and\ \citenamefont
  {Kaminer}(2020)}]{rivera_lightmatter_2020}%
  \BibitemOpen
  \bibfield  {author} {\bibinfo {author} {\bibfnamefont {N.}~\bibnamefont
  {Rivera}}\ and\ \bibinfo {author} {\bibfnamefont {I.}~\bibnamefont
  {Kaminer}},\ }\bibfield  {title} {\bibinfo {title} {Light–matter
  interactions with photonic quasiparticles},\ }\href
  {https://doi.org/10.1038/s42254-020-0224-2} {\bibfield  {journal} {\bibinfo
  {journal} {Nature Reviews Physics}\ }\textbf {\bibinfo {volume} {2}},\
  \bibinfo {pages} {538} (\bibinfo {year} {2020})}\BibitemShut {NoStop}%
\bibitem [{\citenamefont {Varró}(2021)}]{varro_quantum_2021}%
  \BibitemOpen
  \bibfield  {author} {\bibinfo {author} {\bibfnamefont {S.}~\bibnamefont
  {Varró}},\ }\bibfield  {title} {\bibinfo {title} {Quantum {Optical}
  {Aspects} of {High}-{Harmonic} {Generation}},\ }\href
  {https://doi.org/10.3390/photonics8070269} {\bibfield  {journal} {\bibinfo
  {journal} {Photonics}\ }\textbf {\bibinfo {volume} {8}},\ \bibinfo {pages}
  {269} (\bibinfo {year} {2021})}\BibitemShut {NoStop}%
\bibitem [{\citenamefont {F\"oldi}\ \emph {et~al.}(2021)\citenamefont
  {F\"oldi}, \citenamefont {Magashegyi}, \citenamefont {Gombk\"oto},\ and\
  \citenamefont {Varr\'o}}]{foldi_describing_2021}%
  \BibitemOpen
  \bibfield  {author} {\bibinfo {author} {\bibfnamefont {P.}~\bibnamefont
  {F\"oldi}}, \bibinfo {author} {\bibfnamefont {I.}~\bibnamefont {Magashegyi}},
  \bibinfo {author} {\bibfnamefont {A.}~\bibnamefont {Gombk\"oto}},\ and\
  \bibinfo {author} {\bibfnamefont {S.}~\bibnamefont {Varr\'o}},\ }\bibfield
  {title} {\bibinfo {title} {Describing {High}-{Order} {Harmonic} {Generation}
  {Using} {Quantum} {Optical} {Models}},\ }\href
  {https://doi.org/10.3390/photonics8070263} {\bibfield  {journal} {\bibinfo
  {journal} {Photonics}\ }\textbf {\bibinfo {volume} {8}},\ \bibinfo {pages}
  {263} (\bibinfo {year} {2021})}\BibitemShut {NoStop}%
\bibitem [{\citenamefont {Gombk\"oto}\ \emph {et~al.}(2021)\citenamefont
  {Gombk\"oto}, \citenamefont {F\"oldi},\ and\ \citenamefont
  {Varr\'o}}]{gombkoto_quantum-optical_2021}%
  \BibitemOpen
  \bibfield  {author} {\bibinfo {author} {\bibfnamefont {A.}~\bibnamefont
  {Gombk\"oto}}, \bibinfo {author} {\bibfnamefont {P.}~\bibnamefont
  {F\"oldi}},\ and\ \bibinfo {author} {\bibfnamefont {S.}~\bibnamefont
  {Varr\'o}},\ }\bibfield  {title} {\bibinfo {title} {Quantum-optical
  description of photon statistics and cross correlations in high-order
  harmonic generation},\ }\href {https://doi.org/10.1103/PhysRevA.104.033703}
  {\bibfield  {journal} {\bibinfo  {journal} {Phys. Rev. A}\ }\textbf {\bibinfo
  {volume} {104}},\ \bibinfo {pages} {033703} (\bibinfo {year}
  {2021})}\BibitemShut {NoStop}%
\bibitem [{\citenamefont {Rivera-Dean}\ \emph
  {et~al.}(2022{\natexlab{a}})\citenamefont {Rivera-Dean}, \citenamefont
  {Lamprou}, \citenamefont {Pisanty}, \citenamefont {Stammer}, \citenamefont
  {Ordóñez}, \citenamefont {Maxwell}, \citenamefont {Ciappina}, \citenamefont
  {Lewenstein},\ and\ \citenamefont {Tzallas}}]{rivera-dean_strong_2022}%
  \BibitemOpen
  \bibfield  {author} {\bibinfo {author} {\bibfnamefont {J.}~\bibnamefont
  {Rivera-Dean}}, \bibinfo {author} {\bibfnamefont {T.}~\bibnamefont
  {Lamprou}}, \bibinfo {author} {\bibfnamefont {E.}~\bibnamefont {Pisanty}},
  \bibinfo {author} {\bibfnamefont {P.}~\bibnamefont {Stammer}}, \bibinfo
  {author} {\bibfnamefont {A.~F.}\ \bibnamefont {Ordóñez}}, \bibinfo {author}
  {\bibfnamefont {A.~S.}\ \bibnamefont {Maxwell}}, \bibinfo {author}
  {\bibfnamefont {M.~F.}\ \bibnamefont {Ciappina}}, \bibinfo {author}
  {\bibfnamefont {M.}~\bibnamefont {Lewenstein}},\ and\ \bibinfo {author}
  {\bibfnamefont {P.}~\bibnamefont {Tzallas}},\ }\bibfield  {title} {\bibinfo
  {title} {Strong laser fields and their power to generate controllable
  high-photon-number coherent-state superpositions},\ }\href
  {https://doi.org/10.1103/PhysRevA.105.033714} {\bibfield  {journal} {\bibinfo
   {journal} {Physical Review A}\ }\textbf {\bibinfo {volume} {105}},\ \bibinfo
  {pages} {033714} (\bibinfo {year} {2022}{\natexlab{a}})}\BibitemShut
  {NoStop}%
\bibitem [{\citenamefont {Even~Tzur}\ \emph {et~al.}(2022)\citenamefont
  {Even~Tzur}, \citenamefont {Gorlach}, \citenamefont {Birk}, \citenamefont
  {Rivera}, \citenamefont {Krüger}, \citenamefont {Kaminer},\ and\
  \citenamefont {Cohen}}]{Even_ATTO}%
  \BibitemOpen
  \bibfield  {author} {\bibinfo {author} {\bibfnamefont {M.}~\bibnamefont
  {Even~Tzur}}, \bibinfo {author} {\bibfnamefont {A.}~\bibnamefont {Gorlach}},
  \bibinfo {author} {\bibfnamefont {M.}~\bibnamefont {Birk}}, \bibinfo {author}
  {\bibfnamefont {N.}~\bibnamefont {Rivera}}, \bibinfo {author} {\bibfnamefont
  {M.}~\bibnamefont {Krüger}}, \bibinfo {author} {\bibfnamefont
  {I.}~\bibnamefont {Kaminer}},\ and\ \bibinfo {author} {\bibfnamefont
  {O.}~\bibnamefont {Cohen}},\ }\bibinfo {title} {High harmonic generation
  driven by quantum light},\ in\ \href
  {https://sciences.ucf.edu/physics/atto/book-of-abstracts/} {\emph {\bibinfo
  {booktitle} {ATTO 8th International Conference on Attosecond Science and
  Technology (Book of Abstracts)}}}\ (\bibinfo  {publisher} {University of
  Central Florida, Orlando},\ \bibinfo {year} {2022})\ p.~\bibinfo {pages}
  {88}\BibitemShut {NoStop}%
\bibitem [{\citenamefont {Lewenstein}\ \emph {et~al.}(2021)\citenamefont
  {Lewenstein}, \citenamefont {Ciappina}, \citenamefont {Pisanty},
  \citenamefont {Rivera-Dean}, \citenamefont {Stammer}, \citenamefont
  {Lamprou},\ and\ \citenamefont {Tzallas}}]{lewenstein_generation_2021}%
  \BibitemOpen
  \bibfield  {author} {\bibinfo {author} {\bibfnamefont {M.}~\bibnamefont
  {Lewenstein}}, \bibinfo {author} {\bibfnamefont {M.~F.}\ \bibnamefont
  {Ciappina}}, \bibinfo {author} {\bibfnamefont {E.}~\bibnamefont {Pisanty}},
  \bibinfo {author} {\bibfnamefont {J.}~\bibnamefont {Rivera-Dean}}, \bibinfo
  {author} {\bibfnamefont {P.}~\bibnamefont {Stammer}}, \bibinfo {author}
  {\bibfnamefont {T.}~\bibnamefont {Lamprou}},\ and\ \bibinfo {author}
  {\bibfnamefont {P.}~\bibnamefont {Tzallas}},\ }\bibfield  {title} {\bibinfo
  {title} {Generation of optical {Schrödinger} cat states in intense
  laser–matter interactions},\ }\href
  {https://doi.org/10.1038/s41567-021-01317-w} {\bibfield  {journal} {\bibinfo
  {journal} {Nature Physics}\ }\textbf {\bibinfo {volume} {17}},\ \bibinfo
  {pages} {1104} (\bibinfo {year} {2021})}\BibitemShut {NoStop}%
\bibitem [{\citenamefont {Stammer}\ \emph
  {et~al.}(2022{\natexlab{a}})\citenamefont {Stammer}, \citenamefont
  {Rivera-Dean}, \citenamefont {Maxwell}, \citenamefont {Lamprou},
  \citenamefont {Ordóñez}, \citenamefont {Ciappina}, \citenamefont
  {Tzallas},\ and\ \citenamefont {Lewenstein}}]{stammer_quantum_2022}%
  \BibitemOpen
  \bibfield  {author} {\bibinfo {author} {\bibfnamefont {P.}~\bibnamefont
  {Stammer}}, \bibinfo {author} {\bibfnamefont {J.}~\bibnamefont
  {Rivera-Dean}}, \bibinfo {author} {\bibfnamefont {A.}~\bibnamefont
  {Maxwell}}, \bibinfo {author} {\bibfnamefont {T.}~\bibnamefont {Lamprou}},
  \bibinfo {author} {\bibfnamefont {A.}~\bibnamefont {Ordóñez}}, \bibinfo
  {author} {\bibfnamefont {M.~F.}\ \bibnamefont {Ciappina}}, \bibinfo {author}
  {\bibfnamefont {P.}~\bibnamefont {Tzallas}},\ and\ \bibinfo {author}
  {\bibfnamefont {M.}~\bibnamefont {Lewenstein}},\ }\href
  {https://doi.org/10.48550/arXiv.2206.04308} {\bibinfo {title} {Quantum
  electrodynamics of ultra-intense laser-matter interactions}} (\bibinfo {year}
  {2022}{\natexlab{a}}),\ \bibinfo {note} {number: arXiv:2206.04308
  arXiv:2206.04308 [quant-ph]}\BibitemShut {NoStop}%
\bibitem [{\citenamefont {Stammer}\ \emph
  {et~al.}(2022{\natexlab{b}})\citenamefont {Stammer}, \citenamefont
  {Rivera-Dean}, \citenamefont {Lamprou}, \citenamefont {Pisanty},
  \citenamefont {Ciappina}, \citenamefont {Tzallas},\ and\ \citenamefont
  {Lewenstein}}]{stammer_high_2022}%
  \BibitemOpen
  \bibfield  {author} {\bibinfo {author} {\bibfnamefont {P.}~\bibnamefont
  {Stammer}}, \bibinfo {author} {\bibfnamefont {J.}~\bibnamefont
  {Rivera-Dean}}, \bibinfo {author} {\bibfnamefont {T.}~\bibnamefont
  {Lamprou}}, \bibinfo {author} {\bibfnamefont {E.}~\bibnamefont {Pisanty}},
  \bibinfo {author} {\bibfnamefont {M.~F.}\ \bibnamefont {Ciappina}}, \bibinfo
  {author} {\bibfnamefont {P.}~\bibnamefont {Tzallas}},\ and\ \bibinfo {author}
  {\bibfnamefont {M.}~\bibnamefont {Lewenstein}},\ }\bibfield  {title}
  {\bibinfo {title} {High {Photon} {Number} {Entangled} {States} and {Coherent}
  {State} {Superposition} from the {Extreme} {Ultraviolet} to the {Far}
  {Infrared}},\ }\href {https://doi.org/10.1103/PhysRevLett.128.123603}
  {\bibfield  {journal} {\bibinfo  {journal} {Physical Review Letters}\
  }\textbf {\bibinfo {volume} {128}},\ \bibinfo {pages} {123603} (\bibinfo
  {year} {2022}{\natexlab{b}})}\BibitemShut {NoStop}%
\bibitem [{\citenamefont {Tzallas}\ \emph {et~al.}(2022)\citenamefont
  {Tzallas}, \citenamefont {Lamprou}, \citenamefont {Rivera-Dean},
  \citenamefont {Stammer}, \citenamefont {Maxwell}, \citenamefont {Ordóñez},
  \citenamefont {Pisanty}, \citenamefont {Ciappina},\ and\ \citenamefont
  {Lewenstein}}]{Paris_ATTO}%
  \BibitemOpen
  \bibfield  {author} {\bibinfo {author} {\bibfnamefont {P.}~\bibnamefont
  {Tzallas}}, \bibinfo {author} {\bibfnamefont {T.}~\bibnamefont {Lamprou}},
  \bibinfo {author} {\bibfnamefont {J.}~\bibnamefont {Rivera-Dean}}, \bibinfo
  {author} {\bibfnamefont {P.}~\bibnamefont {Stammer}}, \bibinfo {author}
  {\bibfnamefont {A.~S.}\ \bibnamefont {Maxwell}}, \bibinfo {author}
  {\bibfnamefont {A.~F.}\ \bibnamefont {Ordóñez}}, \bibinfo {author}
  {\bibfnamefont {E.}~\bibnamefont {Pisanty}}, \bibinfo {author} {\bibfnamefont
  {M.~F.}\ \bibnamefont {Ciappina}},\ and\ \bibinfo {author} {\bibfnamefont
  {M.}~\bibnamefont {Lewenstein}},\ }\bibinfo {title} {Generation of optical
  ``cat'' states using ``conditioning'' approaches in intense laser-atom
  interactions},\ in\ \href
  {https://sciences.ucf.edu/physics/atto/book-of-abstracts/} {\emph {\bibinfo
  {booktitle} {ATTO 8th International Conference on Attosecond Science and
  Technology (Book of Abstracts)}}}\ (\bibinfo  {publisher} {University of
  Central Florida, Orlando},\ \bibinfo {year} {2022})\ p.~\bibinfo {pages}
  {86}\BibitemShut {NoStop}%
\bibitem [{\citenamefont {Lewenstein}(2022)}]{Maciej_ATTO}%
  \BibitemOpen
  \bibfield  {author} {\bibinfo {author} {\bibfnamefont {M.}~\bibnamefont
  {Lewenstein}},\ }\bibinfo {title} {{Attoscience and Quantum Information}},\
  in\ \href {https://sciences.ucf.edu/physics/atto/book-of-abstracts/} {\emph
  {\bibinfo {booktitle} {ATTO 8th International Conference on Attosecond
  Science and Technology (Book of Abstracts)}}}\ (\bibinfo  {publisher}
  {University of Central Florida, Orlando},\ \bibinfo {year} {2022})\
  p.~\bibinfo {pages} {87}\BibitemShut {NoStop}%
\bibitem [{\citenamefont {Nielsen}\ and\ \citenamefont
  {Chuang}(2010)}]{nielsen_quantum_2010}%
  \BibitemOpen
  \bibfield  {author} {\bibinfo {author} {\bibfnamefont {M.~A.}\ \bibnamefont
  {Nielsen}}\ and\ \bibinfo {author} {\bibfnamefont {I.~L.}\ \bibnamefont
  {Chuang}},\ }\href@noop {} {\emph {\bibinfo {title} {Quantum {Computation}
  and {Quantum} {Information}: 10th {Anniversary} {Edition}}}}\ (\bibinfo
  {publisher} {Cambridge University Press},\ \bibinfo {year}
  {2010})\BibitemShut {NoStop}%
\bibitem [{\citenamefont {Bennett}\ \emph {et~al.}(1993)\citenamefont
  {Bennett}, \citenamefont {Brassard}, \citenamefont {Crépeau}, \citenamefont
  {Jozsa}, \citenamefont {Peres},\ and\ \citenamefont
  {Wootters}}]{bennett_teleporting_1993}%
  \BibitemOpen
  \bibfield  {author} {\bibinfo {author} {\bibfnamefont {C.~H.}\ \bibnamefont
  {Bennett}}, \bibinfo {author} {\bibfnamefont {G.}~\bibnamefont {Brassard}},
  \bibinfo {author} {\bibfnamefont {C.}~\bibnamefont {Crépeau}}, \bibinfo
  {author} {\bibfnamefont {R.}~\bibnamefont {Jozsa}}, \bibinfo {author}
  {\bibfnamefont {A.}~\bibnamefont {Peres}},\ and\ \bibinfo {author}
  {\bibfnamefont {W.~K.}\ \bibnamefont {Wootters}},\ }\bibfield  {title}
  {\bibinfo {title} {Teleporting an unknown quantum state via dual classical
  and {Einstein}-{Podolsky}-{Rosen} channels},\ }\href
  {https://doi.org/10.1103/PhysRevLett.70.1895} {\bibfield  {journal} {\bibinfo
   {journal} {Physical Review Letters}\ }\textbf {\bibinfo {volume} {70}},\
  \bibinfo {pages} {1895} (\bibinfo {year} {1993})}\BibitemShut {NoStop}%
\bibitem [{\citenamefont {Bouwmeester}\ \emph {et~al.}(1997)\citenamefont
  {Bouwmeester}, \citenamefont {Pan}, \citenamefont {Mattle}, \citenamefont
  {Eibl}, \citenamefont {Weinfurter},\ and\ \citenamefont
  {Zeilinger}}]{bouwmeester_experimental_1997}%
  \BibitemOpen
  \bibfield  {author} {\bibinfo {author} {\bibfnamefont {D.}~\bibnamefont
  {Bouwmeester}}, \bibinfo {author} {\bibfnamefont {J.-W.}\ \bibnamefont
  {Pan}}, \bibinfo {author} {\bibfnamefont {K.}~\bibnamefont {Mattle}},
  \bibinfo {author} {\bibfnamefont {M.}~\bibnamefont {Eibl}}, \bibinfo {author}
  {\bibfnamefont {H.}~\bibnamefont {Weinfurter}},\ and\ \bibinfo {author}
  {\bibfnamefont {A.}~\bibnamefont {Zeilinger}},\ }\bibfield  {title} {\bibinfo
  {title} {Experimental quantum teleportation},\ }\href
  {https://doi.org/10.1038/37539} {\bibfield  {journal} {\bibinfo  {journal}
  {Nature}\ }\textbf {\bibinfo {volume} {390}},\ \bibinfo {pages} {575}
  (\bibinfo {year} {1997})}\BibitemShut {NoStop}%
\bibitem [{\citenamefont {Boschi}\ \emph {et~al.}(1998)\citenamefont {Boschi},
  \citenamefont {Branca}, \citenamefont {De~Martini}, \citenamefont {Hardy},\
  and\ \citenamefont {Popescu}}]{boschi_experimental_1998}%
  \BibitemOpen
  \bibfield  {author} {\bibinfo {author} {\bibfnamefont {D.}~\bibnamefont
  {Boschi}}, \bibinfo {author} {\bibfnamefont {S.}~\bibnamefont {Branca}},
  \bibinfo {author} {\bibfnamefont {F.}~\bibnamefont {De~Martini}}, \bibinfo
  {author} {\bibfnamefont {L.}~\bibnamefont {Hardy}},\ and\ \bibinfo {author}
  {\bibfnamefont {S.}~\bibnamefont {Popescu}},\ }\bibfield  {title} {\bibinfo
  {title} {Experimental {Realization} of {Teleporting} an {Unknown} {Pure}
  {Quantum} {State} via {Dual} {Classical} and {Einstein}-{Podolsky}-{Rosen}
  {Channels}},\ }\href {https://doi.org/10.1103/PhysRevLett.80.1121} {\bibfield
   {journal} {\bibinfo  {journal} {Physical Review Letters}\ }\textbf {\bibinfo
  {volume} {80}},\ \bibinfo {pages} {1121} (\bibinfo {year}
  {1998})}\BibitemShut {NoStop}%
\bibitem [{\citenamefont {Gisin}\ and\ \citenamefont
  {Thew}(2007)}]{gisin_quantum_2007}%
  \BibitemOpen
  \bibfield  {author} {\bibinfo {author} {\bibfnamefont {N.}~\bibnamefont
  {Gisin}}\ and\ \bibinfo {author} {\bibfnamefont {R.}~\bibnamefont {Thew}},\
  }\bibfield  {title} {\bibinfo {title} {Quantum communication},\ }\href
  {https://doi.org/10.1038/nphoton.2007.22} {\bibfield  {journal} {\bibinfo
  {journal} {Nature Photonics}\ }\textbf {\bibinfo {volume} {1}},\ \bibinfo
  {pages} {165} (\bibinfo {year} {2007})}\BibitemShut {NoStop}%
\bibitem [{\citenamefont {Liu}\ \emph {et~al.}(1999)\citenamefont {Liu},
  \citenamefont {Eberly}, \citenamefont {Haan},\ and\ \citenamefont
  {Grobe}}]{liu_correlation_1999}%
  \BibitemOpen
  \bibfield  {author} {\bibinfo {author} {\bibfnamefont {W.-C.}\ \bibnamefont
  {Liu}}, \bibinfo {author} {\bibfnamefont {J.~H.}\ \bibnamefont {Eberly}},
  \bibinfo {author} {\bibfnamefont {S.~L.}\ \bibnamefont {Haan}},\ and\
  \bibinfo {author} {\bibfnamefont {R.}~\bibnamefont {Grobe}},\ }\bibfield
  {title} {\bibinfo {title} {Correlation {Effects} in {Two}-{Electron} {Model}
  {Atoms} in {Intense} {Laser} {Fields}},\ }\href
  {https://doi.org/10.1103/PhysRevLett.83.520} {\bibfield  {journal} {\bibinfo
  {journal} {Physical Review Letters}\ }\textbf {\bibinfo {volume} {83}},\
  \bibinfo {pages} {520} (\bibinfo {year} {1999})}\BibitemShut {NoStop}%
\bibitem [{\citenamefont {Christov}(1999)}]{christov_phase-dependent_1999}%
  \BibitemOpen
  \bibfield  {author} {\bibinfo {author} {\bibfnamefont {I.~P.}\ \bibnamefont
  {Christov}},\ }\bibfield  {title} {\bibinfo {title} {Phase-dependent loss due
  to nonadiabatic ionization by sub-10-fs pulses},\ }\href
  {https://doi.org/10.1364/OL.24.001425} {\bibfield  {journal} {\bibinfo
  {journal} {Optics Letters}\ }\textbf {\bibinfo {volume} {24}},\ \bibinfo
  {pages} {1425} (\bibinfo {year} {1999})}\BibitemShut {NoStop}%
\bibitem [{\citenamefont {Christov}(2000)}]{christov_phase-dependent_2000}%
  \BibitemOpen
  \bibfield  {author} {\bibinfo {author} {\bibfnamefont {I.}~\bibnamefont
  {Christov}},\ }\bibfield  {title} {\bibinfo {title} {Phase-dependent
  ionization in the barrier suppression regime},\ }\href
  {https://doi.org/10.1007/PL00021158} {\bibfield  {journal} {\bibinfo
  {journal} {Applied Physics B}\ }\textbf {\bibinfo {volume} {70}},\ \bibinfo
  {pages} {459} (\bibinfo {year} {2000})}\BibitemShut {NoStop}%
\bibitem [{\citenamefont {Omiste}\ and\ \citenamefont
  {Madsen}(2019)}]{omiste_effects_2019}%
  \BibitemOpen
  \bibfield  {author} {\bibinfo {author} {\bibfnamefont {J.~J.}\ \bibnamefont
  {Omiste}}\ and\ \bibinfo {author} {\bibfnamefont {L.~B.}\ \bibnamefont
  {Madsen}},\ }\bibfield  {title} {\bibinfo {title} {Effects of core space and
  excitation levels on ground-state correlation and photoionization dynamics of
  {Be} and {Ne}},\ }\href {https://doi.org/10.1063/1.5082940} {\bibfield
  {journal} {\bibinfo  {journal} {The Journal of Chemical Physics}\ }\textbf
  {\bibinfo {volume} {150}},\ \bibinfo {pages} {084305} (\bibinfo {year}
  {2019})}\BibitemShut {NoStop}%
\bibitem [{\citenamefont {Maxwell}\ \emph {et~al.}(2021)\citenamefont
  {Maxwell}, \citenamefont {Madsen},\ and\ \citenamefont
  {Lewenstein}}]{maxwell_entanglement_2021}%
  \BibitemOpen
  \bibfield  {author} {\bibinfo {author} {\bibfnamefont {A.~S.}\ \bibnamefont
  {Maxwell}}, \bibinfo {author} {\bibfnamefont {L.~B.}\ \bibnamefont
  {Madsen}},\ and\ \bibinfo {author} {\bibfnamefont {M.}~\bibnamefont
  {Lewenstein}},\ }\href {https://doi.org/10.48550/arXiv.2111.10148} {\bibinfo
  {title} {Entanglement of {Orbital} {Angular} {Momentum} in {Non}-{Sequential}
  {Double} {Ionization}}} (\bibinfo {year} {2021}),\ \bibinfo {note} {number:
  arXiv:2111.10148 arXiv:2111.10148 [physics, physics:quant-ph]}\BibitemShut
  {NoStop}%
\bibitem [{\citenamefont {Spanner}\ and\ \citenamefont
  {Brumer}(2007{\natexlab{a}})}]{spanner_coherent_2007}%
  \BibitemOpen
  \bibfield  {author} {\bibinfo {author} {\bibfnamefont {M.}~\bibnamefont
  {Spanner}}\ and\ \bibinfo {author} {\bibfnamefont {P.}~\bibnamefont
  {Brumer}},\ }\bibfield  {title} {\bibinfo {title} {Coherent control and
  entanglement in the attosecond electron-recollision dissociation of
  \${\textbackslash}mathrm\{{D}\}\_\{2\}\{\}{\textasciicircum}\{+\}\$},\ }\href
  {https://doi.org/10.1103/PhysRevA.76.013409} {\bibfield  {journal} {\bibinfo
  {journal} {Physical Review A}\ }\textbf {\bibinfo {volume} {76}},\ \bibinfo
  {pages} {013409} (\bibinfo {year} {2007}{\natexlab{a}})}\BibitemShut
  {NoStop}%
\bibitem [{\citenamefont {Spanner}\ and\ \citenamefont
  {Brumer}(2007{\natexlab{b}})}]{spanner_entanglement_2007}%
  \BibitemOpen
  \bibfield  {author} {\bibinfo {author} {\bibfnamefont {M.}~\bibnamefont
  {Spanner}}\ and\ \bibinfo {author} {\bibfnamefont {P.}~\bibnamefont
  {Brumer}},\ }\bibfield  {title} {\bibinfo {title} {Entanglement and
  timing-based mechanisms in the coherent control of scattering processes},\
  }\href {https://doi.org/10.1103/PhysRevA.76.013408} {\bibfield  {journal}
  {\bibinfo  {journal} {Physical Review A}\ }\textbf {\bibinfo {volume} {76}},\
  \bibinfo {pages} {013408} (\bibinfo {year} {2007}{\natexlab{b}})}\BibitemShut
  {NoStop}%
\bibitem [{\citenamefont {Czirják}\ \emph {et~al.}(2013)\citenamefont
  {Czirják}, \citenamefont {Majorosi}, \citenamefont {Kovács},\ and\
  \citenamefont {Benedict}}]{czirjak_emergence_2013}%
  \BibitemOpen
  \bibfield  {author} {\bibinfo {author} {\bibfnamefont {A.}~\bibnamefont
  {Czirják}}, \bibinfo {author} {\bibfnamefont {S.}~\bibnamefont {Majorosi}},
  \bibinfo {author} {\bibfnamefont {J.}~\bibnamefont {Kovács}},\ and\ \bibinfo
  {author} {\bibfnamefont {M.~G.}\ \bibnamefont {Benedict}},\ }\bibfield
  {title} {\bibinfo {title} {Emergence of oscillations in quantum entanglement
  during rescattering},\ }\href
  {https://doi.org/10.1088/0031-8949/2013/T153/014013} {\bibfield  {journal}
  {\bibinfo  {journal} {Physica Scripta}\ }\textbf {\bibinfo {volume} {T153}},\
  \bibinfo {pages} {014013} (\bibinfo {year} {2013})}\BibitemShut {NoStop}%
\bibitem [{\citenamefont {Majorosi}\ \emph {et~al.}(2017)\citenamefont
  {Majorosi}, \citenamefont {Benedict},\ and\ \citenamefont
  {Czirják}}]{majorosi_quantum_2017}%
  \BibitemOpen
  \bibfield  {author} {\bibinfo {author} {\bibfnamefont {S.}~\bibnamefont
  {Majorosi}}, \bibinfo {author} {\bibfnamefont {M.~G.}\ \bibnamefont
  {Benedict}},\ and\ \bibinfo {author} {\bibfnamefont {A.}~\bibnamefont
  {Czirják}},\ }\bibfield  {title} {\bibinfo {title} {Quantum entanglement in
  strong-field ionization},\ }\href
  {https://doi.org/10.1103/PhysRevA.96.043412} {\bibfield  {journal} {\bibinfo
  {journal} {Physical Review A}\ }\textbf {\bibinfo {volume} {96}},\ \bibinfo
  {pages} {043412} (\bibinfo {year} {2017})}\BibitemShut {NoStop}%
\bibitem [{\citenamefont {Ruberti}(2021)}]{ruberti_quantum_2021}%
  \BibitemOpen
  \bibfield  {author} {\bibinfo {author} {\bibfnamefont {M.}~\bibnamefont
  {Ruberti}},\ }\bibfield  {title} {\bibinfo {title} {Quantum electronic
  coherences by attosecond transient absorption spectroscopy: ab initio
  {B}-spline {RCS}-{ADC} study},\ }\href {https://doi.org/10.1039/D0FD00104J}
  {\bibfield  {journal} {\bibinfo  {journal} {Faraday Discussions}\ }\textbf
  {\bibinfo {volume} {228}},\ \bibinfo {pages} {286} (\bibinfo {year}
  {2021})}\BibitemShut {NoStop}%
\bibitem [{\citenamefont {Vrakking}(2021)}]{vrakking_control_2021}%
  \BibitemOpen
  \bibfield  {author} {\bibinfo {author} {\bibfnamefont {M.~J.}\ \bibnamefont
  {Vrakking}},\ }\bibfield  {title} {\bibinfo {title} {Control of {Attosecond}
  {Entanglement} and {Coherence}},\ }\href
  {https://doi.org/10.1103/PhysRevLett.126.113203} {\bibfield  {journal}
  {\bibinfo  {journal} {Physical Review Letters}\ }\textbf {\bibinfo {volume}
  {126}},\ \bibinfo {pages} {113203} (\bibinfo {year} {2021})}\BibitemShut
  {NoStop}%
\bibitem [{\citenamefont {Koll}\ \emph
  {et~al.}(2022{\natexlab{a}})\citenamefont {Koll}, \citenamefont {Maikowski},
  \citenamefont {Drescher}, \citenamefont {Witting},\ and\ \citenamefont
  {Vrakking}}]{koll_experimental_2022}%
  \BibitemOpen
  \bibfield  {author} {\bibinfo {author} {\bibfnamefont {L.-M.}\ \bibnamefont
  {Koll}}, \bibinfo {author} {\bibfnamefont {L.}~\bibnamefont {Maikowski}},
  \bibinfo {author} {\bibfnamefont {L.}~\bibnamefont {Drescher}}, \bibinfo
  {author} {\bibfnamefont {T.}~\bibnamefont {Witting}},\ and\ \bibinfo {author}
  {\bibfnamefont {M.~J.}\ \bibnamefont {Vrakking}},\ }\bibfield  {title}
  {\bibinfo {title} {Experimental {Control} of {Quantum}-{Mechanical}
  {Entanglement} in an {Attosecond} {Pump}-{Probe} {Experiment}},\ }\href
  {https://doi.org/10.1103/PhysRevLett.128.043201} {\bibfield  {journal}
  {\bibinfo  {journal} {Physical Review Letters}\ }\textbf {\bibinfo {volume}
  {128}},\ \bibinfo {pages} {043201} (\bibinfo {year}
  {2022}{\natexlab{a}})}\BibitemShut {NoStop}%
\bibitem [{\citenamefont {Koll}\ \emph
  {et~al.}(2022{\natexlab{b}})\citenamefont {Koll}, \citenamefont {Maikowski},
  \citenamefont {Drescher}, \citenamefont {Witting},\ and\ \citenamefont
  {Vrakking}}]{Koll_ATTO}%
  \BibitemOpen
  \bibfield  {author} {\bibinfo {author} {\bibfnamefont {L.~M.}\ \bibnamefont
  {Koll}}, \bibinfo {author} {\bibfnamefont {L.}~\bibnamefont {Maikowski}},
  \bibinfo {author} {\bibfnamefont {L.}~\bibnamefont {Drescher}}, \bibinfo
  {author} {\bibfnamefont {T.}~\bibnamefont {Witting}},\ and\ \bibinfo {author}
  {\bibfnamefont {M.~J.~J.}\ \bibnamefont {Vrakking}},\ }\bibinfo {title}
  {Experimental control of quantum-mechanical entanglement in an attosecond
  pumb-probe experiment},\ in\ \href
  {https://sciences.ucf.edu/physics/atto/book-of-abstracts/} {\emph {\bibinfo
  {booktitle} {ATTO 8th International Conference on Attosecond Science and
  Technology (Book of Abstracts)}}}\ (\bibinfo  {publisher} {University of
  Central Florida, Orlando},\ \bibinfo {year} {2022})\ p.~\bibinfo {pages}
  {90}\BibitemShut {NoStop}%
\bibitem [{\citenamefont {Shobeiry}\ \emph {et~al.}(2022)\citenamefont
  {Shobeiry}, \citenamefont {Brunner}, \citenamefont {Fross}, \citenamefont
  {Srinivas}, \citenamefont {Buchleitner}, \citenamefont {Pfeifer},
  \citenamefont {Harth},\ and\ \citenamefont {Moshammer}}]{Shobeiry_ATTO}%
  \BibitemOpen
  \bibfield  {author} {\bibinfo {author} {\bibfnamefont {F.}~\bibnamefont
  {Shobeiry}}, \bibinfo {author} {\bibfnamefont {E.}~\bibnamefont {Brunner}},
  \bibinfo {author} {\bibfnamefont {P.}~\bibnamefont {Fross}}, \bibinfo
  {author} {\bibfnamefont {H.}~\bibnamefont {Srinivas}}, \bibinfo {author}
  {\bibfnamefont {A.}~\bibnamefont {Buchleitner}}, \bibinfo {author}
  {\bibfnamefont {T.}~\bibnamefont {Pfeifer}}, \bibinfo {author} {\bibfnamefont
  {A.}~\bibnamefont {Harth}},\ and\ \bibinfo {author} {\bibfnamefont
  {R.}~\bibnamefont {Moshammer}},\ }\bibinfo {title} {Sub-femtosecond optical
  control of entangled states},\ in\ \href
  {https://sciences.ucf.edu/physics/atto/book-of-abstracts/} {\emph {\bibinfo
  {booktitle} {ATTO 8th International Conference on Attosecond Science and
  Technology (Book of Abstracts)}}}\ (\bibinfo  {publisher} {University of
  Central Florida, Orlando},\ \bibinfo {year} {2022})\ p.~\bibinfo {pages}
  {91}\BibitemShut {NoStop}%
\bibitem [{\citenamefont {Eckart}\ \emph {et~al.}(2021)\citenamefont {Eckart},
  \citenamefont {Trabert}, \citenamefont {Rist}, \citenamefont {Geyer},
  \citenamefont {Schmidt}, \citenamefont {Fehre},\ and\ \citenamefont
  {Kunitski}}]{eckart_ultrafast_2021}%
  \BibitemOpen
  \bibfield  {author} {\bibinfo {author} {\bibfnamefont {S.}~\bibnamefont
  {Eckart}}, \bibinfo {author} {\bibfnamefont {D.}~\bibnamefont {Trabert}},
  \bibinfo {author} {\bibfnamefont {J.}~\bibnamefont {Rist}}, \bibinfo {author}
  {\bibfnamefont {A.}~\bibnamefont {Geyer}}, \bibinfo {author} {\bibfnamefont
  {L.~P.~H.}\ \bibnamefont {Schmidt}}, \bibinfo {author} {\bibfnamefont
  {K.}~\bibnamefont {Fehre}},\ and\ \bibinfo {author} {\bibfnamefont
  {M.}~\bibnamefont {Kunitski}},\ }\href
  {https://doi.org/10.48550/arXiv.2108.10426} {\bibinfo {title} {Ultrafast
  preparation and strong-field ionization of an atomic {Bell}-like state}}
  (\bibinfo {year} {2021}),\ \bibinfo {note} {arXiv:2108.10426 [physics,
  physics:quant-ph]}\BibitemShut {NoStop}%
\bibitem [{\citenamefont {Eckart}\ \emph {et~al.}(2022)\citenamefont {Eckart},
  \citenamefont {Trabert}, \citenamefont {Rist}, \citenamefont {Geyer},
  \citenamefont {Schmidt}, \citenamefont {Fehre},\ and\ \citenamefont
  {Kunitksi}}]{Eckart_ATTO}%
  \BibitemOpen
  \bibfield  {author} {\bibinfo {author} {\bibfnamefont {S.}~\bibnamefont
  {Eckart}}, \bibinfo {author} {\bibfnamefont {D.}~\bibnamefont {Trabert}},
  \bibinfo {author} {\bibfnamefont {J.}~\bibnamefont {Rist}}, \bibinfo {author}
  {\bibfnamefont {A.}~\bibnamefont {Geyer}}, \bibinfo {author} {\bibfnamefont
  {L.~P.~H.}\ \bibnamefont {Schmidt}}, \bibinfo {author} {\bibfnamefont
  {K.}~\bibnamefont {Fehre}},\ and\ \bibinfo {author} {\bibfnamefont
  {M.}~\bibnamefont {Kunitksi}},\ }\bibinfo {title} {Ultrafast preparation and
  detection of entangled atoms},\ in\ \href
  {https://sciences.ucf.edu/physics/atto/book-of-abstracts/} {\emph {\bibinfo
  {booktitle} {ATTO 8th International Conference on Attosecond Science and
  Technology (Book of Abstracts)}}}\ (\bibinfo  {publisher} {University of
  Central Florida, Orlando},\ \bibinfo {year} {2022})\ p.~\bibinfo {pages}
  {92}\BibitemShut {NoStop}%
\bibitem [{\citenamefont {Stammer}(2022)}]{stammer_theory_2022}%
  \BibitemOpen
  \bibfield  {author} {\bibinfo {author} {\bibfnamefont {P.}~\bibnamefont
  {Stammer}},\ }\href {https://doi.org/10.48550/arXiv.2203.04354} {\bibinfo
  {title} {Theory of entanglement and measurement in high harmonic generation}}
  (\bibinfo {year} {2022}),\ \bibinfo {note} {number: arXiv:2203.04354
  arXiv:2203.04354 [physics, physics:quant-ph]}\BibitemShut {NoStop}%
\bibitem [{\citenamefont {Plenio}\ and\ \citenamefont
  {Virmani}(2007)}]{plenio_introduction_2007}%
  \BibitemOpen
  \bibfield  {author} {\bibinfo {author} {\bibfnamefont {M.~B.}\ \bibnamefont
  {Plenio}}\ and\ \bibinfo {author} {\bibfnamefont {S.}~\bibnamefont
  {Virmani}},\ }\bibfield  {title} {\bibinfo {title} {An introduction to
  entanglement measures},\ }\href@noop {} {\bibfield  {journal} {\bibinfo
  {journal} {Quantum Information \& Computation}\ }\textbf {\bibinfo {volume}
  {7}},\ \bibinfo {pages} {1} (\bibinfo {year} {2007})}\BibitemShut {NoStop}%
\bibitem [{Note1()}]{Note1}%
  \BibitemOpen
  \bibinfo {note} {The depletion of the ground state can be easily incorporated
  using, for instance, the ADK or PPT ionization models.}\BibitemShut {Stop}%
\bibitem [{\citenamefont {Milošević}\ \emph
  {et~al.}(2006{\natexlab{b}})\citenamefont {Milošević}, \citenamefont
  {Bauer},\ and\ \citenamefont {Becker}}]{milosevic__quantum-orbit_2006}%
  \BibitemOpen
  \bibfield  {author} {\bibinfo {author} {\bibfnamefont {D.~B.}\ \bibnamefont
  {Milošević}}, \bibinfo {author} {\bibfnamefont {D.}~\bibnamefont {Bauer}},\
  and\ \bibinfo {author} {\bibfnamefont {W.}~\bibnamefont {Becker}},\
  }\bibfield  {title} {\bibinfo {title} {Quantum-orbit theory of high-order
  atomic processes in intense laser fields},\ }\href
  {https://doi.org/10.1080/09500340500186099} {\bibfield  {journal} {\bibinfo
  {journal} {Journal of Modern Optics}\ }\textbf {\bibinfo {volume} {53}},\
  \bibinfo {pages} {125} (\bibinfo {year} {2006}{\natexlab{b}})}\BibitemShut
  {NoStop}%
\bibitem [{Com()}]{Comment1}%
  \BibitemOpen
  \href@noop {} {}\bibinfo {note} {We note that this state does not include
  recombination processes that lead to high-harmonic generation. This is
  because, when solving the differential equation conditioned to the electron
  being in the ground state, we neglect the contributions coming from the
  continuum part of the ansatz. However, if we treat these terms as first order
  perturbation theory terms, we get an extra term in the total state shown in
  Eq.~\eqref{Eq:ATI:state}, which corresponds to the contribution of electrons
  that have recombined with the parent ion, and thus describes HHG processes.
  Since we are not interested in the description of HHG processes, we do not
  include it here.}\BibitemShut {Stop}%
\bibitem [{\citenamefont {Bauer}(2006)}]{bauer2006theory}%
  \BibitemOpen
  \bibfield  {author} {\bibinfo {author} {\bibfnamefont {D.}~\bibnamefont
  {Bauer}},\ }\href@noop {} {\emph {\bibinfo {title} {Lecture notes on Theory
  of intense laser-matter interaction}}}\ (\bibinfo  {publisher}
  {Max-Planck-Institut für Kernhysik, Heilderberg},\ \bibinfo {year}
  {2006})\BibitemShut {NoStop}%
\bibitem [{\citenamefont {Quan}\ \emph {et~al.}(2009)\citenamefont {Quan},
  \citenamefont {Lin}, \citenamefont {Wu}, \citenamefont {Kang}, \citenamefont
  {Liu}, \citenamefont {Liu}, \citenamefont {Chen}, \citenamefont {Liu},
  \citenamefont {He}, \citenamefont {Chen}, \citenamefont {Xiong},
  \citenamefont {Guo}, \citenamefont {Xu}, \citenamefont {Fu}, \citenamefont
  {Cheng},\ and\ \citenamefont {Xu}}]{quan_classical_2009}%
  \BibitemOpen
  \bibfield  {author} {\bibinfo {author} {\bibfnamefont {W.}~\bibnamefont
  {Quan}}, \bibinfo {author} {\bibfnamefont {Z.}~\bibnamefont {Lin}}, \bibinfo
  {author} {\bibfnamefont {M.}~\bibnamefont {Wu}}, \bibinfo {author}
  {\bibfnamefont {H.}~\bibnamefont {Kang}}, \bibinfo {author} {\bibfnamefont
  {H.}~\bibnamefont {Liu}}, \bibinfo {author} {\bibfnamefont {X.}~\bibnamefont
  {Liu}}, \bibinfo {author} {\bibfnamefont {J.}~\bibnamefont {Chen}}, \bibinfo
  {author} {\bibfnamefont {J.}~\bibnamefont {Liu}}, \bibinfo {author}
  {\bibfnamefont {X.~T.}\ \bibnamefont {He}}, \bibinfo {author} {\bibfnamefont
  {S.~G.}\ \bibnamefont {Chen}}, \bibinfo {author} {\bibfnamefont
  {H.}~\bibnamefont {Xiong}}, \bibinfo {author} {\bibfnamefont
  {L.}~\bibnamefont {Guo}}, \bibinfo {author} {\bibfnamefont {H.}~\bibnamefont
  {Xu}}, \bibinfo {author} {\bibfnamefont {Y.}~\bibnamefont {Fu}}, \bibinfo
  {author} {\bibfnamefont {Y.}~\bibnamefont {Cheng}},\ and\ \bibinfo {author}
  {\bibfnamefont {Z.~Z.}\ \bibnamefont {Xu}},\ }\bibfield  {title} {\bibinfo
  {title} {Classical {Aspects} in {Above}-{Threshold} {Ionization} with a
  {Midinfrared} {Strong} {Laser} {Field}},\ }\href
  {https://doi.org/10.1103/PhysRevLett.103.093001} {\bibfield  {journal}
  {\bibinfo  {journal} {Physical Review Letters}\ }\textbf {\bibinfo {volume}
  {103}},\ \bibinfo {pages} {093001} (\bibinfo {year} {2009})}\BibitemShut
  {NoStop}%
\bibitem [{\citenamefont {Blaga}\ \emph {et~al.}(2009)\citenamefont {Blaga},
  \citenamefont {Catoire}, \citenamefont {Colosimo}, \citenamefont {Paulus},
  \citenamefont {Muller}, \citenamefont {Agostini},\ and\ \citenamefont
  {DiMauro}}]{blaga_strong-field_2009}%
  \BibitemOpen
  \bibfield  {author} {\bibinfo {author} {\bibfnamefont {C.~I.}\ \bibnamefont
  {Blaga}}, \bibinfo {author} {\bibfnamefont {F.}~\bibnamefont {Catoire}},
  \bibinfo {author} {\bibfnamefont {P.}~\bibnamefont {Colosimo}}, \bibinfo
  {author} {\bibfnamefont {G.~G.}\ \bibnamefont {Paulus}}, \bibinfo {author}
  {\bibfnamefont {H.~G.}\ \bibnamefont {Muller}}, \bibinfo {author}
  {\bibfnamefont {P.}~\bibnamefont {Agostini}},\ and\ \bibinfo {author}
  {\bibfnamefont {L.~F.}\ \bibnamefont {DiMauro}},\ }\bibfield  {title}
  {\bibinfo {title} {Strong-field photoionization revisited},\ }\href
  {https://doi.org/10.1038/nphys1228} {\bibfield  {journal} {\bibinfo
  {journal} {Nature Physics}\ }\textbf {\bibinfo {volume} {5}},\ \bibinfo
  {pages} {335} (\bibinfo {year} {2009})}\BibitemShut {NoStop}%
\bibitem [{\citenamefont {Liu}\ and\ \citenamefont
  {Hatsagortsyan}(2010)}]{liu_origin_2010}%
  \BibitemOpen
  \bibfield  {author} {\bibinfo {author} {\bibfnamefont {C.}~\bibnamefont
  {Liu}}\ and\ \bibinfo {author} {\bibfnamefont {K.~Z.}\ \bibnamefont
  {Hatsagortsyan}},\ }\bibfield  {title} {\bibinfo {title} {Origin of
  {Unexpected} {Low} {Energy} {Structure} in {Photoelectron} {Spectra}
  {Induced} by {Midinfrared} {Strong} {Laser} {Fields}},\ }\href
  {https://doi.org/10.1103/PhysRevLett.105.113003} {\bibfield  {journal}
  {\bibinfo  {journal} {Physical Review Letters}\ }\textbf {\bibinfo {volume}
  {105}},\ \bibinfo {pages} {113003} (\bibinfo {year} {2010})}\BibitemShut
  {NoStop}%
\bibitem [{\citenamefont {Wigner}(1932)}]{wigner_quantum_1932}%
  \BibitemOpen
  \bibfield  {author} {\bibinfo {author} {\bibfnamefont {E.}~\bibnamefont
  {Wigner}},\ }\bibfield  {title} {\bibinfo {title} {On the {Quantum}
  {Correction} {For} {Thermodynamic} {Equilibrium}},\ }\href
  {https://doi.org/10.1103/PhysRev.40.749} {\bibfield  {journal} {\bibinfo
  {journal} {Physical Review}\ }\textbf {\bibinfo {volume} {40}},\ \bibinfo
  {pages} {749} (\bibinfo {year} {1932})}\BibitemShut {NoStop}%
\bibitem [{\citenamefont {Schleich}(2001)}]{Schleich_Book_2001}%
  \BibitemOpen
  \bibfield  {author} {\bibinfo {author} {\bibfnamefont {W.~P.}\ \bibnamefont
  {Schleich}},\ }\href@noop {} {\emph {\bibinfo {title} {Quantum Optics in
  Phase Space}}}\ (\bibinfo  {publisher} {Wiley-VHC Verlag},\ \bibinfo
  {address} {Weinheim, Germany},\ \bibinfo {year} {2001})\BibitemShut {NoStop}%
\bibitem [{\citenamefont {Hudson}(1974)}]{hudson_when_1974}%
  \BibitemOpen
  \bibfield  {author} {\bibinfo {author} {\bibfnamefont {R.~L.}\ \bibnamefont
  {Hudson}},\ }\bibfield  {title} {\bibinfo {title} {When is the wigner
  quasi-probability density non-negative?},\ }\href
  {https://doi.org/10.1016/0034-4877(74)90007-X} {\bibfield  {journal}
  {\bibinfo  {journal} {Reports on Mathematical Physics}\ }\textbf {\bibinfo
  {volume} {6}},\ \bibinfo {pages} {249} (\bibinfo {year} {1974})}\BibitemShut
  {NoStop}%
\bibitem [{\citenamefont {Smithey}\ \emph {et~al.}(1993)\citenamefont
  {Smithey}, \citenamefont {Beck}, \citenamefont {Raymer},\ and\ \citenamefont
  {Faridani}}]{smithey_measurement_1993}%
  \BibitemOpen
  \bibfield  {author} {\bibinfo {author} {\bibfnamefont {D.~T.}\ \bibnamefont
  {Smithey}}, \bibinfo {author} {\bibfnamefont {M.}~\bibnamefont {Beck}},
  \bibinfo {author} {\bibfnamefont {M.~G.}\ \bibnamefont {Raymer}},\ and\
  \bibinfo {author} {\bibfnamefont {A.}~\bibnamefont {Faridani}},\ }\bibfield
  {title} {\bibinfo {title} {Measurement of the {Wigner} distribution and the
  density matrix of a light mode using optical homodyne tomography:
  {Application} to squeezed states and the vacuum},\ }\href
  {https://doi.org/10.1103/PhysRevLett.70.1244} {\bibfield  {journal} {\bibinfo
   {journal} {Physical Review Letters}\ }\textbf {\bibinfo {volume} {70}},\
  \bibinfo {pages} {1244} (\bibinfo {year} {1993})}\BibitemShut {NoStop}%
\bibitem [{\citenamefont {Royer}(1977)}]{royer_wigner_1977}%
  \BibitemOpen
  \bibfield  {author} {\bibinfo {author} {\bibfnamefont {A.}~\bibnamefont
  {Royer}},\ }\bibfield  {title} {\bibinfo {title} {Wigner function as the
  expectation value of a parity operator},\ }\href
  {https://doi.org/10.1103/PhysRevA.15.449} {\bibfield  {journal} {\bibinfo
  {journal} {Physical Review A}\ }\textbf {\bibinfo {volume} {15}},\ \bibinfo
  {pages} {449} (\bibinfo {year} {1977})}\BibitemShut {NoStop}%
\bibitem [{\citenamefont {Paulus}\ \emph {et~al.}(2003)\citenamefont {Paulus},
  \citenamefont {Lindner}, \citenamefont {Walther}, \citenamefont {Baltuška},
  \citenamefont {Goulielmakis}, \citenamefont {Lezius},\ and\ \citenamefont
  {Krausz}}]{paulus_measurement_2003}%
  \BibitemOpen
  \bibfield  {author} {\bibinfo {author} {\bibfnamefont {G.~G.}\ \bibnamefont
  {Paulus}}, \bibinfo {author} {\bibfnamefont {F.}~\bibnamefont {Lindner}},
  \bibinfo {author} {\bibfnamefont {H.}~\bibnamefont {Walther}}, \bibinfo
  {author} {\bibfnamefont {A.}~\bibnamefont {Baltuška}}, \bibinfo {author}
  {\bibfnamefont {E.}~\bibnamefont {Goulielmakis}}, \bibinfo {author}
  {\bibfnamefont {M.}~\bibnamefont {Lezius}},\ and\ \bibinfo {author}
  {\bibfnamefont {F.}~\bibnamefont {Krausz}},\ }\bibfield  {title} {\bibinfo
  {title} {Measurement of the {Phase} of {Few}-{Cycle} {Laser} {Pulses}},\
  }\href {https://doi.org/10.1103/PhysRevLett.91.253004} {\bibfield  {journal}
  {\bibinfo  {journal} {Phys. Rev. Lett.}\ }\textbf {\bibinfo {volume} {91}},\
  \bibinfo {pages} {253004} (\bibinfo {year} {2003})}\BibitemShut {NoStop}%
\bibitem [{\citenamefont {van Loock}(2011)}]{van_loock_optical_2011}%
  \BibitemOpen
  \bibfield  {author} {\bibinfo {author} {\bibfnamefont {P.}~\bibnamefont {van
  Loock}},\ }\bibfield  {title} {\bibinfo {title} {Optical hybrid approaches to
  quantum information},\ }\href {https://doi.org/10.1002/lpor.201000005}
  {\bibfield  {journal} {\bibinfo  {journal} {Laser \& Photonics Reviews}\
  }\textbf {\bibinfo {volume} {5}},\ \bibinfo {pages} {167} (\bibinfo {year}
  {2011})}\BibitemShut {NoStop}%
\bibitem [{\citenamefont {Kreis}\ and\ \citenamefont {van
  Loock}(2012)}]{kreis_classifying_2012}%
  \BibitemOpen
  \bibfield  {author} {\bibinfo {author} {\bibfnamefont {K.}~\bibnamefont
  {Kreis}}\ and\ \bibinfo {author} {\bibfnamefont {P.}~\bibnamefont {van
  Loock}},\ }\bibfield  {title} {\bibinfo {title} {Classifying, quantifying,
  and witnessing qudit-qumode hybrid entanglement},\ }\href
  {https://doi.org/10.1103/PhysRevA.85.032307} {\bibfield  {journal} {\bibinfo
  {journal} {Physical Review A}\ }\textbf {\bibinfo {volume} {85}},\ \bibinfo
  {pages} {032307} (\bibinfo {year} {2012})}\BibitemShut {NoStop}%
\bibitem [{\citenamefont {Massé}\ \emph {et~al.}(2020)\citenamefont {Massé},
  \citenamefont {Coudreau}, \citenamefont {Keller},\ and\ \citenamefont
  {Milman}}]{masse_implementable_2020}%
  \BibitemOpen
  \bibfield  {author} {\bibinfo {author} {\bibfnamefont {G.}~\bibnamefont
  {Massé}}, \bibinfo {author} {\bibfnamefont {T.}~\bibnamefont {Coudreau}},
  \bibinfo {author} {\bibfnamefont {A.}~\bibnamefont {Keller}},\ and\ \bibinfo
  {author} {\bibfnamefont {P.}~\bibnamefont {Milman}},\ }\bibfield  {title}
  {\bibinfo {title} {Implementable hybrid entanglement witness},\ }\href
  {https://doi.org/10.1103/PhysRevA.102.062406} {\bibfield  {journal} {\bibinfo
   {journal} {Physical Review A}\ }\textbf {\bibinfo {volume} {102}},\ \bibinfo
  {pages} {062406} (\bibinfo {year} {2020})}\BibitemShut {NoStop}%
\bibitem [{\citenamefont {Protopapas}\ \emph {et~al.}(1997)\citenamefont
  {Protopapas}, \citenamefont {Keitel},\ and\ \citenamefont
  {Knight}}]{protopapas_atomic_1997}%
  \BibitemOpen
  \bibfield  {author} {\bibinfo {author} {\bibfnamefont {M.}~\bibnamefont
  {Protopapas}}, \bibinfo {author} {\bibfnamefont {C.~H.}\ \bibnamefont
  {Keitel}},\ and\ \bibinfo {author} {\bibfnamefont {P.~L.}\ \bibnamefont
  {Knight}},\ }\bibfield  {title} {\bibinfo {title} {Atomic physics with
  super-high intensity lasers},\ }\href
  {https://doi.org/10.1088/0034-4885/60/4/001} {\bibfield  {journal} {\bibinfo
  {journal} {Reports on Progress in Physics}\ }\textbf {\bibinfo {volume}
  {60}},\ \bibinfo {pages} {389} (\bibinfo {year} {1997})}\BibitemShut
  {NoStop}%
\bibitem [{\citenamefont {Gilchrist}\ \emph {et~al.}(2004)\citenamefont
  {Gilchrist}, \citenamefont {Nemoto}, \citenamefont {Munro}, \citenamefont
  {Ralph}, \citenamefont {Glancy}, \citenamefont {Braunstein},\ and\
  \citenamefont {Milburn}}]{gilchrist_schrodinger_2004}%
  \BibitemOpen
  \bibfield  {author} {\bibinfo {author} {\bibfnamefont {A.}~\bibnamefont
  {Gilchrist}}, \bibinfo {author} {\bibfnamefont {K.}~\bibnamefont {Nemoto}},
  \bibinfo {author} {\bibfnamefont {W.~J.}\ \bibnamefont {Munro}}, \bibinfo
  {author} {\bibfnamefont {T.~C.}\ \bibnamefont {Ralph}}, \bibinfo {author}
  {\bibfnamefont {S.}~\bibnamefont {Glancy}}, \bibinfo {author} {\bibfnamefont
  {S.~L.}\ \bibnamefont {Braunstein}},\ and\ \bibinfo {author} {\bibfnamefont
  {G.~J.}\ \bibnamefont {Milburn}},\ }\bibfield  {title} {\bibinfo {title}
  {Schrödinger cats and their power for quantum information processing},\
  }\href {https://doi.org/10.1088/1464-4266/6/8/032} {\bibfield  {journal}
  {\bibinfo  {journal} {Journal of Optics B: Quantum and Semiclassical Optics}\
  }\textbf {\bibinfo {volume} {6}},\ \bibinfo {pages} {S828} (\bibinfo {year}
  {2004})}\BibitemShut {NoStop}%
\bibitem [{\citenamefont {Lvovsky}\ \emph {et~al.}(2020)\citenamefont
  {Lvovsky}, \citenamefont {Grangier}, \citenamefont {Ourjoumtsev},
  \citenamefont {Parigi}, \citenamefont {Sasaki},\ and\ \citenamefont
  {Tualle-Brouri}}]{lvovsky_production_2020}%
  \BibitemOpen
  \bibfield  {author} {\bibinfo {author} {\bibfnamefont {A.~I.}\ \bibnamefont
  {Lvovsky}}, \bibinfo {author} {\bibfnamefont {P.}~\bibnamefont {Grangier}},
  \bibinfo {author} {\bibfnamefont {A.}~\bibnamefont {Ourjoumtsev}}, \bibinfo
  {author} {\bibfnamefont {V.}~\bibnamefont {Parigi}}, \bibinfo {author}
  {\bibfnamefont {M.}~\bibnamefont {Sasaki}},\ and\ \bibinfo {author}
  {\bibfnamefont {R.}~\bibnamefont {Tualle-Brouri}},\ }\href
  {https://doi.org/10.48550/arXiv.2006.16985} {\bibinfo {title} {Production and
  applications of non-{Gaussian} quantum states of light}} (\bibinfo {year}
  {2020}),\ \bibinfo {note} {number: arXiv:2006.16985 arXiv:2006.16985
  [physics, physics:quant-ph]}\BibitemShut {NoStop}%
\bibitem [{\citenamefont {Ralph}\ \emph {et~al.}(2003)\citenamefont {Ralph},
  \citenamefont {Gilchrist}, \citenamefont {Milburn}, \citenamefont {Munro},\
  and\ \citenamefont {Glancy}}]{ralph_quantum_2003}%
  \BibitemOpen
  \bibfield  {author} {\bibinfo {author} {\bibfnamefont {T.~C.}\ \bibnamefont
  {Ralph}}, \bibinfo {author} {\bibfnamefont {A.}~\bibnamefont {Gilchrist}},
  \bibinfo {author} {\bibfnamefont {G.~J.}\ \bibnamefont {Milburn}}, \bibinfo
  {author} {\bibfnamefont {W.~J.}\ \bibnamefont {Munro}},\ and\ \bibinfo
  {author} {\bibfnamefont {S.}~\bibnamefont {Glancy}},\ }\bibfield  {title}
  {\bibinfo {title} {Quantum computation with optical coherent states},\ }\href
  {https://doi.org/10.1103/PhysRevA.68.042319} {\bibfield  {journal} {\bibinfo
  {journal} {Physical Review A}\ }\textbf {\bibinfo {volume} {68}},\ \bibinfo
  {pages} {042319} (\bibinfo {year} {2003})}\BibitemShut {NoStop}%
\bibitem [{\citenamefont {Sanders}(1992)}]{sanders_entangled_1992}%
  \BibitemOpen
  \bibfield  {author} {\bibinfo {author} {\bibfnamefont {B.~C.}\ \bibnamefont
  {Sanders}},\ }\bibfield  {title} {\bibinfo {title} {Entangled coherent
  states},\ }\href {https://doi.org/10.1103/PhysRevA.45.6811} {\bibfield
  {journal} {\bibinfo  {journal} {Physical Review A}\ }\textbf {\bibinfo
  {volume} {45}},\ \bibinfo {pages} {6811} (\bibinfo {year}
  {1992})}\BibitemShut {NoStop}%
\bibitem [{\citenamefont {Jeong}\ \emph {et~al.}(2003)\citenamefont {Jeong},
  \citenamefont {Son}, \citenamefont {Kim}, \citenamefont {Ahn},\ and\
  \citenamefont {Brukner}}]{jeong_quantum_2003}%
  \BibitemOpen
  \bibfield  {author} {\bibinfo {author} {\bibfnamefont {H.}~\bibnamefont
  {Jeong}}, \bibinfo {author} {\bibfnamefont {W.}~\bibnamefont {Son}}, \bibinfo
  {author} {\bibfnamefont {M.~S.}\ \bibnamefont {Kim}}, \bibinfo {author}
  {\bibfnamefont {D.}~\bibnamefont {Ahn}},\ and\ \bibinfo {author}
  {\bibfnamefont {c.}~\bibnamefont {Brukner}},\ }\bibfield  {title} {\bibinfo
  {title} {Quantum nonlocality test for continuous-variable states with
  dichotomic observables},\ }\href {https://doi.org/10.1103/PhysRevA.67.012106}
  {\bibfield  {journal} {\bibinfo  {journal} {Physical Review A}\ }\textbf
  {\bibinfo {volume} {67}},\ \bibinfo {pages} {012106} (\bibinfo {year}
  {2003})}\BibitemShut {NoStop}%
\bibitem [{\citenamefont {Stobińska}\ \emph {et~al.}(2007)\citenamefont
  {Stobińska}, \citenamefont {Jeong},\ and\ \citenamefont
  {Ralph}}]{stobinska_violation_2007}%
  \BibitemOpen
  \bibfield  {author} {\bibinfo {author} {\bibfnamefont {M.}~\bibnamefont
  {Stobińska}}, \bibinfo {author} {\bibfnamefont {H.}~\bibnamefont {Jeong}},\
  and\ \bibinfo {author} {\bibfnamefont {T.~C.}\ \bibnamefont {Ralph}},\
  }\bibfield  {title} {\bibinfo {title} {Violation of {Bell}'s inequality using
  classical measurements and nonlinear local operations},\ }\href
  {https://doi.org/10.1103/PhysRevA.75.052105} {\bibfield  {journal} {\bibinfo
  {journal} {Physical Review A}\ }\textbf {\bibinfo {volume} {75}},\ \bibinfo
  {pages} {052105} (\bibinfo {year} {2007})}\BibitemShut {NoStop}%
\bibitem [{\citenamefont {Munro}\ \emph {et~al.}(2002)\citenamefont {Munro},
  \citenamefont {Nemoto}, \citenamefont {Milburn},\ and\ \citenamefont
  {Braunstein}}]{munro_weak-force_2002}%
  \BibitemOpen
  \bibfield  {author} {\bibinfo {author} {\bibfnamefont {W.~J.}\ \bibnamefont
  {Munro}}, \bibinfo {author} {\bibfnamefont {K.}~\bibnamefont {Nemoto}},
  \bibinfo {author} {\bibfnamefont {G.~J.}\ \bibnamefont {Milburn}},\ and\
  \bibinfo {author} {\bibfnamefont {S.~L.}\ \bibnamefont {Braunstein}},\
  }\bibfield  {title} {\bibinfo {title} {Weak-force detection with superposed
  coherent states},\ }\href {https://doi.org/10.1103/PhysRevA.66.023819}
  {\bibfield  {journal} {\bibinfo  {journal} {Physical Review A}\ }\textbf
  {\bibinfo {volume} {66}},\ \bibinfo {pages} {023819} (\bibinfo {year}
  {2002})}\BibitemShut {NoStop}%
\bibitem [{\citenamefont {Schultz}\ and\ \citenamefont
  {Vrakking}(2014)}]{SchulzVrakking_Book}%
  \BibitemOpen
  \bibfield  {author} {\bibinfo {author} {\bibfnamefont {T.}~\bibnamefont
  {Schultz}}\ and\ \bibinfo {author} {\bibfnamefont {M.}~\bibnamefont
  {Vrakking}},\ }\href {https://doi.org/https://doi.org/10.1002/9783527677689}
  {\emph {\bibinfo {title} {Attosecond and XUV Physics}}}\ (\bibinfo
  {publisher} {John Wiley \& Sons, Ltd},\ \bibinfo {year} {2014})\BibitemShut
  {NoStop}%
\bibitem [{\citenamefont {Rivera-Dean}\ \emph
  {et~al.}(2022{\natexlab{b}})\citenamefont {Rivera-Dean}, \citenamefont
  {Stammer}, \citenamefont {Maxwell}, \citenamefont {Lamprou}, \citenamefont
  {Tzallas}, \citenamefont {Lewenstein},\ and\ \citenamefont
  {Ciappina}}]{ZenodoLink}%
  \BibitemOpen
  \bibfield  {author} {\bibinfo {author} {\bibfnamefont {J.}~\bibnamefont
  {Rivera-Dean}}, \bibinfo {author} {\bibfnamefont {P.}~\bibnamefont
  {Stammer}}, \bibinfo {author} {\bibfnamefont {A.~S.}\ \bibnamefont
  {Maxwell}}, \bibinfo {author} {\bibfnamefont {T.}~\bibnamefont {Lamprou}},
  \bibinfo {author} {\bibfnamefont {P.}~\bibnamefont {Tzallas}}, \bibinfo
  {author} {\bibfnamefont {M.}~\bibnamefont {Lewenstein}},\ and\ \bibinfo
  {author} {\bibfnamefont {M.~F.}\ \bibnamefont {Ciappina}},\ }\href
  {https://doi.org/10.5281/zenodo.6907735} {\bibinfo {title} {Light-matter
  entanglement after above-threshold ionization processes in atoms}},\ \bibinfo
  {howpublished} {Zenodo} (\bibinfo {year} {2022}{\natexlab{b}})\BibitemShut
  {NoStop}%
\bibitem [{\citenamefont {Scully}\ and\ \citenamefont
  {Zubairy}(2001)}]{ScullyBook}%
  \BibitemOpen
  \bibfield  {author} {\bibinfo {author} {\bibfnamefont {M.~O.}\ \bibnamefont
  {Scully}}\ and\ \bibinfo {author} {\bibfnamefont {M.~S.}\ \bibnamefont
  {Zubairy}},\ }\href@noop {} {\emph {\bibinfo {title} {Quantum optics}}}\
  (\bibinfo  {publisher} {Cambridge University Press, Cambridge},\ \bibinfo
  {year} {2001})\BibitemShut {NoStop}%
\bibitem [{\citenamefont {Gerry}\ and\ \citenamefont
  {Knight}(2005)}]{Gerry__Book_2001}%
  \BibitemOpen
  \bibfield  {author} {\bibinfo {author} {\bibfnamefont {C.}~\bibnamefont
  {Gerry}}\ and\ \bibinfo {author} {\bibfnamefont {P.}~\bibnamefont {Knight}},\
  }\href@noop {} {\emph {\bibinfo {title} {Introductory Quantum Optics}}}\
  (\bibinfo  {publisher} {Cambridge University Press},\ \bibinfo {address}
  {Cambridge, UK},\ \bibinfo {year} {2005})\BibitemShut {NoStop}%
\bibitem [{\citenamefont {Virtanen}\ \emph {et~al.}(2020)\citenamefont
  {Virtanen}, \citenamefont {Gommers}, \citenamefont {Oliphant}, \citenamefont
  {Haberland}, \citenamefont {Reddy}, \citenamefont {Cournapeau}, \citenamefont
  {Burovski}, \citenamefont {Peterson}, \citenamefont {Weckesser},
  \citenamefont {Bright}, \citenamefont {{van der Walt}}, \citenamefont
  {Brett}, \citenamefont {Wilson}, \citenamefont {Millman}, \citenamefont
  {Mayorov}, \citenamefont {Nelson}, \citenamefont {Jones}, \citenamefont
  {Kern}, \citenamefont {Larson}, \citenamefont {Carey}, \citenamefont {Polat},
  \citenamefont {Feng}, \citenamefont {Moore}, \citenamefont {{VanderPlas}},
  \citenamefont {Laxalde}, \citenamefont {Perktold}, \citenamefont {Cimrman},
  \citenamefont {Henriksen}, \citenamefont {Quintero}, \citenamefont {Harris},
  \citenamefont {Archibald}, \citenamefont {Ribeiro}, \citenamefont
  {Pedregosa}, \citenamefont {{van Mulbregt}},\ and\ \citenamefont {{SciPy 1.0
  Contributors}}}]{2020SciPy-NMeth}%
  \BibitemOpen
  \bibfield  {author} {\bibinfo {author} {\bibfnamefont {P.}~\bibnamefont
  {Virtanen}}, \bibinfo {author} {\bibfnamefont {R.}~\bibnamefont {Gommers}},
  \bibinfo {author} {\bibfnamefont {T.~E.}\ \bibnamefont {Oliphant}}, \bibinfo
  {author} {\bibfnamefont {M.}~\bibnamefont {Haberland}}, \bibinfo {author}
  {\bibfnamefont {T.}~\bibnamefont {Reddy}}, \bibinfo {author} {\bibfnamefont
  {D.}~\bibnamefont {Cournapeau}}, \bibinfo {author} {\bibfnamefont
  {E.}~\bibnamefont {Burovski}}, \bibinfo {author} {\bibfnamefont
  {P.}~\bibnamefont {Peterson}}, \bibinfo {author} {\bibfnamefont
  {W.}~\bibnamefont {Weckesser}}, \bibinfo {author} {\bibfnamefont
  {J.}~\bibnamefont {Bright}}, \bibinfo {author} {\bibfnamefont {S.~J.}\
  \bibnamefont {{van der Walt}}}, \bibinfo {author} {\bibfnamefont
  {M.}~\bibnamefont {Brett}}, \bibinfo {author} {\bibfnamefont
  {J.}~\bibnamefont {Wilson}}, \bibinfo {author} {\bibfnamefont {K.~J.}\
  \bibnamefont {Millman}}, \bibinfo {author} {\bibfnamefont {N.}~\bibnamefont
  {Mayorov}}, \bibinfo {author} {\bibfnamefont {A.~R.~J.}\ \bibnamefont
  {Nelson}}, \bibinfo {author} {\bibfnamefont {E.}~\bibnamefont {Jones}},
  \bibinfo {author} {\bibfnamefont {R.}~\bibnamefont {Kern}}, \bibinfo {author}
  {\bibfnamefont {E.}~\bibnamefont {Larson}}, \bibinfo {author} {\bibfnamefont
  {C.~J.}\ \bibnamefont {Carey}}, \bibinfo {author} {\bibfnamefont
  {{\.I}.}~\bibnamefont {Polat}}, \bibinfo {author} {\bibfnamefont
  {Y.}~\bibnamefont {Feng}}, \bibinfo {author} {\bibfnamefont {E.~W.}\
  \bibnamefont {Moore}}, \bibinfo {author} {\bibfnamefont {J.}~\bibnamefont
  {{VanderPlas}}}, \bibinfo {author} {\bibfnamefont {D.}~\bibnamefont
  {Laxalde}}, \bibinfo {author} {\bibfnamefont {J.}~\bibnamefont {Perktold}},
  \bibinfo {author} {\bibfnamefont {R.}~\bibnamefont {Cimrman}}, \bibinfo
  {author} {\bibfnamefont {I.}~\bibnamefont {Henriksen}}, \bibinfo {author}
  {\bibfnamefont {E.~A.}\ \bibnamefont {Quintero}}, \bibinfo {author}
  {\bibfnamefont {C.~R.}\ \bibnamefont {Harris}}, \bibinfo {author}
  {\bibfnamefont {A.~M.}\ \bibnamefont {Archibald}}, \bibinfo {author}
  {\bibfnamefont {A.~H.}\ \bibnamefont {Ribeiro}}, \bibinfo {author}
  {\bibfnamefont {F.}~\bibnamefont {Pedregosa}}, \bibinfo {author}
  {\bibfnamefont {P.}~\bibnamefont {{van Mulbregt}}},\ and\ \bibinfo {author}
  {\bibnamefont {{SciPy 1.0 Contributors}}},\ }\bibfield  {title} {\bibinfo
  {title} {{{SciPy} 1.0: Fundamental Algorithms for Scientific Computing in
  Python}},\ }\href {https://doi.org/10.1038/s41592-019-0686-2} {\bibfield
  {journal} {\bibinfo  {journal} {Nat. Meth.}\ }\textbf {\bibinfo {volume}
  {17}},\ \bibinfo {pages} {261} (\bibinfo {year} {2020})}\BibitemShut
  {NoStop}%
\bibitem [{\citenamefont {Lam}\ \emph {et~al.}(2015)\citenamefont {Lam},
  \citenamefont {Pitrou},\ and\ \citenamefont {Seibert}}]{NumbaRed}%
  \BibitemOpen
  \bibfield  {author} {\bibinfo {author} {\bibfnamefont {S.~K.}\ \bibnamefont
  {Lam}}, \bibinfo {author} {\bibfnamefont {A.}~\bibnamefont {Pitrou}},\ and\
  \bibinfo {author} {\bibfnamefont {S.}~\bibnamefont {Seibert}},\ }\bibfield
  {title} {\bibinfo {title} {Numba: A llvm-based python jit compiler},\ }in\
  \href@noop {} {\emph {\bibinfo {booktitle} {Proceedings of the Second
  Workshop on the LLVM Compiler Infrastructure in HPC}}}\ (\bibinfo {year}
  {2015})\ pp.\ \bibinfo {pages} {1--6}\BibitemShut {NoStop}%
\end{thebibliography}%

\newpage
\onecolumngrid
\appendix
\vspace{0.5cm}
\begin{center}
    \textbf{APPENDIX}
\end{center}
\section{Solving the time-dependent Schrödinger equation}\label{App:TDSE}
In this appendix, we explicitly solve the time-dependent Schrödinger equation presented in the main text, considering the ansatz shown in Eq.~\eqref{Eq:ansatz}. We also introduce the approximations that we consider in order to evaluate this differential equation.

\subsubsection{Conditioning onto a continuum state}
In order to find the different coefficients appearing in the considered ansatz, we first condition the above Schrödinger equation on finding the electron in a continuum state $\ket{\vb{v}}$, such that Eq.~\eqref{schrodinger_1} reads
\begin{equation}\label{eq:app:continuum:Sch:eq}
    \begin{aligned}
	i\hbar 
		\pdv{}{t}
		\big(
			b(\vb{v},t) \ket{\Phi(\vb{v},t)}
		\big)&=
		 \dfrac{\vb{v}^2}{2m}
		    b(\vb{v},t)
		    \ket{\Phi(\vb{v},t)}
		            + e\big(
				            \vb{E}_\text{cl}(t)
		                    + \hat{\vb{E}}(t)
				        \big)\cdot
				\mel{\vb{v}}{\hat{\vb{R}}}{\text{g}}a(t) \ket{\Phi_{\text{g}}(t)}
		\\&\quad
	    + i \hbar e\big(
				   	   \vb{E}_\text{cl}(t) + \hat{\vb{E}}(t)
					\big)\cdot
				\nabla_{\vb{v}}
					\big( 
						b(\vb{v},t) \ket{\Phi(\vb{v},t)}
					\big),
    \end{aligned}
\end{equation}
where we have written the continuum-continuum transition term from $\vb{v}'$ to $\vb{v}$ as $\mel{\vb{v}}{\vb{R}}{\vb{v}'} = i\hbar \nabla_{\vb{v}}\delta(\vb{v}-\vb{v}') + (\hbar/e)\vb{g}(\vb{v},\vb{v}')$ \cite{lewenstein_theory_1994,amini_symphony_2019}, where the second term includes the effects due to the rescattering with the core center, and that can be treated perturbatively. This is because along the manuscript, we work with photoelectron energies $\lesssim 3U_p$, where $U_p = e^2E_0^2/4m\omega_L^2$ is the ponderomotive energy, for which rescattering events that lead to high-order ATI processes (HATI) \cite{milosevic_above-threshold_2006,milosevic__quantum-orbit_2006}
do not play an important role. Thus, we neglect them in the following.

The differential equation presented in \eqref{eq:app:continuum:Sch:eq} has a well-defined homogeneous and inhomogeneous part. Since the solution of this equation can be written as the solution of the homogeneous part plus a solution to the inhomogeneous one, we first focus on the former. In order to solve the homogeneous equation, we expand it as
\begin{equation}\label{Eq:app:exact:Sch:equation}
    \begin{aligned}
	i\hbar \pdv{b(\vb{v},t)}{t} \ket{\Phi(\vb{v},t)}
		+	i\hbar b(\vb{v},t) \pdv{\ket{\Phi(\vb{v},t)}}{t}
		&= 
	    \dfrac{\vb{v}^2}{2m}b(\vb{v},t)\ket{\Phi(\vb{v},t)}
	    + i \hbar e
	    	   \vb{E}_{\text{cl}}(t)
				\cdot 
				\big[
				    (\nabla_{\vb{v}}
						b(\vb{v},t)) \ket{\Phi(\vb{v},t)}
						+ b(\vb{v},t) (\nabla_{\vb{v}}
							 \ket{\Phi(\vb{v},t)})
				\big]
			\\&\quad
			+ i \hbar e
				   	   \hat{\vb{E}}(t)
				\cdot 
				\big[
				    (\nabla_{\vb{v}}
						b(\vb{v},t)) \ket{\Phi(\vb{v},t)}
						+ b(\vb{v},t) (\nabla_{\vb{v}}
			    	 \ket{\Phi(\vb{v},t)})
				\big].
	\end{aligned}
\end{equation}

At the right hand side, we have the sum of three different terms. The first one introduces the energy of the scattered electron; the second term defines the influence of the average value of the field on the electron's trajectory; finally the third one characterizes the quantum fluctuations. In particular, for the third term we find different contributions. In first place, the first one introduces the back-action of the electron's trajectory on the quantum optical state of the field, while the second governs the back-action of the quantum optical perturbations in the semiclassical trajectories. In the following, we work under the assumption that the electron trajectory does not get affected by the quantum optical perturbations, so that the last term we have just described can be omitted. Therefore, hereupon we work with the following approximated version of the Schrödinger equation shown in Eq.~\eqref{Eq:app:exact:Sch:equation}
\begin{equation}
    \begin{aligned}
	i\hbar \pdv{b(\vb{v},t)}{t} \ket{\Phi(\vb{v},t)}
		+	i\hbar b(\vb{v},t) \pdv{\ket{\Phi(\vb{v},t)}}{t}
		&=
	\dfrac{\vb{v}^2}{2m}
		    b(\vb{v},t)
		    \ket{\Phi(\vb{v},t)}
	   + i \hbar e
				   	   \vb{E}_{\text{cl}}(t)
				\cdot 
				\big[
				    (\nabla_{\vb{v}}
						b(\vb{v},t)) \ket{\Phi(\vb{v},t)}
						+b(\vb{v},t) (\nabla_{\vb{v}}
						 \ket{\Phi(\vb{v},t)})
				\big]
			\\&\quad
			+ i \hbar e
				   	   \hat{\vb{E}}(t)
				\cdot 
				    (\nabla_{\vb{v}}
						b(\vb{v},t)) \ket{\Phi(\vb{v},t)},
	\end{aligned}
\end{equation}
such that we write the differential equation as a sum of two contributions
\begin{equation}\label{eq:app:approx:Sch:joined}
    \begin{aligned}
        &\bigg[
            i\hbar \pdv{b(\vb{v},t)}{t}
		    - i \hbar e \vb{E}_{\text{cl}}(t)
				    \cdot
				    \big(
					    \nabla_{\vb{v}} b(\vb{v},t)
				    \big)
		    - \dfrac{\vb{v}^2}{2m}b(\vb{v},t)
		\bigg]
		    \ket{\Phi(\vb{v},t)}
		\\
		&\hspace{1cm}
		+
		\bigg[
		    i\hbar b(\vb{v},t) \pdv{\ket{\Phi(\vb{v},t)}}{t}
		    - i \hbar e
			    	   \vb{E}_{\text{cl}}(t) \cdot\big(\nabla_{\vb{v}}\ket{\Phi(\vb{v},t)}\big) b(\vb{v},t)
		    - i\hbar e\hat{\vb{E}}(t)
			    \cdot \big(
				    \nabla_{\vb{v}} b(\vb{v},t)
			    \big) \ket{\Phi(\vb{v},t)}
		\bigg]
		= 0.
    \end{aligned}
\end{equation}

In order to solve this equation, we first solve it for $b(\vb{v},t)$ by setting the first bracket to zero. By doing this, we recover the Schrödinger equation which describes the evolution of the scattered electron in the continuum, that is
\begin{equation}
    i\hbar \pdv{b(\vb{v},t)}{t}
		    - i \hbar e\vb{E}_{\text{cl}}(t)
				    \cdot
				    \big(
					    \nabla_{\vb{v}} b(\vb{v},t)
				    \big)
		    - \dfrac{\vb{v}^2}{2m}b(\vb{v},t) = 0,
\end{equation}
which is solved by
\begin{equation}\label{Eq:app:semiclassical:action}
\begin{aligned}
	b(\vb{v},t)
		&=
		    b(\vb{v},t_0)
		   \exp[-\dfrac{i}{\hbar}\int^t_{t_0} \dd \tau
				\dfrac{1}{2m}
				\bigg(
					\vb{p} + \dfrac{e}{c}\vb{A}(\tau)
				\bigg)^2
			]
\end{aligned}
\end{equation}
where $\vb{p} = \vb{v} - (e/c)\vb{A}(t)$ is the canonical momentum and $\vb{A}(t)$ the classical vector potential of the applied field. By implementing Eq.~\eqref{Eq:app:semiclassical:action} into Eq.~\eqref{eq:app:approx:Sch:joined}, the latter gets simplified to
\begin{equation}\label{Eq:after:b(v,t)}
    \begin{aligned}
		    i\hbar b(\vb{v},t) \pdv{\ket{\Phi(\vb{v},t)}}{t}
		    &- i \hbar e
			    	   \vb{E}_{\text{cl}}(t) \cdot\big(\nabla_{\vb{v}}\ket{\Phi(\vb{v},t)}\big) b(\vb{v},t)
		    - i\hbar e \hat{\vb{E}}(t)
			    \cdot \big(
				    \nabla_{\vb{v}} b(\vb{v},t)
			    \big) \ket{\Phi(\vb{v},t)}
		= 0.
	\end{aligned}
\end{equation}

Before proceeding further, let us first compute the gradient with respect to $\vb{v}$ of $b(\vb{v},t)$
\begin{equation}
	\nabla_{\vb{v}} b(\vb{v},t)
		= -\dfrac{i}{\hbar} \Delta \vb{r}(\vb{v},\tau,t_0)
			b(\vb{v},t),
\end{equation}
where $\Delta \vb{r}(\vb{v},t,t_0)$ denotes the electronic displacement in the continuum between times $t_0$ and $t$, and is given by
\begin{equation}
	\Delta \vb{r}(\vb{v},t,t_0)
		= \dfrac{1}{m} \int^t_{t_0}
			\dd \tau \Big(
					    \vb{v}
					    -\dfrac{e}{c}\vb{A}(t)
					    + \dfrac{e}{c}\vb{A}(\tau)
				     \Big).
\end{equation}

Introducing the above functions in Eq.~\eqref{Eq:after:b(v,t)}, we find
\begin{equation}
\begin{aligned}
	i \hbar\pdv{\ket{\Phi(\vb{v},t)}}{t}
		&- i\hbar e\vb{E}_{\text{cl}}(t)
			\cdot \nabla_{\vb{v}}\ket{\Phi(\vb{v},t)}
			=e \hat{\vb{E}}(t) \cdot \Delta \vb{r}(\vb{v},t,t_0) \ket{\Phi(\vb{v},t)},
\end{aligned}
\end{equation}
which, after writing the kinetic momentum in terms as the canonical momentum as we did before, leads to a linear equation in the creation and annihilation operators that can be solved by (see for instance \cite{rivera-dean_strong_2022,stammer_quantum_2022})
\begin{equation}
    \begin{aligned}
	\ket{\Phi(\vb{v},t)}
		&= \Tilde{D}
			\big(
				\boldsymbol{\delta}(\vb{v},t,t_0)
			\big)\ket{\Phi(\vb{v},t_0)}
		= \prod_{\vb{k},\mu} e^{i\varphi_{\vb{k},\mu}(\vb{v},t)}
			D\big(
				\delta_{\vb{k},\mu}(\vb{v},t,t_0)
			   \big)\ket{\Phi(\vb{v},t_0)},
	\end{aligned}
\end{equation}
where $\varphi_{\vb{k,\mu}}(\vb{v},t)$ is a phase prefactor that arises when solving the quantum optical part of the Schrödinger equation (see for instance \cite{stammer_quantum_2022}), and $\delta_{\vb{k},\mu}(\vb{v},t,t_0)$ is the Fourier transform of the electronic displacement from the initial time $t_0$ up to $t$. Both quantities are given by
\begin{equation}
	\delta_{\vb{k},\mu}(\vb{v},t,t_0)
		= -\dfrac{e}{\hbar} \sqrt{\dfrac{\hbar \omega_k}{2\epsilon_0 V}}
			\int^t_{t_0} \dd t' 
				\Delta\vb{r}(\vb{v},t',t_0)e^{i\omega_k t'},
\end{equation}
\begin{equation}
    \begin{aligned}
    \varphi_{\vb{k},\mu}(\vb{v},t)
        &= \dfrac{e^2}{\hbar^2}
            \dfrac{\hbar \omega_k}{2\epsilon_0 V}
            \int^t_{t_0}\dd t_1
            \int^{t_1}_{t_0}\dd t_2
            \Big(
                \boldsymbol{\epsilon}_{\vb{k},\mu}\cdot \Delta \vb{r}(\vb{v},t_1,t_0)
            \Big)
              \Big( \boldsymbol{\epsilon}_{\vb{k},\mu}\cdot \Delta \vb{r}(\vb{v},t_2,t_0)
            \Big)
            \sin(\omega_k(t_1-t_2)).
    \end{aligned}
\end{equation}

According to the above expression, the main effect we can observe on the quantum optical state comes from the radiation generated by the electron when it freely oscillates in the field, described by $\delta_{\vb{k},\mu}(\vb{v},t,t_0)$. We now use this solution to the homogeneous equation in order to find the solution to the inhomogeneous one shown in Eq.~\eqref{eq:app:continuum:Sch:eq}. After introducing the initial conditions, we find
\begin{equation}
    \begin{aligned}
    b(\vb{p},t) \ket{\Phi(\vb{p},t)}
		&= - \dfrac{i}{\hbar} \int^t_{t_0} \dd t'
			e^{-\tfrac{i}{\hbar}\int^t_{t_0} \dd \tau
				\tfrac{1}{2m}
				\Big(
					\vb{p} + \tfrac{e}{c}\vb{A}(\tau)
				\Big)^2}
			\Tilde{D}
			\big(
				\boldsymbol{\delta}(\vb{p},t,t')
			\big)
			\big(\vb{E}_{\text{cl}}(t') + \hat{\vb{E}}(t')\big)
			\cdot \vb{d}
				\Big(
					{\vb{p}+\dfrac{e}{c}\vb{A}(t')}
				\Big)
			a(t')
		    \ket{\Phi_\text{g}(t')},
    \end{aligned}
\end{equation}
where we have expressed the kinetic momentum of the electron in terms of the canonical momentum. We have also expressed the dipole matrix element connecting the ground state with a continuum state as $\vb{d}(\vb{p}+(e/c)\vb{A}(t))$.

\subsubsection{Conditioning onto the ground state and characterization of the final state}	
Similarly to what we had before, the differential equation we get after projecting onto the ground state is given by
\begin{equation}
    \begin{aligned}
	i\hbar \pdv{}{t}
		\big(
			a(t) \ket{\Phi_{\text{g}}(t)}
		\big)
		&= -I_p a(t) \ket{\Phi_{\text{g}}(t)}
			+ \int \dd^3 p \ b(\vb{p},t) 
			\big(\vb{E}_{\text{cl}}(t) + \hat{\vb{E}}(t)\big)
				\cdot \vb{d}^*\Big(\vb{p}+\dfrac{e}{c}\vb{A}(t)\Big)
				\ket{\Phi(\vb{p},t)},
	\end{aligned}
\end{equation}
and assuming that the depletion of the ground state is very small such that it remains almost unperturbed, we approximate the previous differential equation by
\begin{equation}
	i\hbar \pdv{}{t}\Big(a(t)\ket{\Phi_{\text{g}}(t)}\Big)
		\approx -I_p a_{\text{g}}(t) \ket{\Phi_{\text{g}}(t)},
\end{equation}
which is solved by
\begin{equation}
	a_{\text{g}}(t)\ket{\Phi_{\text{g}}(t)}
		= e^{-\tfrac{i}{\hbar}I_p(t-t_0)}
			\ket{\bar{0}}.
\end{equation}

Thus, we finally find for our initial ansatz
\begin{align}\label{eq:app:Final:total:state}
    \ket{\psi(t)}
		&= e^{-\tfrac{i}{\hbar}I_p(t-t_0)}
			\ket{\text{g}}\ket{\bar{0}}
			\nonumber\\
			& \quad
				- \dfrac{i}{\hbar} \int \dd^3 \vb{p} \int^t_{t_0} \dd t'
				e^{-\tfrac{i}{\hbar}S(\vb{p},t,t')}
				\Tilde{D}
				\big(
					\boldsymbol{\delta}(\vb{p},t,t')
				\big)
				\big(\vb{E}_{\text{cl}}(t') + \hat{\vb{E}}(t')\big)
				\cdot \vb{d}
					\Big(
						{\vb{p}+\dfrac{e}{c}\vb{A}(t')}
					\Big)
				\ket{\vb{p}+\dfrac{e}{c}\vb{A}(t)}\ket{\bar{0}},
\end{align}
where $S(p,t,t')$ is the semiclassical action defined in Eq.~\eqref{Eq:semiclassical:action}.

\subsubsection{Undoing the displacement operation}
The state presented in Eq.~\eqref{Eq:Final:total:state} is defined in a displaced frame of reference. Thus, we now proceed to undo the initial transformation and look for the state of the system in the original frame of reference. In the following, we consider the transformation on the part of such state that has been already ionized, since the transformation acting upon the part for which the electron remains in the ground state is trivial. This way, we get
\begin{equation}\label{eq:app:Approx:ionized}
	\begin{aligned}
	\ket{\psi_\text{ion}(t)}
		&= - \dfrac{i}{\hbar} \int \dd^3 p \int^{t}_{t_0} \dd t'
			e^{-\tfrac{i}{\hbar}S(\vb{p},t,t')}
				\Tilde{D}
				\big(
					\boldsymbol{\delta}(\vb{p},t,t')
				\big)
				\Big(
					\prod_{\vb{k},\mu}
						e^{i \Im{\alpha\delta^*_{\vb{k},\mu}(\vb{p},t,t')}}
				\Big)
				\\
				&\quad\hspace{2cm}
				\times
				\vb{d}
					\Big(
						{\vb{p}+\dfrac{e}{c}\vb{A}(t')}
					\Big)
				\cdot
				\Bigg[
					\vb{E}_{\text{cl}}(t')
					\ket{\vb{p}+\dfrac{e}{c}\vb{A}(t)}
						\bigotimes_{\vb{k},\mu\in \text{IR}}
						\ket{{\alpha}_{\vb{k},\mu}}
						\bigotimes_{\vb{k},\mu\in \text{HH}}
						\ket{0_{\vb{k},\mu}}
					\\& \hspace{5cm}
					+ \sum_{\vb{k},\mu} \sqrt{\dfrac{\hbar\omega_k}{2\epsilon_0 V}} 
						\boldsymbol{\epsilon}_{\vb{k},\mu}e^{i\omega_k t'}
						\bigg[
							\prod_{\vb{k}',\mu'\in \text{IR}}D(\alpha_{\vb{k}',\mu'})
						\bigg]
						\ket{1_{\vb{k},\mu}}
						\bigotimes_{\vb{k}'',\mu''\neq\vb{k},\mu}
						\ket{0_{\vb{k''},\mu''}}
				\Bigg],
	\end{aligned}
\end{equation}
where we have a first contribution (inside the brackets of the second line) coming from the input electric field acting at the ionization time, while the other terms incorporate weak quantum optical fluctuations. In the context of this document, we are working within the strong-field regime, where the amplitude of the input electric field is in the order of $10^{7}$ V/cm or larger. For this reason, we expect the first term to be the dominant one and, consequently, we approximate the previous state by
\begin{equation}\label{Eq:App:ATI:state}
	\begin{aligned}
	\ket{\psi_\text{ion}(t)} &\simeq
		- \dfrac{i}{\hbar}
		\bigg[
			\prod_{\vb{k}',\mu'\in \text{IR}}D(\alpha_{\vb{k}',\mu'})
		\bigg]
		\int \dd^3 p \int^{t}_{t_0} \dd t'
			e^{-\tfrac{i}{\hbar}S(\vb{p},t,t')}
				\Tilde{D}
				\big(
					\boldsymbol{\delta}(\vb{p},t,t')
				\big)
				\vb{d}
					\Big(
						{\vb{p}+\dfrac{e}{c}\vb{A}(t')}
					\Big)
				\cdot
					\vb{E}_{\text{cl}}(t')
					\ket{\vb{p}+\dfrac{e}{c}\vb{A}(t)}
						\bigotimes_{\vb{k},\mu}
						\ket{0_{\vb{k},\mu}},
	\end{aligned}
\end{equation}
where in this last expression we have further moved the displacement characterizing the amplitude of the input field in front of the state.

\section{Computing the ionization times according to the semiclassical framework}\label{App:Ionization:Times}
\begin{figure}
    \centering
    \includegraphics[width = 0.6\columnwidth]{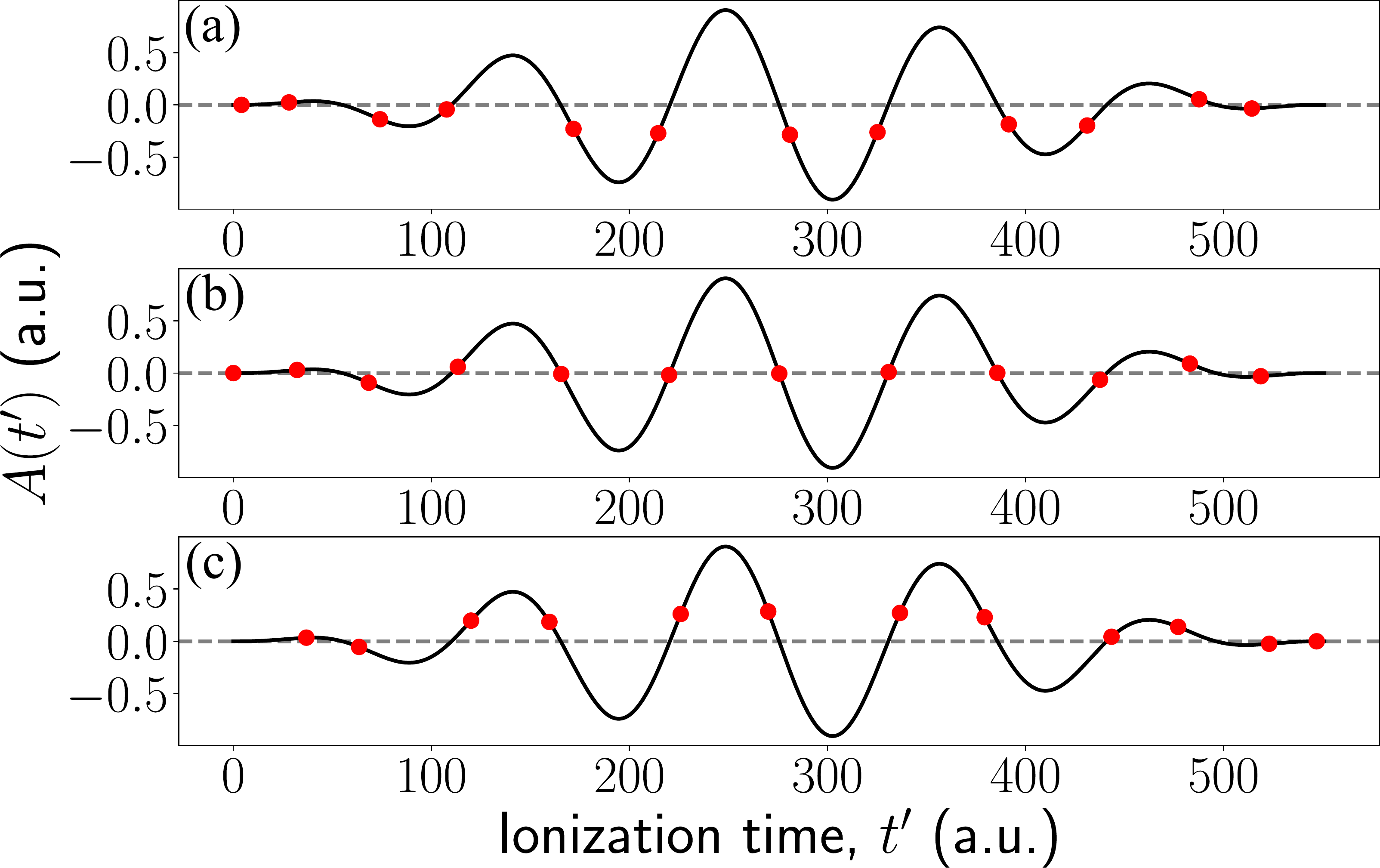}
    \caption{Real part of the ionization times (red circles), found by solving numerically Eq.~\eqref{Eq:app:ionization:time} for a 1D hydrogen system with $I_p = 0.5$ a.u., which is excited by a 5-cycle electromagnetic field with $\omega_L = 0.057$ a.u. and $E_0 = 0.053$ a.u. for the field's amplitude. In particular, we show three cases, (a) $p= 0.43$ a.u., (b) $p = 0$ a.u. and (c) $p=-0.43$ a.u., in the different subplots. The black solid line shows the vector potential of the field $A(t)$ evaluated at all possible ionization times.}
    \label{Fig:Ionization:times}
\end{figure}

The semiclassical expressions can be obtained from the quantum optical description provided in Sec.~\ref{Sec:Theory:background} by setting $\delta_{\vb{k},\mu}(\vb{p},t,t') = 0$ and tracing out the quantum optical degrees of freedom, which would lie in a vacuum state. Thus, the probability amplitude of finding an electron in the continuum is then given by
\begin{equation}
    \Tilde{M}(\vb{p},t) 
    = \int \dd t'     
        e^{-\tfrac{i}{\hbar}S(\vb{p},t,t')}
			\vb{d}
					\Big(
						{\vb{p}+\dfrac{e}{c}\vb{A}(t')}
					\Big)
				\cdot
					\vb{E}_{\text{cl}}(t'),
\end{equation}
and therefore the associated probability can be found by computing $\lvert \Tilde{M}(\vb{p},t)\rvert^2$. In order to evaluate this expression, we take into account that the phase factor is a highly oscillating function, which motivates the use of the saddle-point approximation \cite{lewenstein_theory_1994}. In order to apply this method, we need to solve the saddle-point equation defined by
\begin{equation}\label{Eq:app:ionization:time}
    \pdv{S(\vb{p},t,t')}{t'}\bigg\lvert_{t'=t_\text{ion}}
        \hspace{-0.5cm}= 0
        \Rightarrow 
            \bigg[
                \vb{p}
                + \dfrac{e}{c}\vb{A}(t_\text{ion})
            \bigg]^2 + I_p = 0,
\end{equation}
which defines the ionization time of an electron that undergoes a tunneling process. In particular, we note that Eq.~\eqref{Eq:app:ionization:time} can only be solved for $t_\text{ion} \in \mathbbm{C}$, such that $t_\text{ion}$ corresponds to the time where the electron \emph{enters the barrier}, and its real part corresponds to the time at which the electron appears in the continuum, i.e. it \emph{exits the barrier}~\cite{olga_simpleman}. While for monochromatic laser fields this equation can be solved exactly, for fields with limited duration in time (pulses) it needs to be tackled numerically. The real part of the ionization times found for the system we consider (hydrogen atom with $I_p = 0.5$ a.u. excited by a 5-cycle electromagnetic field with $\omega_L = 0.057$ a.u. and $E_0 = 0.053$ a.u.) is shown in Fig.~\ref{Fig:Ionization:times} (red circles). In particular, we consider three cases (a) $p= 0.43$ a.u., (b) $p = 0.00$ a.u. and (c) $p=-0.43$ a.u., in the different subplots. As we can see, for positive values of the canonical momentum the real part of the ionization time is located in regions where $A(\Re{t_\text{ion}}) < 0$, while for negative values it is located in regions where $A(\Re{t_\text{ion}})>0$. Finally, when $p=0.00$ a.u. the ionization times are located at $A(\Re{t_\text{ion}}) \approx 0$.

\section{Wigner function computation: analytical expression and numerical procedure}\label{App:Sec:Wigner}
In this appendix, we compute the analytic expression of the Wigner function for the states shown in Eqs.~\eqref{Eq:QO:parts}, when conditioned to a single value of the electron momentum. Furthermore, we present the details of the numerical analysis for obtaining the plots shown in Figs.~\ref{Fig:Wigner:pn:oe} and \ref{Fig:Wigner:diff:mom}. The numerical implementation has been entirely performed in Python and can be found in \cite{ZenodoLink}.

\subsection{Analytical expression}
According to~\cite{royer_wigner_1977}, the Wigner function of a quantum state $\hat{\rho}$ can be written as follows
\begin{equation}\label{eq:def:wigner}
	W(\beta) = \dfrac{2}{\pi}\tr(D(\beta)\Pi D(-\beta) \hat{\rho}),
\end{equation}
where $\Pi$ is the parity operator, and $\beta$ is a complex quantity whose real and imaginary parts characterize each of the quadratures in the photonic phase space. In our case, we are interested in the case where the quantum state $\hat{\rho}$ is given by
\begin{equation}
	\hat{\rho} = \dyad{\psi_\text{ATI}(p,t)},
\end{equation}
where $\ket{\psi_\text{ATI}(p,t)}$ is the quantum state of the field when conditioned to ATI processes, and when looking at a single value of the canonical electron momentum $p$. More explicitly, under the approximations considered in the main text and restricting ourselves to a single mode analysis for the input coherent state and a linearly polarized field, this state is given by
\begin{equation}
	\ket{\psi_\text{ATI}(p,t)}
		= D(\alpha)\int^t_{t_0} \dd t'
				\Tilde{M}(p,t')
				\bigotimes_{\vb{k}}
				  e^{i\varphi_{\vb{k},\mu}(p,t')} \ket{\delta_{\vb{k}}(p,t,t')}.
\end{equation}

Introducing this last expression in the definition of the Wigner function given in \eqref{eq:def:wigner}, we get
\begin{equation}
	\begin{aligned}
	W(\beta) 
		&= \int^t_{t_0} \dd t_1\int^t_{t_0} \dd t_2
			M^*(p,t_1)M(p,t_2) C_\text{HH}(p,t,t_1,t_2)
			e^{i(\varphi_{\vb{k}_L,\mu}(p,t_2)-\varphi_{\vb{k}_L,\mu}(p,t_1))}
			\\&\hspace{2.5cm}\times
			\mel{0}{D^\dagger\big(\delta(p,t,t_1)\big) D^\dagger(\alpha) D(\beta)
					\Pi
					D(-\beta)D(\alpha)D\big(\delta(p,t,t_2)\big)}{0},
	\end{aligned}
\end{equation}
where we have defined
\begin{equation}
    \begin{aligned}
    C_\text{HH}(p,t,t_1,t_2)
        = \prod_{\vb{k}\neq \vb{k}_L}
            \braket{\delta_{\vb{k}}(p,t,t_1)}{\delta_{\vb{k}}(p,t,t_2)}e^{i(\varphi_{\vb{k},\mu}(p,t_2)-\varphi_{\vb{k},\mu}(p,t_1))},
    \end{aligned}
\end{equation}
and $\delta_{\vb{k}_L}(p,t,t_i) \equiv \delta(p,t,t_i)$. If we now introduce in our expressions $\Tilde{\beta} = \beta-\alpha$, we can write our Wigner function as 
\begin{equation}\label{Eq:wigner:intermediate}
	\begin{aligned}
	W(\Tilde{\beta}) 
		&= \int^t_{t_0} \dd t_1\int^t_{t_0} \dd t_2
			M^*(p,t_1)M(p,t_2)
			C_\text{HH}(p,t,t_1,t_2)e^{i(\varphi_{\vb{k}_L,\mu}(p,t_2)-\varphi_{\vb{k}_L,\mu}(p,t_1))}
			\\ &\hspace{3cm}\times
			\mel{0}{D^\dagger\big(\delta(p,t,t_1)\big)D(\Tilde{\beta}
					\Pi
					D(-\Tilde{\beta})D\big(\delta(p,t,t_2)\big)}{0}
		\\
		&= \int^t_{t_0} \dd t_1\int^t_{t_0} \dd t_2
			M^*(p,t_1)M(p,t_2)
			C_\text{HH}(p,t,t_1,t_2)e^{i(\varphi_{\vb{k}_L,\mu}(p,t_2)-\varphi_{\vb{k}_L,\mu}(p,t_1))}
			w(\Tilde{\beta}, \delta_1,\delta_2),
	\end{aligned}
\end{equation}
where $\delta_i$ is a shorthand notation for $\delta(p,t,t_i)$, and $w(\Tilde{\beta},\delta_1,\delta_2)$ another simplified notation for the matrix element shown after the first equality in \eqref{Eq:wigner:intermediate}. Furthermore, this expression show us that the Wigner function shape remains unperturbed upon the performance of an unitary operation acting over the whole quantum state. This is not the case of other quantum optical observables like the photon number probability distribution.

We present now some properties of the displacement operator \cite{ScullyBook,Gerry__Book_2001}
\begin{equation}
	D(\beta)D(\delta) = e^{\tfrac12(\beta \delta^*-\beta^*\delta)}
					D(\beta + \delta),
\end{equation}
which allow us to express the matrix element in $w(\Tilde{\beta},\delta_1,\delta_2)$ as
\begin{equation}
\label{wbeta}
	\begin{aligned}
	w(\Tilde{\beta},\delta_1,\delta_2)
		&= e^{\tfrac12[\Tilde{\beta}^*(\delta_2 - \delta_1)-\Tilde{\beta}(\delta_2 - \delta_1)^*]}
		   \mel{0}{D(\Tilde{\beta}-\delta_1)\Pi D(-\Tilde{\beta}+\delta_2)}{0},
	\end{aligned}
\end{equation}
and introducing the following properties of the parity operator $\Pi$
\begin{equation}
	\Pi\Pi = \mathbbm{1},
	\quad \quad\Pi D(\alpha)\Pi = D(-\alpha),
	\quad\quad \Pi \ket{0} = \ket{0},
\end{equation}
we can write Eq.~\eqref{wbeta} as
\begin{equation}
	\begin{aligned}
	w(\Tilde{\beta},\delta_1,\delta_2)
		&= e^{\tfrac12[\Tilde{\beta}^*(\delta_2 - \delta_1)-\Tilde{\beta}(\delta_2 - \delta_1)^*]}
		   \mel{0}{D(\Tilde{\beta}-\delta_1) D(\Tilde{\beta}-\delta_2)}{0}\\
		& = e^{\tfrac12[\Tilde{\beta}^*(\delta_2 - \delta_1)-\Tilde{\beta}(\delta_2 - \delta_1)^*]}
			e^{\tfrac12[(\Tilde{\beta}-\delta_1)(\Tilde{\beta}-\delta_2)^*
				- (\Tilde{\beta}-\delta_1)^*(\Tilde{\beta}-\delta_2)]}
			\braket{0}{2\Tilde{\beta} - \delta_1-\delta_2}\\
		& = e^{\Tilde{\beta}^*(\delta_2 - \delta_1)-\Tilde{\beta}(\delta_2 - \delta_1)^*}
			e^{\tfrac12(\delta_1\delta_2^* -\delta_1^*\delta_2)}
			e^{-\tfrac12 \lvert 2 \Tilde{\beta} - \delta_1-\delta_2\rvert^2}.
	\end{aligned}
\end{equation}

Thus, writing everything together, we find for the final expression of the Wigner function as
\begin{equation}\label{App:Final:Wigner:Expression}
    \begin{aligned}
	W(\Tilde{\beta})
		&= \int^t_{t_0} \dd t_1\int^t_{t_0} \dd t_2
			M^*(p,t_1)M(p,t_2)
			C_\text{HH}(p,t,t_1,t_2)e^{i(\varphi_{i,\vb{k}_L,\mu}(p,t_2)-\varphi_{i,\vb{k}_L,\mu}(p,t_1))}
			\\&\hspace{2.5cm}\times
			e^{\Tilde{\beta}^*(\delta_2 - \delta_1)-\Tilde{\beta}(\delta_2 - \delta_1)^*}
			e^{\tfrac12(\delta_1\delta_2^* -\delta_1^*\delta_2)}
			e^{-\tfrac12 \lvert 2 \Tilde{\beta} - \delta_1-\delta_2\rvert^2}.
	\end{aligned}
\end{equation}

\subsection{Numerical procedure: the saddle-point approximation}\label{App:Sec:Wigner:Saddles}
In the Wigner function presented in Eq.~\eqref{App:Final:Wigner:Expression}, we have contributions from two kind of terms. On the one hand, we have the semiclassical terms which are provided by the probability amplitudes $M(p,t)$. On the other hand, we have the quantum optical terms which are provided by the other terms appearing in the expression. Both of them contribute with a certain phase to the integrals. In particular, the semiclassical terms provide a phase which depends on the semiclassical action $S(\vb{p},t,t')$ and that scales with $\sqrt{U_p}$, while the quantum optical ones provide a phase that depends on the displacement and which scales as $\lvert\delta_{\vb{k},\mu}(p,t,t')\rvert^2$. For the range of laser parameters we work with in Fig.~\ref{Fig:Wigner:pn:oe}, we have that $\sqrt{U_p}\sim 10$ while $\lvert\delta_{\vb{k},\mu}(p,t,t')\rvert \sim 10^{-1}$. Thus, we expect the semiclassical phase to play a dominant role in the phase of the integrand. Thus, for the sake of simplicity, we rewrite our integral as
\begin{equation}
    \begin{aligned}
	W(\Tilde{\beta})
		&= \int^t_{t_0} \dd t_1\int^t_{t_0} \dd t_2
			\Tilde{M}(p,t_1,t_2,\Tilde{\beta})
			e^{\tfrac{i}{\hbar}S(p,t,t_1)}
		    e^{-\tfrac{i}{\hbar}S(p,t,t_2)},
	\end{aligned}
\end{equation}
where we have explicitly separated the dominant phase terms from the rest, which has been compressed in the complex function $\Tilde{M}(p,t_1,t_2,\Tilde{\beta})$. Thus, since this function changes slowly in comparison to the highly oscillatory term, we perform the saddle-point approximation in order to compute the integrals, such that we write
\begin{equation}
   W(\Tilde{\beta})
        \simeq
        \sum_{t_{1,\text{ion}},t_{2,\text{ion}}}
            \sqrt{\dfrac{2\pi}{i\abs{S''(p,t,t_{1,\text{ion}})}}}
            \sqrt{\dfrac{2\pi i}{\abs{S''(p,t,t_{2,\text{ion}})}}}
            \Tilde{M}(p,t_{1,\text{ion}},t_{2,\text{ion}},\Tilde{\beta})
           e^{\tfrac{i}{\hbar}S(p,t,t_{1,\text{ion}})}
		    e^{-\tfrac{i}{\hbar}S(p,t,t_{2,\text{ion}})}.
\end{equation}

In this last expression, $t_{i,\text{ion}}$ are the ionization times computed from the evaluation of the saddle-points as presented in Eq.~\eqref{Eq:app:ionization:time}. Since the two phases are exactly the same, the ionization times $t_{i,\text{ion}}$ coincide for $i=1$ and $i=2$, although in the sum we have to consider all possible combinations.

We note that the same approach can be done for the evaluation of the overlap between the quantum optical states shown in Eq.~\eqref{Eq:QO:parts}. The only difference appears in the definition of the $\Tilde{M}(p,t_1,t_2,\Tilde{\beta})$, which instead of having a Wigner element for the fundamental mode, this function is replaced by an overlap between two coherent states.

\subsection{Numerical procedure: using a numerical integrator}\label{App:Sec:Wigner:Numerical}
Unlike the Wigner function plots presented in Fig.~\eqref{Fig:Wigner:pn:oe}, in Fig.~\eqref{Fig:Wigner:diff:mom} we are working with a different regime of laser parameters and, more importantly, we are considering the phenomenological contribution of $N$ atoms participating in the process. Therefore, the requirements for applying the saddle-point approximation as done in the previous subsection are not met now. Thus, in order to perform these plots, we instead opted for a full numerical approach where the integration is done with numerical approaches. In particular, we used the \texttt{nquad} integration routine defined in \texttt{SciPy} \cite{2020SciPy-NMeth} in order to perform the double integration shown in Eq.~\eqref{App:Final:Wigner:Expression}. With the aim of speeding up the code, we used the \texttt{numba} package \cite{NumbaRed} which accelerates the evaluation of the different functions needed for computing the integral. 

\begin{figure}
    \centering
    \includegraphics[width=1\columnwidth]{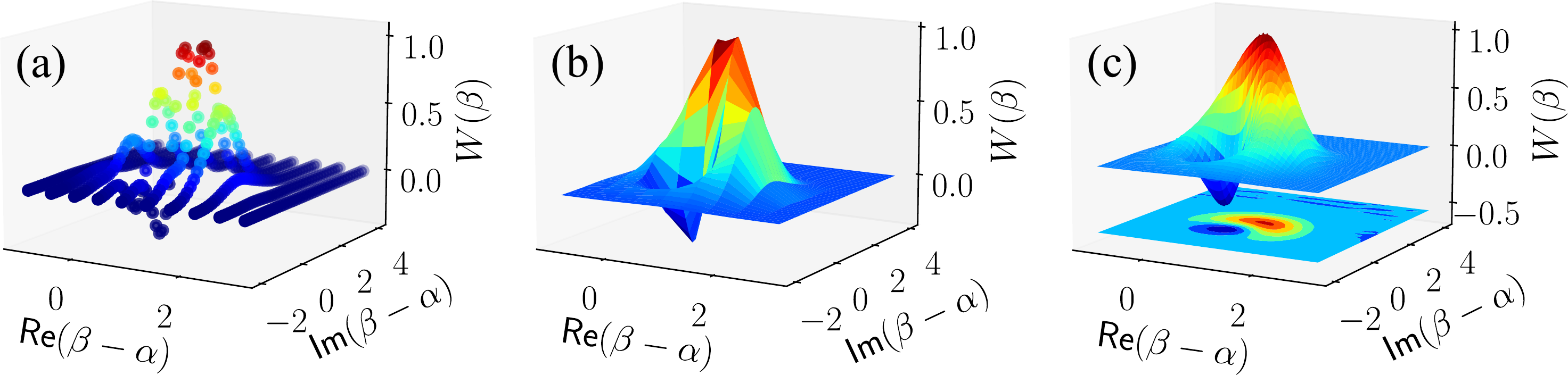}
    \caption{Evolution of the Wigner function (for $p=0.00$ a.u. and $N\sim10^4$) throughout each of the different processing steps. We initially considered a grid of size $20\times 40$ for which we evaluated the Wigner function according to Eq.~\eqref{App:Final:Wigner:Expression}. This leads to subplots (a) and (b), which respectively show a scatter and a surface plot of the obtained data. Afterwards, in order to smooth the obtained functions, we use an interpolation scheme which is evaluated over a grid of size $500 \times 500$ over the range defined by the initial grid. The corresponding output is shown in subplot (c). In these plots, we have normalized the Wigner function to its maximum value.}
    \label{Fig:App:Wigner:process}
\end{figure}

In general, the functions that appear within the integral are complex and in principle, the \texttt{nquad} function does not admit the evaluation of complex integrands, which means that the real and imaginary parts of the integrand have to be evaluated separately. However, since the Wigner function is real as it describes a quasiprobability distribution, for the numerical analysis we first checked within a grid that this statement was satisfied (as a sanity check), but in order to generate the plots we avoid the integration over the complex part. With this said, the generation of the plots consists of two parts:
\begin{enumerate}
    \item First, we generate a 2D grid of points, namely $\{x_0, \dots, x_{m}\}$ and $\{y_0,\dots,y_{n}\}$, where $n\neq m$ in general, which define the real and imaginary parts of $\beta$, i.e. $\beta \equiv x + i y$. For each of these points, we numerically perform the integral shown in Eq.~\eqref{App:Final:Wigner:Expression}. The evaluation has been done in a single 2 GHz-CPU core, and the same process has been performed in parallel for different values of the canonical momentum in each of the remaining cores. For an initial grid of size $10\times40$, the evaluation of the Wigner function takes around a day (for $p=0.00$ a.u. and $N=10^4$). Note that the size of the grid, as well as its limits, has to be adapted accordingly depending on the value of the canonical momentum $p$ and the number of atoms $N$ that are considered, such that increasing the values of these two quantities requires bigger matrices with larger limits, and therefore more computational resources. For the considered grid, and using $p=0.00$ a.u. and $N=10^4$ together with a 5-cycle linearly polarized pulse with $E_0 = 0.053$ a.u., $\omega_L = 0.057$ a.u. and a sinusoidal squared envelope, we find Fig.~\ref{Fig:App:Wigner:process}~(a) which shows the value of the Wigner function in each of the evaluated points such that the corresponding surface plot is shown in Fig.~\ref{Fig:App:Wigner:process}~(b). Note that we have normalized the Wigner function to its maximum in the studied region.
    
    \item The second part consists of smoothing the Wigner function plot. This can be done in the \emph{hard} but exact way, or in the \emph{easy} but less accurate way. The first implies increasing the integration grid, such that more points are introduced in the evaluation of the Wigner function itself. Thus, we get more points in the plot and thus we obtain an exact way of smoothing the plot. However, the main drawback here is that a larger number of points implies more computational resources. Thus, the less exact alternative, but more flexible, approach is to perform an interpolation of the points we have already calculated for the initial grid. This way, and within the considered range, we can artificially increase the number of evaluation points (in our case we move from a grid of $10\times40$ to another one of size $500\times500$ within the same limits) without the need of performing again the numerical integration in Eq.~\eqref{App:Final:Wigner:Expression}. In order to implement this feature, we used the \texttt{interpolate.griddata} function provided by the \texttt{SciPy} package, which allows to perform this interpolation according to different methods for two dimensional data. Note that, the finer the initial grid is, the more exact would be the interpolation scheme. Thus, different values of momentum $p$ and number of atoms $N$ require a different number of initial evaluation points in order for this approach to be valid. After the smoothing, we get Fig.~\ref{Fig:App:Wigner:process}~(c). 
\end{enumerate}

\section{Entropy of entanglement}\label{App:Entanglement:Expressions}
In this section of the appendix, we explicitly compute the analytic expression for the entropy of entanglement presented in the main text. Our starting point is the light-matter state given in Eq.~\eqref{Eq:State:Orthonormal}, which is written in the orthonormal basis $\{\ket{p},\ket{-p}\}\otimes \{\ket{u},\ket{v}\}$. Note that this basis representation allow us to treat our state as effectively lying in a $2\otimes2$ Hilbert space, which then allow us for a simple characterization of the entanglement by means of the entropy of entanglement measure. In order to do so, we first compute the reduced density matrix with respect to one of the subsystems. Here, we trace out the $\{\ket{u},\ket{\nu}\}$ modes, such that reduced state reads
\begin{equation}
    \hat{\rho} 
        = \dfrac{\mathcal{N}_+}{\mathcal{N}} \dyad{p}
            + \dfrac{\mathcal{N}_-}{\mathcal{N}} \dyad{-p}
            + e^{-i\theta}(\mu^2 - \nu^2)
                \dfrac{\sqrt{\mathcal{N}_+\mathcal{N}_-}}{\mathcal{N}}
                \dyad{p}{-p}
            + e^{i\theta}(\mu^2 - \nu^2)
                \dfrac{\sqrt{\mathcal{N}_+\mathcal{N}_-}}{\mathcal{N}}
                \dyad{-p}{p}.
\end{equation}

Here, the associated Schmidt matrix is given by
\begin{equation}
    \mathcal{S}
        = \dfrac{1}{\mathcal{N}}
            \mqty(\mathcal{N}_+ & e^{-i\theta}(\mu^2 - \nu^2)
                \sqrt{\mathcal{N}_+\mathcal{N}_-}\\
                e^{-i\theta}(\mu^2 - \nu^2)
                \sqrt{\mathcal{N}_+\mathcal{N}_-} & \mathcal{N}_-),
\end{equation}
and whose eigenvalues are given by
\begin{equation}
    \lambda_{\pm}
        = \dfrac12
            \Bigg[
                1 \pm \sqrt{1+16\nu^2(\nu^2-1)\dfrac{\mathcal{N}_+\mathcal{N}_-}{\mathcal{N}}}
            \Bigg].
\end{equation}

From this expression, we see that the amount of entanglement in our state depends on two elements: the overlap among the distinct continuous variable components, and on the relative population between the $\ket{+}$ and $\ket{-}$ states. In particular, if the overlap between the two states tends to one, then $\nu \to 0$ which leads to $\lambda_+ = 1$ and $\lambda_- = 0$. On the other hand, in the case they do not overlap at all, but the relative population is completely unbalanced, then we recover the definition of a pure state and we get $\lambda_+ = 1$ and $\lambda_- = 0$. Likewise, if the superposition is completely balanced, the generated state is maximally entangled. 

\end{document}